\documentclass[
aps,physrev,twocolumn,groupedaddress,
nofootinbib,amsmath,amssymb
]{revtex4-2}

\usepackage{orcidlink}
\usepackage{bm}
\usepackage{dcolumn}
\usepackage{graphicx}

\usepackage{adsmacros}

\newcommand{\ddif}{\mathrm{d}}
\newcommand{\lya}{Ly$\alpha$}
\newcommand{\lyaf}{Ly$\alpha$ forest}

\newcommand{\poned}{$P_{\mathrm{1D}}$}
\newcommand{\pthreed}{$P_{\mathrm{3D}}$}

\newcommand{\skm}{s~km$^{-1}$}
\newcommand{\kms}{km~s$^{-1}$}
\newcommand{\hmpc}{$h^{-1}$~Mpc}

\newcommand{\impc}{Mpc$^{-1}$}

\begin{document}

\title{Unified reconstruction of the Lyman-alpha power spectrum \\with Hamiltonian Monte Carlo}

\author{Naim~G\"oksel \surname{Kara\c cayl\i}\orcidlink{0000-0001-7336-8912}}
\email{karacayli.1@osu.edu}
\affiliation{Center for Cosmology and AstroParticle Physics, The Ohio State University, 191 West Woodruff Avenue, Columbus, OH 43210, USA}
\affiliation{Department of Astronomy, The Ohio State University, 4055 McPherson Laboratory, 140 W 18th Avenue, Columbus, OH 43210, USA}
\affiliation{Department of Physics, The Ohio State University, 191 West Woodruff Avenue, Columbus, OH 43210, USA}
\author{Peter L. Taylor}
\affiliation{Center for Cosmology and AstroParticle Physics, The Ohio State University, 191 West Woodruff Avenue, Columbus, OH 43210, USA}
\affiliation{Department of Astronomy, The Ohio State University, 4055 McPherson Laboratory, 140 W 18th Avenue, Columbus, OH 43210, USA}
\affiliation{Department of Physics, The Ohio State University, 191 West Woodruff Avenue, Columbus, OH 43210, USA}

\date{\today}

\begin{abstract}
The complex geometry of the Ly$\alpha$ forest data has motivated the use of various two-point statistics as alternatives to the three-dimensional power spectrum ($P_{\mathrm{3D}}$), which carries cosmological information in Fourier space. On large scales, the three-dimensional correlation function ($\xi_\mathrm{3D}$) has provided robust measurements of the baryon acoustic oscillation (BAO) scale at 150~Mpc. On smaller scales, the one-dimensional power spectrum, $P_{\mathrm{1D}}(k_\|)$, has been the primary tool for extracting information. At the same time, the cross-spectrum, $P_\times(\theta, k_\|)$, has been introduced to incorporate angular information without the complications caused by survey window functions. We propose an analytical forward-modeling framework to reconstruct $P_{\mathrm{3D}}$ from all these observables, based on the mathematical relation between them and $P_{\mathrm{3D}}$. We demonstrate the performance of our method using a hypothetical mock data vector representative of future Dark Energy Spectroscopic Instrument (DESI) measurements. We show that the monopole of $P_{\mathrm{3D}}$ can be reconstructed in 25 $k$ bins between $0.07~\mathrm{Mpc}^{-1}$ and $1.8~\mathrm{Mpc}^{-1}$, achieving an average precision of $\sigma_P/P=13\%$ across the bins. Our method can serve as an intermediary for consistency checks, though it is not intended to replace direct $P_{\mathrm{3D}}$ estimation.
\end{abstract}

\maketitle

\section{Introduction}
The \lyaf\ is a collection of absorption lines in quasar spectra that trace the neutral hydrogen gas in the intergalactic medium. This powerful technique is capable of mapping vast volumes of the universe between redshifts two and five with kiloparsec resolution in the line-of-sight direction. However, the \lyaf\ data also has a strongly anisotropic geometry, making the optimal extraction of two-point statistics difficult \cite{slosarMeasurementBaryonAcoustic2013, fontriberaEstimate3DPowerLya2018, belsunce3dLymanAlphaPowerSpectru2024, karacayliOptimal3dEstimator2025}. The lines of sight are densely sampled in the radial direction at kiloparsec resolution, yet sparsely sampled in the transverse direction, limited by the number of targeted quasars per square degree. 

Due to this complex geometry, statistics that are easier to estimate, specifically, the 1D power spectrum (\poned, \cite{croftRecoveryPowerSpectrum1998, mcdonaldLyUpalphaForest2006, palanque-delabrouilleOnedimensionalLyalphaForest2013, chabanierOnedimensionalPowerSpectrum2019, karacayliOptimal1dDesiEdr2023, ravouxFFTP1dEDR2023, karacayliQmleP1dDesiDr12024, ravouxFFTP1dDesiDr12024}) and the 3D correlation function ($\xi_\mathrm{3D}$, \cite{bautistajuliane.MeasurementBaryonAcoustic2017, bourbouxCompletedSDSSIVExtended2020, desiKp6BaoLya2024, desiY3LyaBAO2025}) have been the most prominent measurements, which are sensitive to different scales. \poned\ probes scales smaller than one Mpc, while $\xi_\mathrm{3D}$ can robustly determine the baryon acoustic oscillations (BAO) scale of 150~Mpc. A hybrid statistics called cross-spectrum, $P_\times(\theta, k_\|)$, is proposed to partially overcome the difficulties in survey geometry by being in configuration space in the transverse direction and in Fourier space in the line-of-sight direction \cite{fontriberaEstimate3DPowerLya2018, abdulkarimMeasurementofPcross2024}.
This hybrid statistic precedes a (sub-optimal) \pthreed\ measurement from the extended Baryon Oscillation Spectroscopic Survey (eBOSS) \cite{belsunce3dLymanAlphaPowerSpectru2024} and a recent development of an efficient optimal \pthreed\ estimator \cite{karacayliOptimal3dEstimator2025}. Ultimately, all these statistics are analytically connected to the \lyaf\ \pthreed.
It is then reasonable to ask whether these measured statistics are consistent.

The first measurements of \poned\ were converted to a \pthreed\ measurement using the relation \cite{croftRecoveryPowerSpectrum1998, croftPowerSpectrumMass1999}:
\begin{equation}
    P_\mathrm{3D}^\mathrm{iso}(k) = - \frac{2\pi}{k} \frac{\ddif P_\mathrm{1D} }{\ddif k}.
\end{equation}
Since this differentiation amplifies noise and uncertainties, \citet{chabanierMatterPowerSpectrum2019} employed the total variation regularization technique to stabilize the \pthreed\ reconstruction with the ultimate goal of reconstructing the linear matter power spectrum similar to \citet{tegmarkMatterPowerReconstruction2002}. This approach has three shortcomings: (1) Differentiating noisy data is greatly undesirable. (2) This relation requires a 3D power spectrum to be defined in the same way, yielding an ``isotropic" $P_\mathrm{3D}^\mathrm{iso}(k)$ even though the actual \pthreed$(k, \mu)$ is anisotropic. (3) This intermediate product complicates the recovery of the linear matter power spectrum by coupling different scales.

In this work, we revisit the \lyaf\ \pthreed\ reconstruction using fully descriptive analytical expressions and expand it to utilize all \lyaf\ two-point statistics (except \pthreed\ of course). We first develop our methodology and assess its efficiency on mock data, leaving the application to real data for future work, since contaminants and distortions in the correlation function must be taken into account for a meaningful reconstruction of \pthreed. We base our performance tests loosely on the  Dark Energy Spectroscopic Instrument (DESI, \cite{leviDESIExperimentWhitepaper2013}) \lyaf\ measurements \cite{karacayliOptimal1dDesiEdr2023, desiKp6BaoLya2024}.

Continuing the \poned\ example, instead of using the differential relation, we employ the integral relation between \poned\ and \pthreed:
\begin{equation}
    P_\mathrm{1D}(k_\|) = \int^\infty_{k_\|}\mathrm{d} k ~\frac{k P_{\mathrm{3D}}(k, \mu=k_\| / k)}{2\pi},
\end{equation}
with the additional key realization that only the first few multipoles dominate \pthreed, considerably narrowing down the degrees of freedom. With the advances in auto-differentiation \cite{baydin_automatic_2018} and Hamiltonian Monte Carlo sampling \cite{duaneHybridMonteCarlo1987, betancourt_conceptual_2017}, it is now trivial to forward model this relation and efficiently sample the posterior of \pthreed\ for a given \poned\ observation.
An additional benefit of forward modeling using analytical relations is that the other two-point statistics of the \lya\ forest can be jointly analyzed to reconstruct \pthreed\ across a wide range of scales. These relations are model-independent in the absence of systematics and contaminants, but mildly model-dependent in the presence of such complications. Once reconstructed, this \pthreed\ can serve as a cross-check amongst observations and a future \pthreed\ measurement, or be used for linear matter power spectrum reconstruction or the usual cosmological inference.

The outline of this paper is as follows. Section~\ref{sec:p3dmodel} describes our input model for the \lyaf\ power spectrum based on a fitting function. We develop formulas representing the quadrupole-to-monopole and hexadecapole-to-monopole ratios to gain even more precision in reconstruction. Section~\ref{sec:forward} describes our methodology to reconstruct \pthreed\ and test the reconstruction performance starting from \poned\ and building up to a joint analysis using mock data. We reflect on our method and examine directions for future work in Section~\ref{sec:discuss}.

\section{Modelling the Lyman-alpha forest power spectrum\label{sec:p3dmodel}}
We use a fitting function for the input \lya\ forest power spectrum, $P_\mathrm{3D}(k, \mu)$, to develop our framework, noting that the reconstructed power spectrum does not depend on the details of the model adopted here. We employ the fitting function of \citet{arinyoNonLinearPowerLya2015} for \pthreed, which is tested on simulations and improves upon the work of \citet{mcdonaldPredictingLyaPower2003} by reducing the number of free parameters to five.

This function is a standard extension with a non-linear correction term $(F_\mathrm{NL})$ to the linear theory with a bias term and a Kaiser term for the redshift space distortions:
\begin{equation}
    \label{eq:p3d_lss}P_\mathrm{3D}(k, \mu) = b_F^2 (1 + \beta_F \mu^2)^2 P_L(k) F_\mathrm{NL}(k, \mu),
\end{equation}
where $P_L(k)$ is the linear matter power spectrum at redshift $z$. We calculate $P_L(k)$ using \texttt{camb} \cite{lewisCAMB2000} with the best-fit cosmological parameters of \citet{collaborationPlanck2018Results2020}.

Empirical $F_\mathrm{NL}$ functions aim to capture three major deviations from the linear theory: non-linear enhancement due to gravitational collapse, pressure smoothing that suppresses the power at a characteristic scale known as the filtering scale or Jeans scale \cite{gnedinProbingUniverseLya1998}, and thermal broadening of line profiles and non-linear peculiar velocities that smooth the power spectrum anisotropically:
\begin{equation}
    \label{eq:arinyo} \ln F_\mathrm{NL} = q_1 \Delta^2(k) \left[1 - A_\nu^{-1}\frac{(k_\|/k_{\nu, 0})^{b_\nu}}{(k/k_{\nu, 0})^{c_\nu}} \right] - \left( \frac{k}{k_p} \right)^2,
\end{equation}
where $\Delta^2(k) \equiv k^3 P_L(k) / 2\pi^2$, $k_\|\equiv k\mu$, and $k_{\nu, 0} = 1~$\impc.\footnote{We have applied the following transformations to the original fitting function: $c_\nu=b_\nu-a_\nu$ and $A_\nu = k_\nu^{a_\nu}$.} $q_1$ controls the non-linear growth. The pressure smoothing is captured by $k_p$. The thermal broadening and non-linear peculiar velocities are described by parameters $A_\nu, b_\nu,$ and $c_\nu$. Our adopted values are tabulated in Table~\ref{tab:accel2_poly}, which are based on ref.~\cite{karacayliQmleP1dDesiDr12024}. They derive these values from state-of-the-art simulations \cite{chabanierAccel2Simulations2024} and supplement them by best-fitting values to measured \poned\ from DESI DR1.

\begin{table}
    \caption{\label{tab:accel2_poly}Our nominal values for the \pthreed\ fitting function as reported in ref.~\cite{karacayliQmleP1dDesiDr12024}. These are obtained by fitting a polynomial of $\mathcal{Z}=\ln ((1+z)/3.4)$ to the natural logarithm of each parameter reported in ref.~\cite{chabanierAccel2Simulations2024}, except for the $A_\nu$ parameter. This is replaced with the reported best-fitting expression to DESI DR1 measurements in ref.~\cite{karacayliQmleP1dDesiDr12024}.}
    \begin{ruledtabular}
    \begin{tabular}{c|l}
    Parameter & Best-fitting power law \\
    \hline
     $b_F$ & $0.12 \times \exp(3.38 \mathcal{Z})$ \\
     $\beta_F$ & $1.62 \times \exp(-1.34 \mathcal{Z}) $ \\
     $q_1$ & $0.77\times \exp(0.547 \mathcal{Z} + 5.12 \mathcal{Z}^2)$ \\
     $A_\nu$ & $0.40 \times \exp(0.53 \mathcal{Z})$ \\
     $b_\nu$ & $1.64$ \\
     $c_\nu$ &  $1.27 \times \exp(0.38 \mathcal{Z}) $ \\
     $k_p$ & $23.1 \times \exp(-1.60 \mathcal{Z})$ \\
    \end{tabular}
    \end{ruledtabular}
\end{table}

\begin{figure}
    \includegraphics[width=\columnwidth]{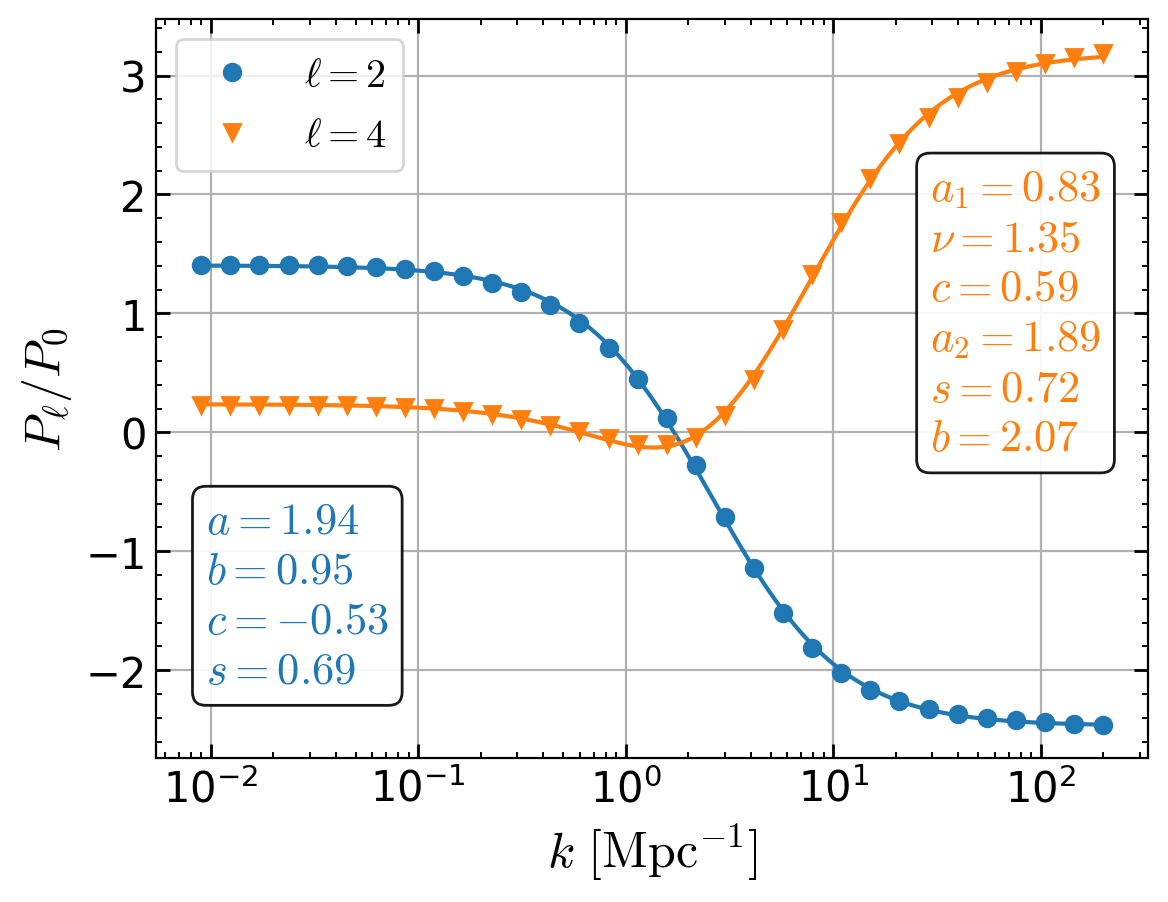}
    \caption{The ratio between higher order multipoles and the monopole based on the fitting function of \lya\ \pthreed. The quadrupole-to-monopole ratio is blue, whereas the hexadecapole-to-monopole ratio is orange. Solid lines follow our proposed fitting functions with the best-fit values, which can be found in the main text.}
    \label{fig:multipole_monopole_ratio}
\end{figure}

\pthreed\ is dominated by the first three even multipoles $\ell=0,2,4$ \cite{kirkbyFittingMethodsBaryon2013, belsunce3dLymanAlphaPowerSpectru2024, karacayliOptimal3dEstimator2025}. Therefore, we focus on reconstructing these multipoles instead of $P_\mathrm{3D}(k, \mu)$ in arbitrary $k$ and $\mu$ bins. The even multipoles are given by an integration over the Legendre polynomials $\mathcal{L}_\ell(\mu)$:
\begin{equation}
    P_\ell(k) = (2\ell + 1) \int_0^1 \ddif \mu ~ \mathcal{L}_\ell(\mu) P_\mathrm{3D}(k, \mu).
\end{equation}
Looking into the ratios $P_2/P_0$ and $P_4/P_0$, we find the following functions describe them exceptionally well for the range of non-linear correction parameters measured from simulations:
\begin{equation}
    \label{eq:quad_mono_ratio}\frac{P_2(k)}{P_0(k)} = -a \tanh[s (\ln k - b)] + c
\end{equation}
for the quadrupole-to-monopole ratio and
\begin{equation}
    \label{eq:hexa_mono_ratio}\frac{P_4(k)}{P_0(k)} = \left(a_1 \mathrm{e}^{-k^\nu} - c\right) + a_2 \{1 + \tanh[s (\ln k - b)]\}
\end{equation}
for the hexadecapole-to-monopole ratio. We calculate this ratio using Eqs.~(\ref{eq:p3d_lss}) and (\ref{eq:arinyo}) with parameters in Table~\ref{tab:accel2_poly} at $z=2.4$ for 32 loglinearly spaced $k$ bins between $9\times10^{-3}\text{~\impc}\leq k \leq 2\times 10^2\text{~\impc}$. Fig.~\ref{fig:multipole_monopole_ratio} demonstrates this ratio and the best-fit values to the two fitting functions above.
The validity of these functional forms breaks down when $b_\nu$ approaches or becomes less than $c_\nu$, which is not the case according to simulation analyses \cite{chabanierAccel2Simulations2024, arinyoNonLinearPowerLya2015}. We present the exact analytical expressions in Appendix~\ref{app:analytic_multipoles}.

\begin{figure}
    \includegraphics[width=\columnwidth]{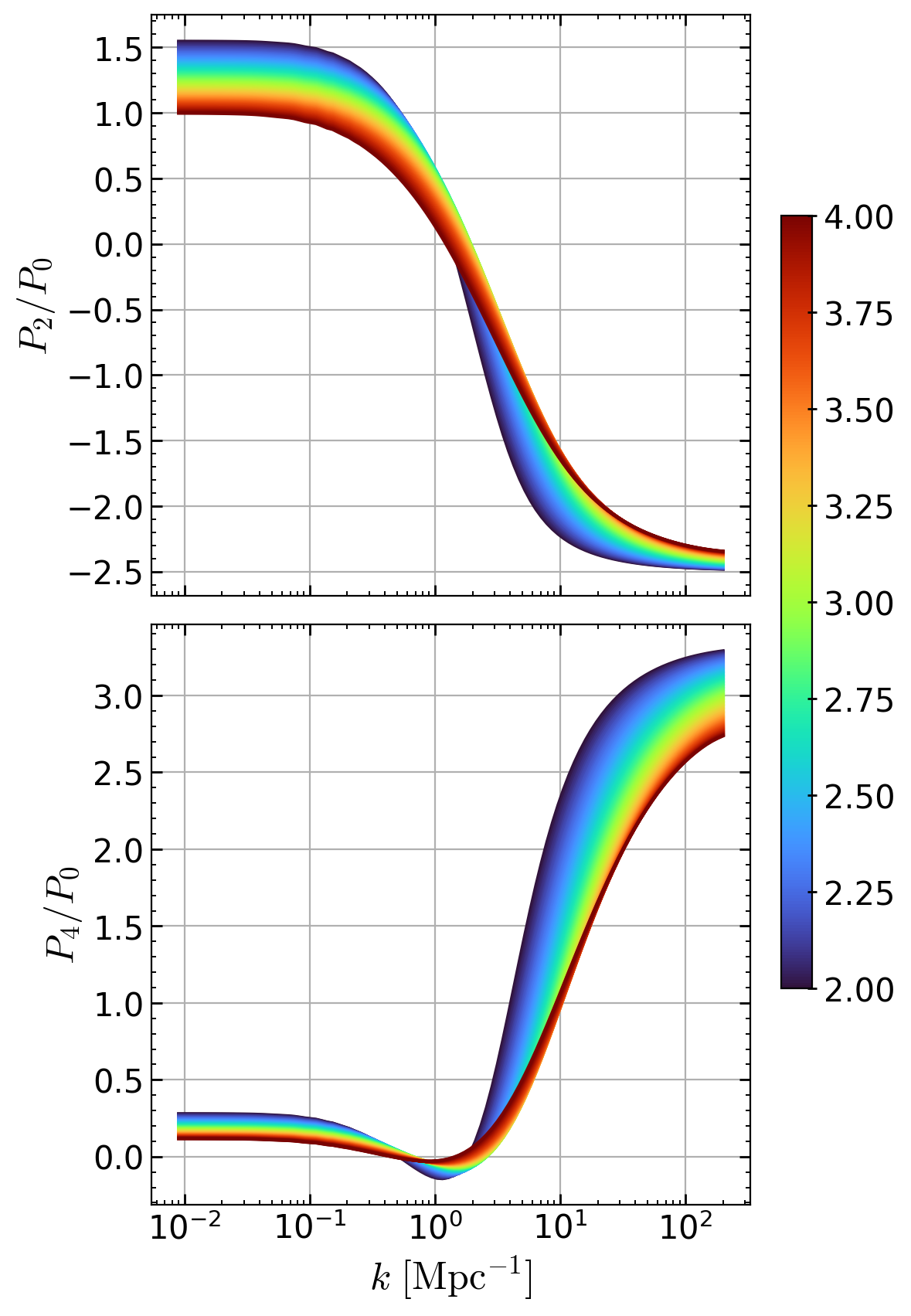}
    \caption{The quadrupole-to-monopole ratio ({\it top}) and the hexadecapole-to-monopole ratio  ({\it bottom})  for the redshift range $2<z<4$. 
    Our fitting functions hold for the entire redshift range.}
    \label{fig:quad_hexa_mono_ratios_z}
\end{figure}

Assuming that these relations generalize to all physical scenarios, we will test leveraging these relations by scaling the monopole contribution instead of estimating the quadrupole and hexadecapole in arbitrary $k$ bins. This reduces the degrees of freedom and improves the constraining power of our forward model. Of course, our formulation has more parameters than Eq.~(\ref{eq:arinyo}). Noting only the $\beta_F, q_1A_\nu^{-1}, b_\nu, ~\text{and}~c_\nu$ parameters source the anisotropy, there must be constant surfaces in our parameter space ($f(\bm\theta)=\text{const}$) that can be used to reduce the degrees of freedom further. Let us illustrate this point with two examples: (1) At low $k$, the quadrupole-to-monopole ratio becomes $a+c$, which goes to the linear theory expectation that only depends on $\beta_F$. The same limit applies to the hexadecapole-to-monopole ratio. (2) At high $k$, both ratios approach a constant value across all redshifts, which means $c-a=\text{const}$ for Eq.~(\ref{eq:quad_mono_ratio}) and $2a_2 - c=\text{const}$ for Eq.~(\ref{eq:hexa_mono_ratio}). These can be seen in Fig.~\ref{fig:quad_hexa_mono_ratios_z}. We do not attempt to trim down our ratio parameter space using these limits, which would incrementally become equivalent to using the fitting function of five free parameters in Eq.~\ref{eq:arinyo}.

The major observed quantities are the one-dimensional power spectrum \poned, the three-dimensional correlation function $\xi_\mathrm{3D}$, and the hybrid quantity $P_\times(\theta, k_\|)$. We now formulate their connection to \pthreed.

\subsection{One-dimensional power spectrum \texorpdfstring{$P_\mathrm{1D}$}{P1D}}
Theoretically, $P_\mathrm{1D}$ is given by an integral of \pthreed\ in the transverse direction:
\begin{equation}
    P_\mathrm{1D}(k_\|) = \int^\infty_{k_\|}\mathrm{d}\ln k ~\frac{k^2 P_{\mathrm{3D}}(k, \mu)}{2\pi},
\end{equation}
where $\mu = k_\| / k$. Using the multipole expansion, \pthreed\ can be written in terms of $P_\ell$s:
\begin{equation}
    \label{eq:p1d_integ}P_\mathrm{1D}(k_\|) = \sum_{\ell} \int^\infty_{k_\|}\mathrm{d}\ln k ~\mathcal{L}_\ell(\mu) \frac{k^2 P_\ell(k)}{2\pi}.
\end{equation}

\poned\ is measured between the velocity scales of $10^{-3}~\text{\skm} < k \lesssim 0.045~\text{\skm}$ using DESI data \cite{karacayliOptimal1dDesiEdr2023, ravouxFFTP1dEDR2023, karacayliQmleP1dDesiDr12024, ravouxFFTP1dDesiDr12024}, which can be converted to \impc\ units by multiplying with $H(z)/(1+z) \approx 70-85 ~\text{\kms~\impc}$ in the redshift range $2.2<z<4.0$. Additionally, the integration reaches its 99.7\% cumulative value at $k=50~\text{\impc}$ for the highest $k_\|$ value of $4~\text{\impc}$. This informs us that the integration starts at $k_\|=0.07~\text{\impc}$ and only scales $0.07~\text{\impc} < k < 50~\text{\impc}$ of $P_\ell$s contribute to \poned.

\begin{figure}
    \includegraphics[width=\columnwidth]{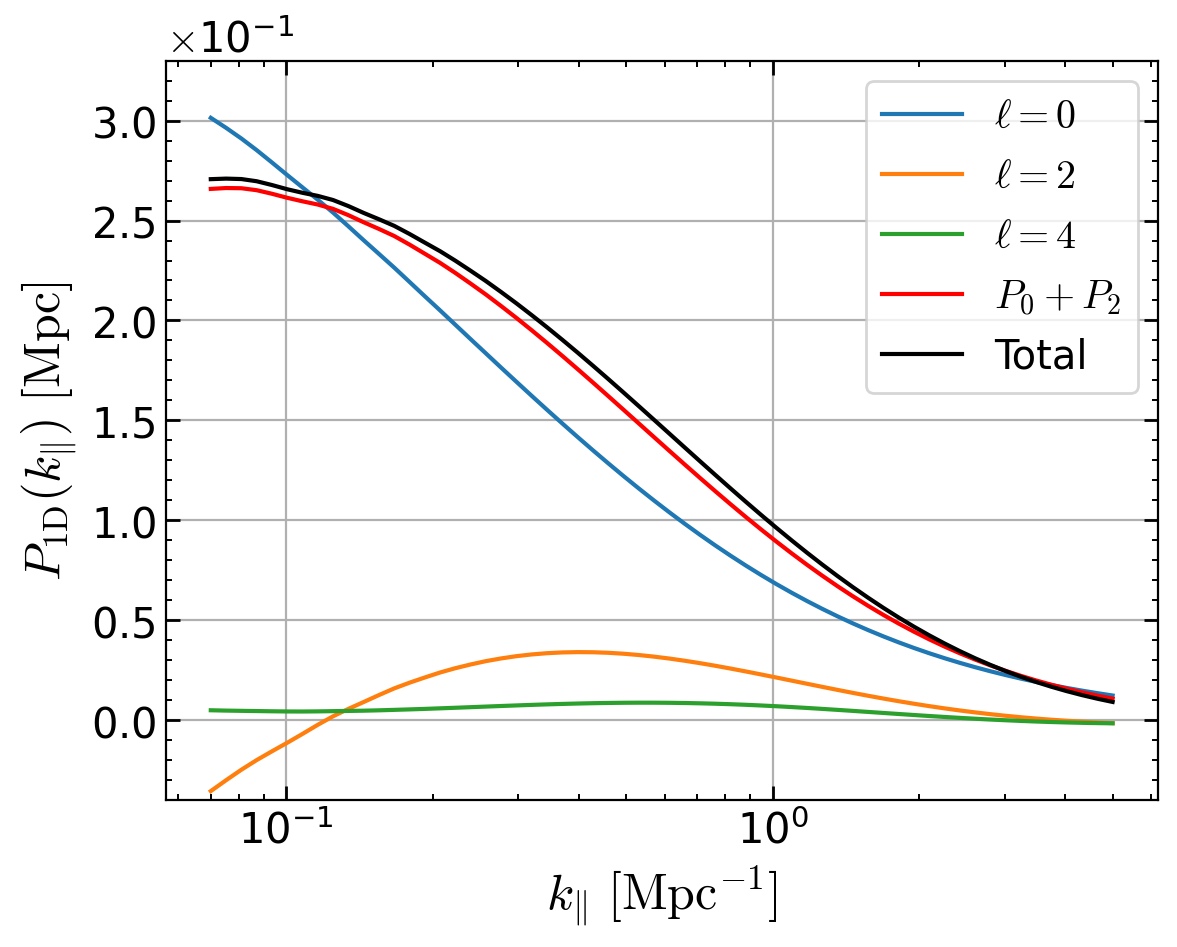}
    \caption{Each multipole's contribution to \poned. The monopole and quadrupole compose the majority of \poned. We ignore the hexadecapole contribution for \poned\ and $P_\times$ analyses.}
    \label{fig:p1d_multipoles}
\end{figure}

Fig.~\ref{fig:p1d_multipoles} illustrates that the first two even multipoles of the power spectrum contribute the most to \poned. For simplicity, we ignore the hexadecapole term for both \poned\ and $P_\times$ analyses.

\subsection{Cross-spectrum \texorpdfstring{$P_\times(\theta, k_\|)$}{Pcross}}
The hydrid statistics $P_\times(\theta, k_\|)$ is given by
\begin{align}
    P_\times(\theta, k_\|) &= \int \frac{\tilde{k}_\bot \ddif \tilde{k}_\bot}{2\pi} J_0(\tilde{k}_\bot\theta) P_\mathrm{3D}(\tilde{k}_\bot, k_\|) \\
    &= \chi^2 \int \frac{k_\bot \ddif k_\bot}{2\pi} J_0(k_\bot \chi \theta) P_\mathrm{3D}(k_\bot, k_\|),
\end{align}
where $\theta$ is the angular separation between sightlines and $\tilde{k}_\bot$ is in inverse angular units \cite{fontriberaEstimate3DPowerLya2018, abdulkarimMeasurementofPcross2024}. In the second line, we used the distant observer and small angular separation approximation to write $\tilde{k}_\bot = k_\bot \chi$, where $\chi$ is the comoving distance to redshift $z$ in Mpc such that $k_\bot$ is in \impc. We write this with multipoles in the FFTLog formalism \cite{talmanNumericalFourierBessel1978, hamiltonUncorrelatedModesNonlinear2000}:
\begin{equation}
    P_\times = \frac{\chi^2}{2\pi r_\bot} \int r_\bot \ddif  k_\bot J_0(k_\bot r_\bot)  k_\bot \sum_{\ell} \mathcal{L}_\ell(\mu)  P_\ell(k),
\end{equation}
defining $r_\bot \equiv \chi \theta$, $k=\sqrt{k_\bot^2 + k_\|^2}$, and $\mu=k_\|/k$.

Note that in the limit of zero angular separation, this statistic becomes equivalent to \poned: $P_\times(\theta\rightarrow0, k_\|) = \text{\poned}$. As such, it is ideally suited to be a close companion to \poned\ that captures small-scale anisotropic information and to disentangle the monopole and quadrupole contributions to \pthreed.

Quasar continuum errors distort the $P_\times$ measurements in the transverse direction in the absence of marginalization. However, unlike correlation function measurements, these distortions have not yet been calculated and modeled using a robust formalism, such as the distortion matrix. We will ignore this complication in the forward modeling of $P_\times$, assuming the optimal estimator formalism presented in \citet{fontriberaEstimate3DPowerLya2018} successfully removes these distortions, as is the case for optimal estimators of \poned\ and \pthreed\ \cite{karacayliOptimal1DLy2020, karacayliOptimal3dEstimator2025}.

\subsection{Correlation function multipoles \texorpdfstring{$\xi_\ell(r)$}{XiEll}}
The multipoles of the correlation function are given by
\begin{equation}
    \xi_\ell(r) = i^\ell \int_0^\infty \ddif k ~ j_\ell(kr) \frac{k^2 P_\ell(k)}{2\pi^2},
\end{equation}
where $j_\ell(x)$ is the spherical Bessel function. This can be written in FFTLog formalism for fast calculations using $j_\ell(x) = J_{\ell + 1/2}(x) \sqrt{\pi/2x}$ where $J_\ell$ are the Bessel functions of the first kind \cite{talmanNumericalFourierBessel1978, hamiltonUncorrelatedModesNonlinear2000}.
\begin{align}
\label{eq:xiell_fftlog}
    \xi_\ell(r) &= i^\ell \int_0^\infty \ddif k ~ \sqrt{\frac{\pi}{2kr}} J_{\ell + 1/2}(kr) \frac{k^2 P_\ell(k)}{2\pi^2} \\
    &= \frac{i^\ell }{(2\pi r)^{3/2}} \int_0^\infty r \ddif k ~ J_{\ell + 1/2}(kr) k^{3/2} P_\ell(k)
\end{align}

The large-scale structure surveys do not measure the \lya\ forest correlation function in multipoles due to complications caused by quasar continuum errors. These errors distort the correlation function by modifying the signal, mixing multipoles, and leaking signal to higher-order multipoles. Even though these complications can be forward modeled as we will show shortly,
we advocate for optimally estimating the correlation function \cite{slosarMeasurementBaryonAcoustic2013} or using machine-learning-based continuum fitting techniques \cite{turnerLyaForestMeanFluxFromDesiY12024} to remove these distortions for real data applications.

The mathematical link between undistorted and distorted correlation function multipoles is as follows.
The measurements of the correlation function are done in transverse-radial $(tr)$ direction binning $(r_\bot, r_\|)$. The connection between this scheme and multipoles is a linear operation: $\bm{\xi}_{tr} = \mathbf{P}_m \bm{\xi}_\ell$ for a chosen set of $n_\ell$ multipoles of $\xi_\ell$ in some $n_r$ bins. The matrix $\mathbf{P}_m$ includes each multipole model's linear interpolation kernel and associated Legendre polynomial. The observed $\bm{\xi}^*_{tr}$ is distorted by a distortion matrix, also a linear operation,
\begin{equation}
    \label{eq:xitr_obs_dist_model}\bm{\xi}^*_{tr} = \mathbf{D}_{tr} \mathbf{P}_m \bm{\xi}_\ell.
\end{equation}
Now we convert $\bm{\xi}^*_{tr}$ to observed multipoles $\bm{\xi}^*_\ell$. The formulation is the same, but without the distortion matrix. The interpolation matrix has to be constructed using the data points, so we denote it as $\mathbf{P}_d$ to differentiate. The weighted least-squares solution using the covariance matrix $\mathbf{C}_{tr}$ is the following:
\begin{equation}
    \hat{\bm{\xi}}^*_\ell = [(\mathbf{P}^\mathrm{T}_d \mathbf{C}^{-1}_{tr} \mathbf{P}_d)^{-1} \mathbf{P}^\mathrm{T}_d \mathbf{C}_{tr}^{-1}] \bm{\xi}^*_{tr} = \mathbf{S}_{\ell,tr} \bm{\xi}^*_{tr}.
\end{equation}
Using this solution matrix $\mathbf{S}_{\ell,tr}$, the covariance matrix of observed multipoles $\mathbf{C}_\ell$ and the distortion matrix in multipole space $\mathbf{D}_{\ell}$ are given by
\begin{align}
    \mathbf{C}_\ell &= \mathbf{S}_{\ell,tr} \mathbf{C}_{tr} \mathbf{S}^\mathrm{T}_{\ell,tr},\\
    \mathbf{D}_{\ell} &= \mathbf{S}_{\ell,tr} \mathbf{D}_{tr} \mathbf{P}_m.
\end{align}
This $\mathbf{D}_{\ell}$ matrix needs to be calculated once and can be repeatedly applied in forward modeling: $\bm{\xi}^*_\ell = \mathbf{D}_{\ell} \bm{\xi}_\ell$.

There is a caveat. We attempted to directly reconstruct the undistorted multipoles from real data using the least-squares solution to Eq.~(\ref{eq:xitr_obs_dist_model}). This does not work, most likely because of errors in the distortion matrix, in which case, the forward model will not magically solve the problem. However, the forward model can be helpful if the errors in the distortion matrix can also be included in the model.
We confirmed that a valid least-squares solution can be constructed for the equation $\bm{\xi}^*_\ell = \mathbf{D}_{\ell} \bm{\xi}_\ell$ for noiseless $\bm{\xi}^*_\ell$ vectors. Hence, we strongly recommend that the distortion matrix be eliminated through available means in the literature.

\section{Forward modeling\label{sec:forward}}

We first briefly describe our core methodology, Hamiltonian Monte Carlo, then test the efficiency of knot reconstruction using noiseless mock data vectors that approximate the statistical properties of ``future" real data.

\subsection{Hamiltonian Monte Carlo}
Hamiltonian Monte Carlo (also referred to as Hybrid Monte Carlo, HMC \cite{duaneHybridMonteCarlo1987}) exploits the gradient of the log-likelihood function to find the high-likelihood regions and then samples the posterior efficiently with a low rejection rate. This is particularly important in high-dimensional spaces. However, HMC's performance is highly dependent on the step size and the number of steps. The No-U-Turn Sampler (NUTS) addresses this problem of HMC by eliminating the need to fine-tune the number of steps \cite{hoffman_no-u-turn_2014}. We use the NUTS implementation in the \texttt{NumPyro} software package \cite{bingham_pyro_2019, phan_numpyro_2019}. \texttt{NumPyro} itself depends on the automatic differentiation software package \texttt{JAX} \cite{jax2018github}, which redefines the semantics of computational operations by incorporating derivative values for a given function and applying the chain rule to propagate these values \cite{baydin_automatic_2018}. 

We adopt a Gaussian likelihood for a data vector $\bm d$ and its covariance matrix $\mathbf{C}$:
\begin{equation}
    \ln \mathcal{L}(\bm d|\bm \theta) = -(\bm d -\bm m(\bm \theta))^\mathrm{T} \mathbf{C}^{-1} (\bm d -\bm m(\bm \theta)),
\end{equation}
where $\bm d$ represents one or a combination of summary statistics, $\bm m(\bm \theta)$ is the model reconstruction of those statistics for a set of parameters $\bm \theta \sim \text{\pthreed}$. We make two simplifications: we ignore the cross-covariance between different summary statistics and assume diagonal covariance matrices for \poned\ and $P_\times$. As mentioned, we apply two strategies for \pthreed\ parameterization: 
\begin{enumerate}
    \item Infer the multipoles of \pthreed\ in some bins (knots) of $k_n$ such that $\bm\theta\rightarrow P_\ell(k_n)$.
    \item Utilize our fitting functions in Eqs.~(\ref{eq:quad_mono_ratio}) and (\ref{eq:hexa_mono_ratio}) for the ratio of multipoles, such that $\bm \theta \rightarrow \{P_{\ell=0}(k_n), a, b, c, s,...\}$.
\end{enumerate}
The posterior distribution is $\mathcal{U}(\bm\theta)=\pi(\bm\theta)\mathcal{L}(\bm d|\bm \theta)$, where $\pi(\bm\theta)$ is the prior and $-\ln \mathcal{U}(\bm\theta)$ is the ``potential energy" in HMC formalism \cite{nealMcmcHamiltonianMonteCarlo2011}.

\texttt{JAX} does not support all numerical algorithms, such as dynamic refinement techniques for integration. This programmatically constrains us to implement definite integrals using fixed quadrature rules. Fortunately, the Fast Fourier Transform (FFT) algorithms are incorporated into \texttt{JAX}, which enables us to transform between the correlation function and the power spectrum without hindrance.

Additionally, we use the following commonly-used \texttt{python} packages in our analysis: \texttt{numpy}\footnote{\url{https://numpy.org}}
an open source project aiming to enable numerical computing with \texttt{python} \citep{numpy},
\texttt{scipy}\footnote{\url{https://scipy.org}}
an open-source project with algorithms for scientific computing. We make plots using
\texttt{matplotlib}\footnote{\url{https://matplotlib.org}}
a comprehensive library for creating static, animated, and interactive visualizations in \texttt{python}
\citep{matplotlib}.

\subsection{One-dimensional power spectrum}
Let us first investigate the performance of forward modeling for \poned. We define 20 linearly-interpolated loglinear bins between $k=0.07~$\impc\ and $50~$\impc\ for $k^2P_\ell(k)$, limiting ourselves to only the monopole and quadrupole terms. Note that we estimate the knots of $k^2P_\ell(k)$ instead of $P_\ell(k)$, as this quantity is smoother in $\ln k$ and better suited for linear interpolation. The integrations in Eq.~(\ref{eq:p1d_integ}) are evaluated using Gauss-Legendre quadrature with 30 points for the isotropic $P_0-0.5P_2$ term and Gauss-Laguerre quadrature with ten points for the $1.5\mu^2P_2$ term. The integration accuracy improves with an increasing number of $k$ bins (not as much with quadrature evaluation points), but this reduces the constraining power in each $k$ bin. Our parameter choices optimize these two competing objectives.

Intuitively, we do not expect to reconstruct the quadrupole from \poned\ alone since the transverse information is integrated out from \poned. The uncertainty should flow to the monopole term, such that the forward model ``fails" to gain information beyond the prior. To test this, we match the forward model to the input model exactly, i.e., creating a noiseless \poned\ data vector using two multipoles with the integration technique described above. The mock \poned\ data vector is calculated on 50 linearly spaced points between $k=0.07~$Mpc and $1.9~$Mpc, and assigned a 3\% precision on all points, approximating an improved DESI measurement beyond the early data release measurement at $z=2.4$ \cite{karacayliOptimal1dDesiEdr2023}. We impose a loose Gaussian prior on each bin centered at the true value $k^2P_\ell^\mathrm{true}(k)$ with standard deviation equal to $\sigma = 0.5 |k^2P_\ell^\mathrm{true}(k)| + 0.2$.
After 5,000 warmup steps and 5,000 samples drawn, the forward model marginally improves upon the prior \emph{only} for the monopole.

This means, as expected, that we need external information to reconstruct \pthreed\ from \poned\ alone. Before including additional data sets, we assess the usefulness of our quadrupole-to-monopole ratio formula in Eq.~(\ref{eq:quad_mono_ratio}). Assuming this ratio holds for all physical scenarios, we can marginalize only four fitting parameters within a reasonable range instead of arbitrary $k$ bins of the quadrupole. We impose a uniform prior\footnote{Using a Gaussian prior instead of a uniform prior for $\alpha_i$ parameters produces similar results.}  between $\alpha^*_i\pm0.1$ for each parameter, where $a^*_i$ are the same best-fit values from Fig.~\ref{fig:multipole_monopole_ratio}.

\begin{figure}
    \includegraphics[width=\columnwidth]{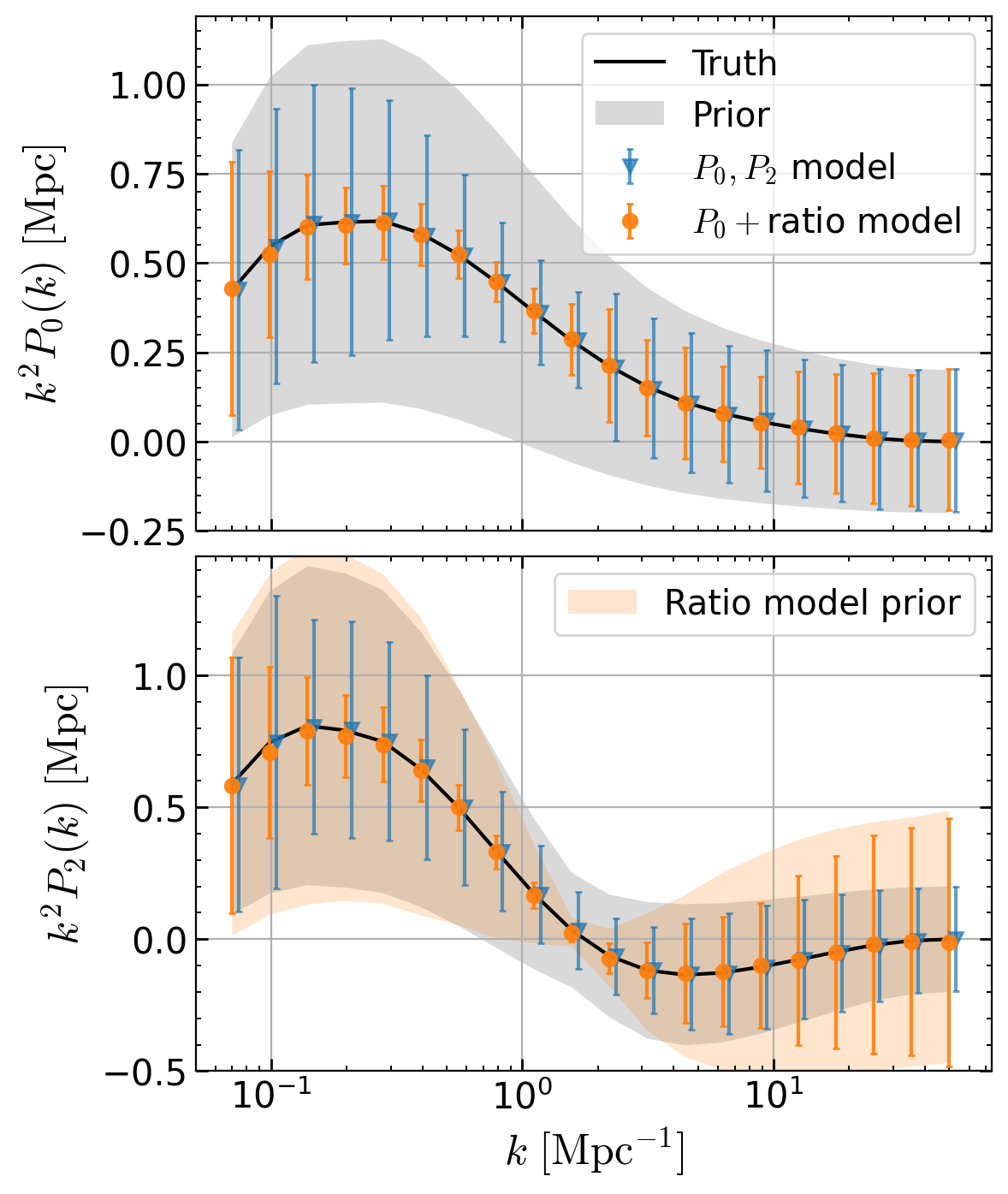}
    \caption{The reconstructed monopole ({\it top}) and quadrupole ({\it bottom}) from a noiseless \poned\ data vector. While estimating the quadrupole independently ({\it blue triangles}), the forward model cannot improve upon the priors. Leveraging the quadrupole-to-monopole ratio ({\it orange circles}) can reconstruct the monopole in eight $k$ bins by yielding three times smaller error bars than the prior.}
    \label{fig:rec_p1d_alone}
\end{figure}

Fig.~\ref{fig:rec_p1d_alone} shows the reconstructed monopole on the top panel and quadrupole on the bottom panel for both models.
The results when both $P_0$ and $P_2$ are independently modeled are shown in blue triangles. As discussed, they do not surpass the prior (grey-shaded region). When the quadrupole is modeled as a scaling of the monopole, the reconstructed $P_0$ improves in signal, as shown by the orange squares. We define the scales at which the forward model improves upon the prior when the error bars shrink by a factor of three. These scales are limited to eight $k$ bins between $0.14~\text{\impc}\lesssim k \lesssim 1.6~\text{\impc}$.

It is instructive to inspect the derived quadrupole for the ratio model, which is subject to a different prior (light orange-shaded region) and displays the same trend on the surface. However, we find that the four parameters describing the ratio model uniformly fill the prior volume, so the quadrupole contains no additional information beyond the monopole.

\subsection{Cross-spectrum}
$P_\times$ is a natural companion to \poned\ as it is measured at the same line-of-sight scales while also incorporating information at small angular scales.
In this subsection, we introduce a hypothetical $P_\times$ measurement into the knot reconstruction and quantify its merit in combination with \poned.

\begin{table}
    \caption{\label{tab:px_numpairs}Number of total pairs from \citet{abdulkarimMeasurementofPcross2024} used to estimate errors on our mock $P_\times$ vector. These numbers reflect the SDSS DR16 (eBOSS) data sample. We assume their relative importance carries over to the current DESI survey.}
    \begin{ruledtabular}
    \begin{tabular}{l|ccccc}
    $r_\perp~[\text{Mpc}]$ & 0 & 0.6 & 1.4 & 3.0 & 4.0 \\
    $N$ & 10286 & 133 & 418 & 420 & 544\\
    \end{tabular}
    \end{ruledtabular}
\end{table}

The only measurement of $P_\times$ so far is by \citet{abdulkarimMeasurementofPcross2024}, which is distorted by continuum fitting errors, which we ignore as discussed. We base our calculation on their measurement points by creating mock $P_\times$ measurements at $z=2.4$ in six linear $k_\|$ bins starting at $k_\|=0.07~\text{\impc}$ with $\Delta k_\|=0.14~\text{\impc}$ for four $r_\bot$ bins at $0.6, 1.4, 3.0,~\text{and}~4.0~\text{Mpc}$. The significant reduction in the number of $k_\|$ bins compared to \poned\ suppresses noise due to a limited number of pairs in $P_\times$. We use their reported number of pairs in each $\theta$ bin as a proxy for the number of modes available in that bin. Table~\ref{tab:px_numpairs} presents these values.
We assign error to each bin using $\sigma = \epsilon P_\times$, where $\epsilon$ converts the 3\% error we assigned to \poned\ measurements to $P_\times$ measurements by conserving the number of modes in each $k_\|$ bin:
\begin{equation}
    \epsilon = 3\% \sqrt{(N_\mathrm{1D}/N_\mathrm{pairs})(\Delta k^\mathrm{1D}_\| / \Delta k^\times_\| )}.
\end{equation}

Again, we focus only on the monopole and quadrupole. The addition of $P_\times$ helps constrain the monopole and the quadrupole on four $k$ knots between $0.56~\text{\impc} < k < 2.22~\text{\impc}$
using the threshold of a factor of three improvement upon priors. To reconstruct these multipoles in more $k$ bins, the $P_\times$ measurement needs to be improved either by measuring it in additional angular bins or higher $k$ bins or by increasing precision with optimal estimators. Our prescription for measurement precision is approximate and likely misrepresents DESI's current capabilities. The data will ultimately determine what is achievable, so we refrain from speculating further on alternative choices.

\begin{figure}
    \centering
    \includegraphics[width=\linewidth]{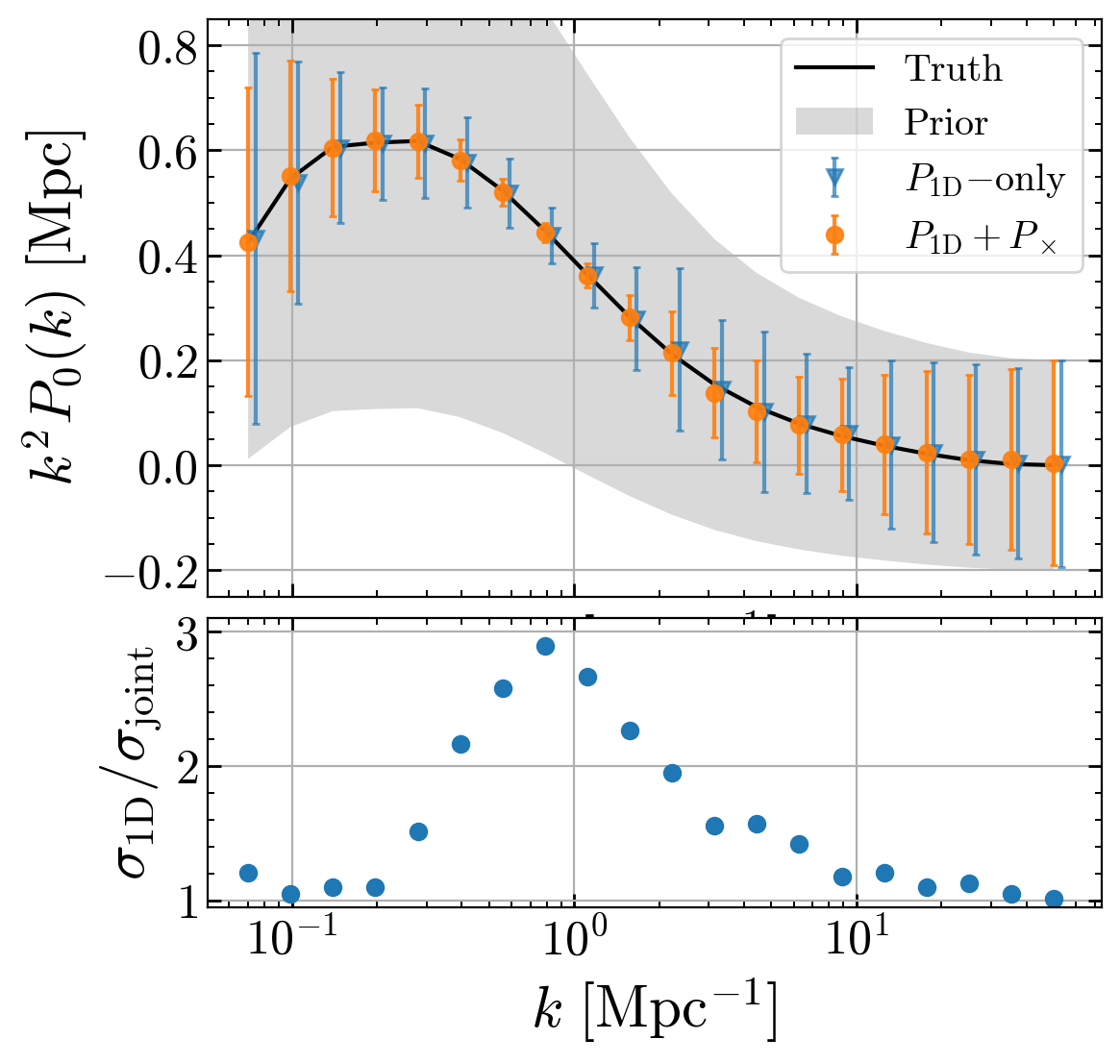}
    \caption{({\it Top}) The reconstructed monopoles leveraging the quadrupole-to-monopole ratio from a \poned-only analysis ({\it blue triangles}) and \poned$+P_\times$ joint analysis ({\it orange circles}). The quadrupole is omitted because it contains no additional information beyond the monopole. ({\it Bottom}) The error improvement factor between a \poned-only analysis and \poned$+P_\times$ joint analysis.}
    \label{fig:pxp1d_ratio}
\end{figure}
We now introduce the ratio equation to extract more information, which has already proved fruitful for the \poned-only analysis. The joint analysis reconstructs the monopole in nine $k$ bins between $0.14~\text{\impc}\lesssim k \lesssim 2.22~\text{\impc}$ with significantly higher precision.  The reconstructed monopoles for a \poned-only analysis and \poned$+P_\times$ joint analysis are shown in Fig.~\ref{fig:pxp1d_ratio}. The ratio parameters for the quadrupole still cannot be constrained beyond their priors, so the derived quadrupole is omitted from this plot. The error improvement factor in reconstructed bins ranges from $1.2$ to $2.9$ and averages $2.1$ compared to the \poned-only analysis using the ratio equation. This is shown in the bottom panel of the same figure.

We have focused on recovering the 3D power spectrum from \poned\ and $P_\times$ data. However, our conclusions can be generalized to any cosmological inference using these statistics. Even though a wide range of scales contributes to \poned\ and $P_\times$, the cosmological information is concentrated at $k \sim 1~\text{\impc}$. Large $(k \lesssim 0.1~\text{\impc})$ and small scales $(k \gtrsim 2~\text{\impc})$ cannot be exploited and are sources of uncertainty. This parallels the existing \poned\ literature that most of the cosmological information in \poned\ is compressed into the amplitude and logarithmic slope at a pivot scale $k_p=0.7\text{~\impc}$ \cite{mcdonaldLinearTheoryLyaSdss2005, pedersenEmulator2021}.

\subsection{Correlation function multipoles}

We gauge the efficiency of our forward model based on the DESI DR1 correlation function measurement \cite{desiKp6BaoLya2024}. We fit five even multipoles to this data vector using 47 linearly interpolated $r$ bins with $4~\text{\hmpc}$ bin size. The first bin is centered at $14~\text{\hmpc}$; and the range $12~\text{\hmpc}<r<200~\text{\hmpc}$ is used in the fit, yielding a reduced chi-squared value of one. The last $r$ bin is susceptible to edge effects, so it is removed, leaving us with 46 radial bins. DESI DR1 reports the correlation function with \hmpc\ units, which we convert to Mpc units using $h=0.6766$.

\begin{figure}
    \includegraphics[width=0.95\linewidth]{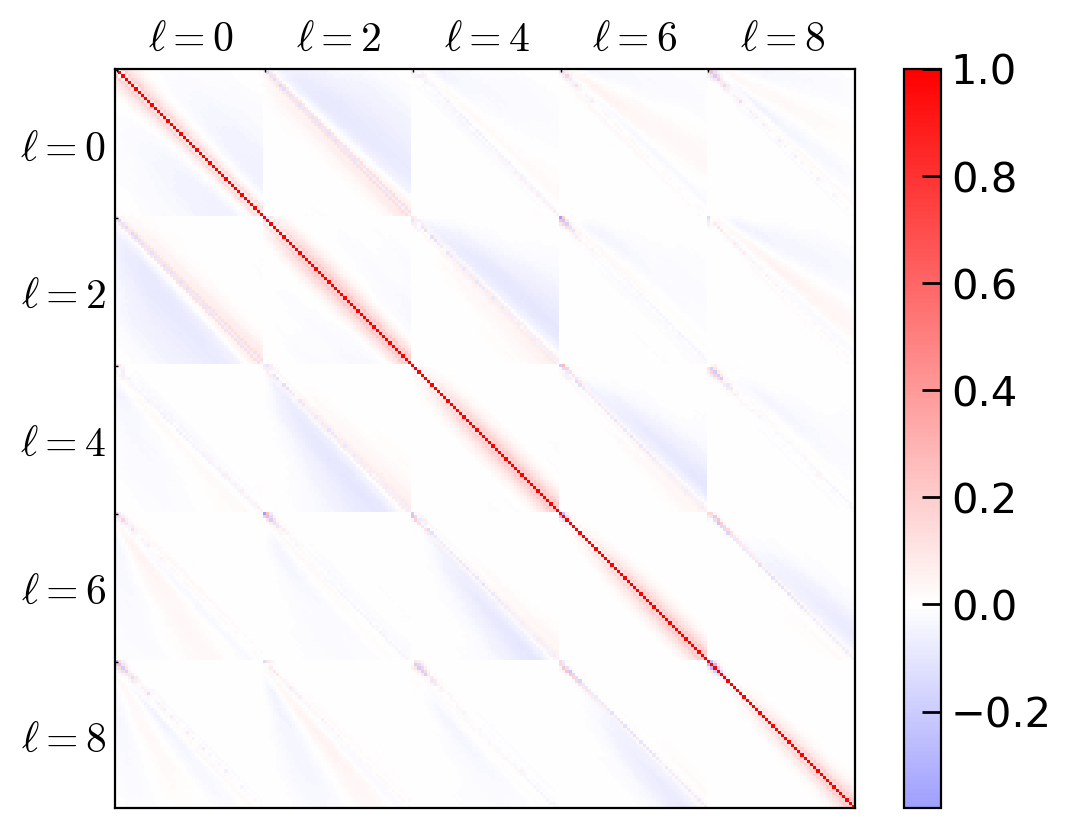}
    \caption{The covariance correlation matrix between multipoles based on DESI DR1 for five even multipoles. There are weak correlations between neighboring multipoles.}
    \label{fig:dr1_cov_ell}
\end{figure}

\begin{figure}
    \includegraphics[width=0.9\linewidth]{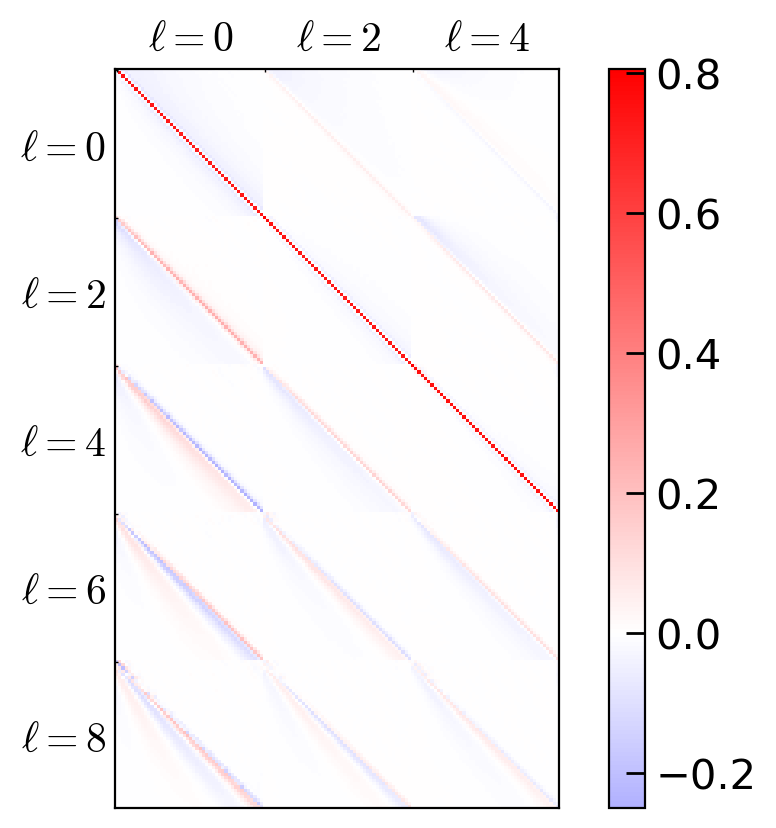}
    \caption{The distortion matrix that multiplies the first three even model correlation function multipoles to yield the observed five multipoles based on DESI DR1 \lya\ correlation function measurement.}
    \label{fig:dr1_dist_ell}
\end{figure}

We do not apply our method to data yet, as it is complicated by inaccuracies in the distortion matrix and anisotropic effects such as high-column density systems that extend to large scales and metals that produce additional peaks in the correlation function \cite{desiKp6BaoLya2024}. The covariance and distortion matrices are of interest to us in creating a mock data vector without inaccuracy. These matrices are illustrated in Fig.~\ref{fig:dr1_cov_ell} and Fig.~\ref{fig:dr1_dist_ell}. We refer the reader to the Appendix~\ref{app:dr1_cf} for the multipole representation of the DESI DR1 correlation function measurement.

We reconstruct the power spectrum multipoles in the range of $0.001~\text{\impc} < k < 1~\text{\impc}$ while supplementing the choice of $k$ bins by the BAO locations. We densely sample the range $0.02~\text{\impc} < k < 0.2~\text{\impc}$ by placing our knots at the extremum points induced by BAO and the midpoints between those extrema, resulting in 20 knots. The remaining knots are log-linearly placed. There are nine bins between $0.001~\text{\impc} < k < 0.02~\text{\impc}$ and five bins between $0.2~\text{\impc} < k < 1~\text{\impc}$, both approximating a bin size of $\Delta\ln k \approx 0.4$:

Let us start with an undistorted data vector. The distortions remove the large-scale signal from the data, which, in addition to complicating the analysis, also reduces the correlations in the covariance matrix. So, by keeping the large-scale signal while removing its contribution to the covariance, we will overestimate the reconstruction capability in this scenario. As before, we match the forward model to the input model exactly by creating a noiseless $\bm{\xi}_\ell$ data vector using three multipoles with FFTLog. Although the hexadecapole term remains weak and unconstrained, it is weakly correlated with other multipoles. By modelling it and propagating the covariance matrix in the forward model, we marginalize out its effects from the knot reconstructed monopole and quadrupole. The correlation function multipoles are smoothed out by the linear interpolation kernel, which we model with an additional $\text{sinc}^2(k\Delta r/2)$ term in Eq.~(\ref{eq:xiell_fftlog}).

We impose the same Gaussian prior on each knot centered at the true value $k^2P_\ell^\mathrm{true}(k)$ with standard deviation equal to $\sigma = 0.5 |k^2P_\ell^\mathrm{true}(k)| + 0.2$, and perform 5,000 warmup steps and draw 5,000 samples. The reconstructed monopole improves upon the prior by a factor of three in 12 $k$ bins between $0.07~\text{\impc} \leq k \leq 0.019~\text{\impc}$, while the reconstructed quadrupole improves in 12 $k$ bins between $0.08~\text{\impc} \leq k \leq 0.02~\text{\impc}$.

\begin{figure}
    \includegraphics[width=0.9\linewidth]{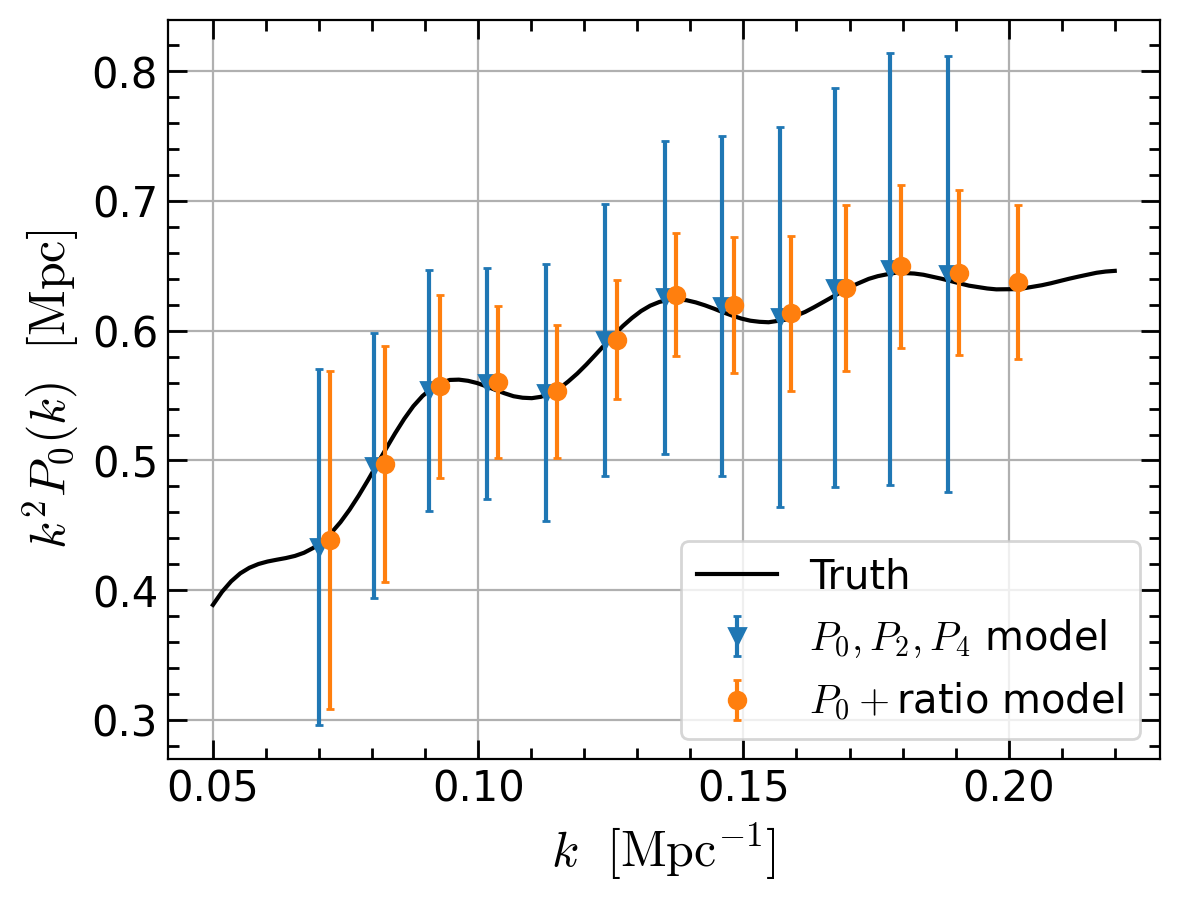}
    \caption{The reconstructed monopole from a noiseless, undistorted correlation function data vector. Estimating all three multipoles ({\it blue triangles}) yields larger error bars compared to leveraging the ratio between multipoles ({\it orange circles}).}
    \label{fig:rec_xiell_alone}
\end{figure}

Now, let us introduce the ratios in Eqs.~(\ref{eq:quad_mono_ratio}) and (\ref{eq:hexa_mono_ratio}) instead of reconstructing the quadrupole and hexadecapole in arbitrary $k$ knots. We impose the same uniform prior of $\alpha_i^*\pm 0.1$ for each parameter $\alpha_i^*$ centered at the best-fit value as we have done for \poned. Fig.~\ref{fig:rec_xiell_alone} demonstrates that the reconstructed monopole is significantly more precise thanks to this relation.
Thirteen $k$ bins between $0.07~\text{\impc} \leq k \leq 0.02~\text{\impc}$ are reconstructed with error bars improved by a factor of 2.4 compared to the model not utilizing the ratio relations. The uniform prior space is filled within the $\alpha_i^* \pm 0.1$ interval. Of course, the gain in precision is also driven by this choice. We enlarge the interval to $\alpha_i^*\pm 0.5$ and find that $\alpha_i$ parameters remain unconstrained and the ratio model becomes less precise (by $30\%$) but still outperforms reconstructing every multipole independently.

Incorporating the distortion matrix is a straightforward linear algebra application. Specifically, we multiply the true model, which consists of three multipoles, by the distortion matrix to obtain the correlation function in five multipoles. Now that the large-scale signal is removed from the data vector, we expect a decline in efficiency. Indeed, that is the outcome: a $9.4\%$ increase in error bars in the reconstructed $k$ range. The forward model completely deconvolves the effect of the distortion matrix as expected from a simple linear relation. We note that this is an ``easy" test that lacks complications due to metal oscillations and noise that could be leaking in from the higher-order multipoles. These will have to be investigated for real data applications.

\subsection{Unified reconstruction}
We lastly showcase the strength of combining all statistics by performing a unified forward modelling. The setup for each statistic remains the same. Only the monopole and quadrupole are considered for \poned\ and $P_\times$, while the hexadecapole is introduced for $\xi_\ell$.

\begin{figure}
    \includegraphics[width=0.9\columnwidth]{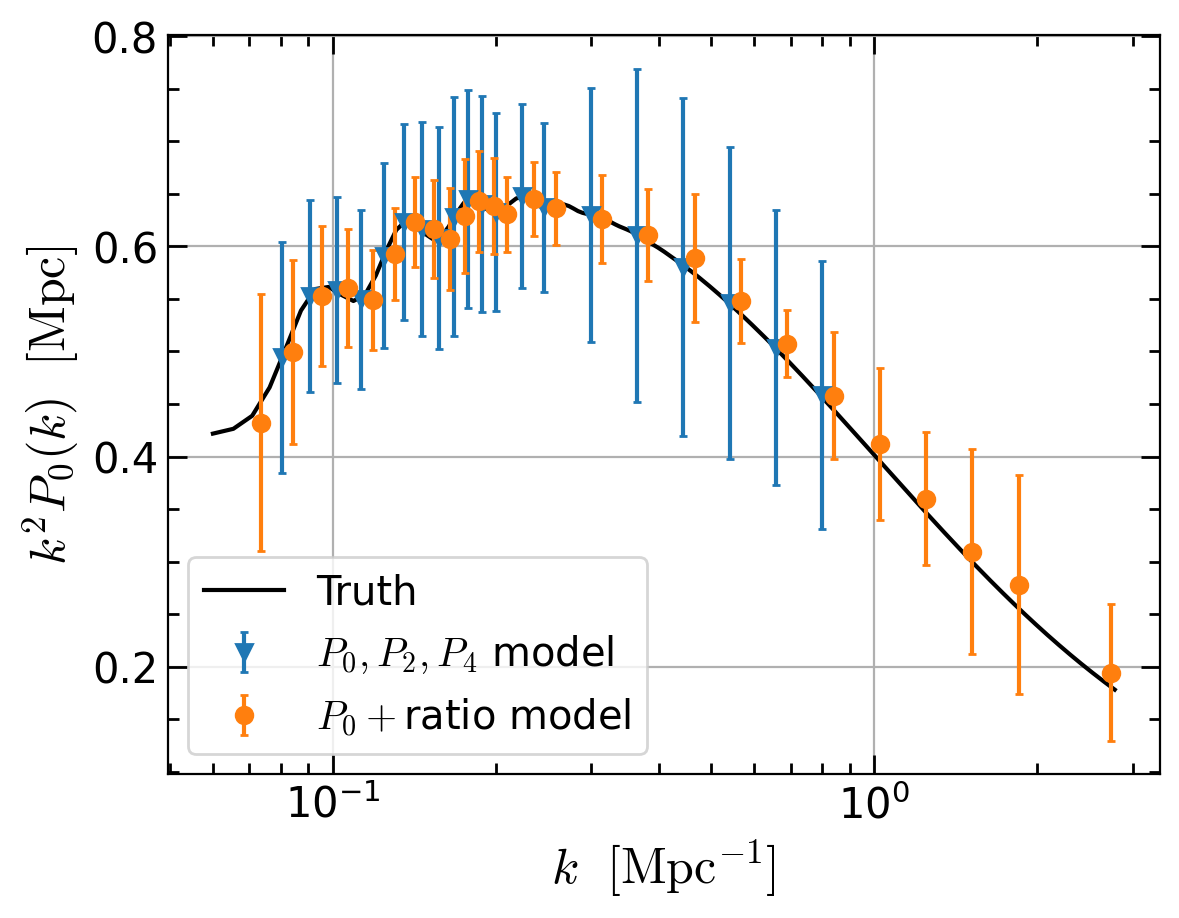}
    \caption{The reconstructed monopole from the unified reconstruction of all three two-point statistics. Leveraging the ratio between multipoles ({\it orange circles}) sizably improves upon estimating all three multipoles ({\it blue triangles}). The joint analysis with the ratio relation reconstructs the monopole between $0.07~\text{\impc} \leq k \leq 1.8~\text{\impc}$. Note that the orange circles are shifted for clarity, creating an illusion of a minor bias at high $k$.}
    \label{fig:joint_rec_mono}
\end{figure}

We start by measuring each multipole independently. This reconstructs the monopole in 20 knots between $0.08~\text{\impc} \leq k \leq 0.8~\text{\impc}$. The quadrupole reconstruction is not contiguous: the first 14 reconstructed bins are between $0.08~\text{\impc} \leq k \leq 0.025~\text{\impc}$, and after a large gap in $k$, there are three more bins at $k = 0.80, 0.97, 1.19~\text{\impc}$. Including the ratio relation significantly improves the monopole reconstruction. The uniform prior of $\alpha_i^*\pm 0.1$ is the same as before.
We recover 25 bins between $0.07~\text{\impc} \leq k \leq 1.8~\text{\impc}$ and another bin at $k=2.6~\text{\impc}$ with considerably higher precision. These are shown in Fig.~\ref{fig:joint_rec_mono}.

\begin{figure}
    \centering
    \includegraphics[width=\linewidth]{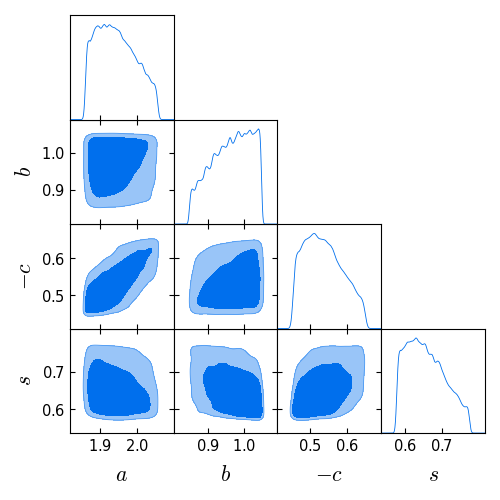}
    \caption{The posterior of the quadrupole-monopole ratio parameters. It is predominantly defined by the uniform prior. A correlation between the $a-c$ parameters seems to emerge, which could justify a minor parameterization of Eq.~(\ref{eq:quad_mono_ratio}).}
    \label{fig:corner_plots_quad}
\end{figure}

\begin{figure}
    \centering
    \includegraphics[width=\linewidth]{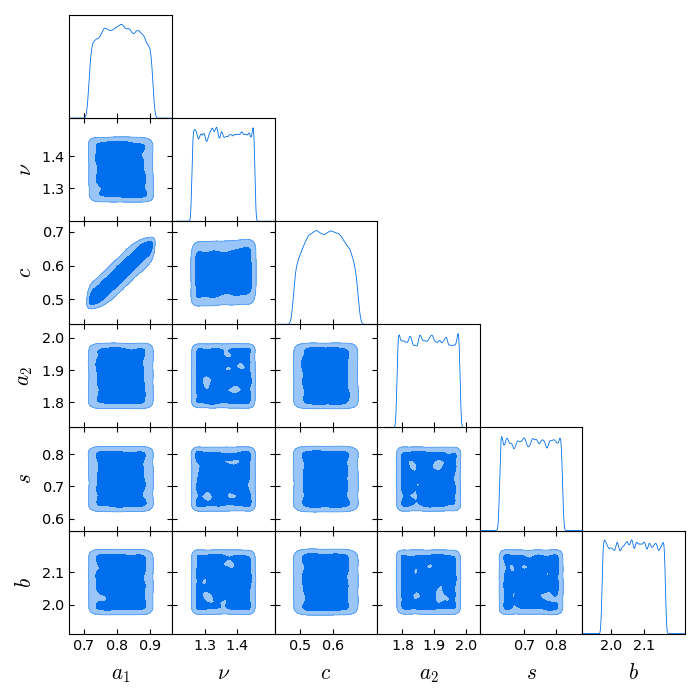}
    \caption{The posterior of the hexadecapole-monopole ratio parameters. Similar to the quadrupole-monopole ratio parameters, it mainly demonstrates the uniform prior. A correlation between the $a_1-c$ parameters strongly appears in this case, which could justify a minor parameterization of Eq.~(\ref{eq:hexa_mono_ratio}).}
    \label{fig:corner_plots_hexa}
\end{figure}

Fig.~\ref{fig:corner_plots_quad} shows the posterior of the quadrupole-monopole ratio parameters, while Fig.~\ref{fig:corner_plots_hexa} shows the posterior of the hexadecapole-monopole ratio parameters. These parameters still fill the prior volume. However, a correlation between the $a-c$ parameters of the quadrupole-monopole ratio and another correlation between the $a_1-c$ parameters of the hexadecapole-monopole ratio seems to emerge. These correlations can justify a minor parameterization of Eqs.~(\ref{eq:quad_mono_ratio}) and (\ref{eq:hexa_mono_ratio}).

\begin{figure}
    \includegraphics[width=0.9\linewidth]{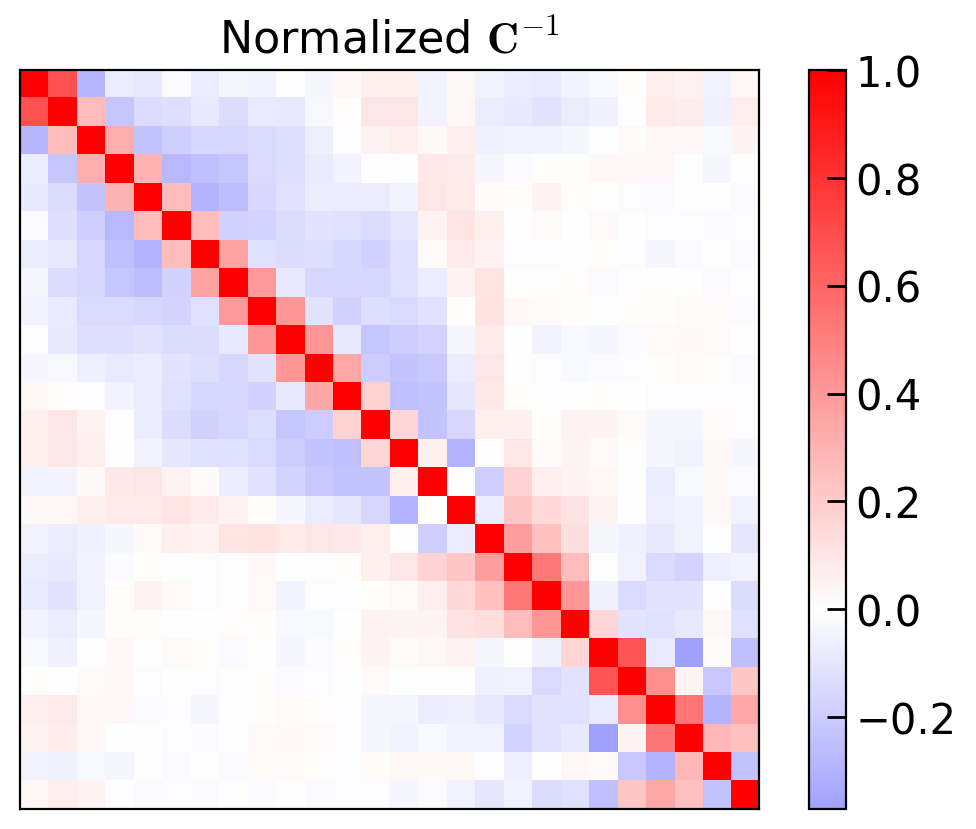}
    \caption{The inverse covariance correlations of the reconstructed monopole from the joint analysis using the ratio relation. Correlations between neighboring $k$ bins are due to linear interpolation in forward modelling.}
    \label{fig:joint_rec_cov}
\end{figure}

Lastly, Fig.~\ref{fig:joint_rec_cov} illustrates the estimated inverse covariance, which reveals correlations between neighboring bins of the reconstructed power spectrum. This is expected due to linear interpolation in forward modelling. The weaker correlations between more distant bins result from a more complex interplay between the data covariance matrix and the mapping of two-point statistics to the multipoles of \pthreed.  

\begin{figure*}
    \centering
    \includegraphics[width=\linewidth]{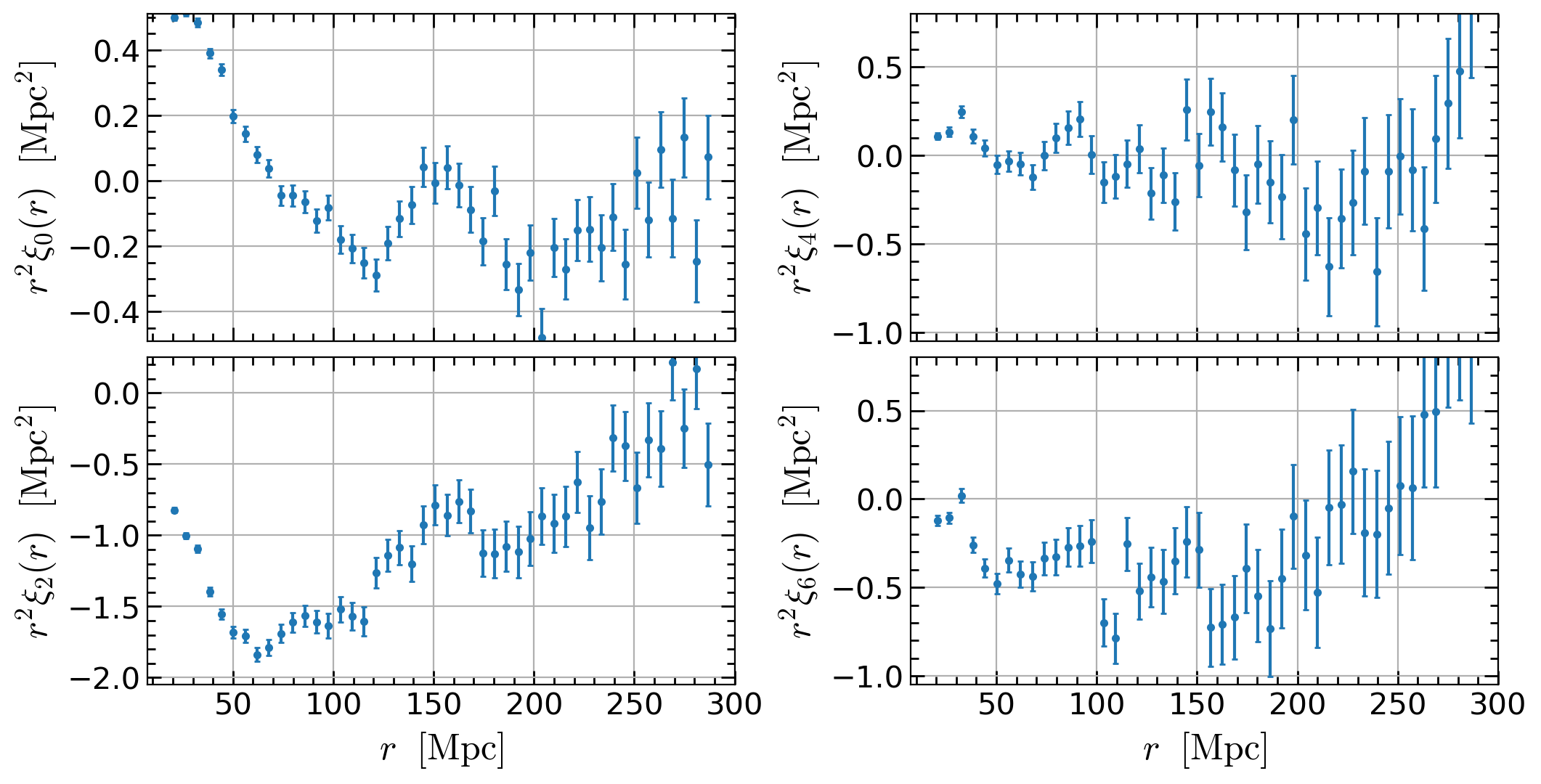}
    \caption{DESI DR1 correlation function multipoles. The BAO peak is beautifully illustrated in the monopole at $r=150~$Mpc.}
    \label{fig:dr1_cf_ell}
\end{figure*}

\section{Discussion\label{sec:discuss}}
We have developed a forward model framework to reconstruct the \lyaf\ three-dimensional power spectrum from the other two-point statistics of the \lyaf. Our first key ingredient is that the first three even multipoles contribute the most to \pthreed. This realization considerably narrows down the degrees of freedom by instead reconstructing $P_\ell(k)$ for three $\ell$ values. The second key ingredient concerns the one-dimensional power spectrum specifically. Historically, the differential equation relation between \poned\ and \pthreed\ has been used to recover an ``isotropic" \pthreed$(k)$. This differentiation of noisy \poned\ measurements is numerically unstable and undesirable. We instead use the integral relation between the two and sample the posterior of \pthreed\ using Hamiltonian Monte Carlo, which preserves the statistical properties of the observed \poned. Even though our method does not replace the direct estimation of \pthreed\ from data, it can still be used for cosmological inference from all observations or as a consistency check at minimum.

Our work carries similarities with the minimally parametric reconstruction of the primordial power spectrum ($P_\mathrm{pri}$) of ref.~\cite{birdMinimallyParametricReconstructionLya2011}, in which $P_\mathrm{pri}$ is reconstructed from \poned\ data in four knots between $0.4~\text{\impc} \lesssim k \lesssim 1.9~\text{\impc}$ using anchor points at large and small scales. In our formalism, these anchor points translate to strong priors on \pthreed\ in $k < 0.4~\text{\impc}$  and $k>2~\text{\impc}$. Although we have not pursued this avenue, it is certainly a reasonable prior and straightforward approach to implement in HMC. The major difference between our work and this, of course, is that we do not reconstruct $P_\mathrm{pri}$. As ref.~\cite{birdMinimallyParametricReconstructionLya2011} demonstrates, making this connection requires hydrodynamical simulations seeded with fluctuations using non-parametric $P_\mathrm{pri}$. However, there is an additional merit to our \pthreed\ reconstruction approach: it enables other methods for cosmological inference (not limited to $P_\mathrm{pri}$), including effective field theory of the \lyaf\ \cite{ivanovEffectiveLya2024}.

We choose to linearly interpolate the reconstructed $k^2P_\ell(k)$ bins, which ensures continuity, albeit it correlates neighboring bins. Smoothness can be imposed with a higher-order interpolation, such as a cubic spline or a Gaussian process. However, it is hard to foresee the benefits of this approach as more and more bins become correlated, diffusing the information that could otherwise be isolated to specific scales.

The observations are contaminated by metal transitions and high-column density systems. These contaminations can be included in the model and removed from the reconstructed \pthreed, simplifying the inference that follows. Theoretically, these contaminations are common to all observables and can be jointly modelled. However, we expect this joint modelling of systematics to be more susceptible to modeling errors. It is likely better to handle these individually for each observable.

\begin{acknowledgments}
We thank David Weinberg for the fruitful discussion.

N.G.K. acknowledges support from the United States Department of Energy, Office of High Energy Physics under Award Number DE-SC0011726.
\end{acknowledgments}

\appendix
\section{Analytic expressions for the multipoles\label{app:analytic_multipoles}}
The multipoles can be exactly integrated and expressed in terms of the lower incomplete gamma function $\gamma(z, c) \equiv\int_0^c t^{z-1}\mathrm{e}^{-t} \ddif t$. We start by separating the isotropic and anisotropic parts of in the multipole integration:
\begin{equation}
    P_\ell(k) = P_\mathrm{iso}(k) \mathcal{I}_\ell(k),
\end{equation}
where the isotropic component is defined as
\begin{equation}
    P_\mathrm{iso}(k) = b_F^2 P_L(k) \mathrm{e}^{q_1 \Delta^2(k) - (k/k_p)^2},
\end{equation}
and the anisotropic integration as
\begin{equation}
    \label{eq:multi_exact}\mathcal{I}_\ell(k) = (2\ell + 1) \int_0^1 \ddif\mu ~\mathcal{L}_\ell(\mu) (1+\beta\mu^2)^2 \mathrm{e}^{-\phi(k) \mu^{b_\nu}},
\end{equation}
where $\phi(k) = q_1 A_\nu^{-1} \Delta^2(k) (k/k_{\nu,0})^{a_\nu}$ and $a_\nu = b_\nu - c_\nu$.

To simplify the notation in expressions following the integration in Eq.~(\ref{eq:multi_exact}), we first define $\tilde{\gamma}(z, c)=c^{-z}\gamma(z, c)$ and use the short-hand notation $\tilde{\gamma}_m \equiv \tilde{\gamma}(m/b_\nu, \phi)$. The anisotropic integration yields the following expressions for the first three even multipoles:
\begin{align}
    \mathcal{I}_0(k) &= \frac{1}{b_\nu} \left(
    \tilde{\gamma}_1 + 2\beta \tilde{\gamma}_3 + \beta^2 \tilde{\gamma}_5 \right), \\
    \mathcal{I}_2(k) &= \frac{5}{2b_\nu} \left(-\tilde{\gamma}_1 - (2\beta-3) \tilde{\gamma}_3 + (6-\beta)\beta \tilde{\gamma}_5 + 3\beta^2 \tilde{\gamma}_7 \right), \\
    \mathcal{I}_4(k) &= \frac{9}{8b_\nu} \Big(  3\tilde{\gamma}_1 - 6(5-\beta) \tilde{\gamma}_3 \nonumber \\
    &\qquad\qquad + (35 - 3\beta(20 -\beta)) \tilde{\gamma}_5 \nonumber\\
    &\qquad\qquad + 10\beta(7-3\beta) \tilde{\gamma}_7  + 35\beta^2 \tilde{\gamma}_9\Big).
\end{align}
These expressions ultimately describe the multipole ratios for the \citet{arinyoNonLinearPowerLya2015} fitting function. Noting that $\tilde{\gamma}(z, c)\rightarrow1/z$ as $c\rightarrow0$, we recover the Kaiser relations on large scales:

\begin{align}
\mathcal{I}_0(k\rightarrow0) &\rightarrow 1 + \frac{2\beta}{3} + \frac{\beta^2}{5}, \\
\mathcal{I}_2(k\rightarrow0) &\rightarrow \frac{4\beta}{3} + \frac{4\beta^2}{7}, \\
\mathcal{I}_4(k\rightarrow0) &\rightarrow \frac{8\beta^2}{35}.\\
\end{align}


\section{DESI DR1 correlation function multipoles\label{app:dr1_cf}}
Fig.~\ref{fig:dr1_cf_ell} shows the first four correlation function multipoles. The fifth multipole is omitted for clarity.

\bibliography{references,additional_references}

\begin{thebibliography}{46}%
\makeatletter
\providecommand \@ifxundefined [1]{%
 \@ifx{#1\undefined}
}%
\providecommand \@ifnum [1]{%
 \ifnum #1\expandafter \@firstoftwo
 \else \expandafter \@secondoftwo
 \fi
}%
\providecommand \@ifx [1]{%
 \ifx #1\expandafter \@firstoftwo
 \else \expandafter \@secondoftwo
 \fi
}%
\providecommand \natexlab [1]{#1}%
\providecommand \enquote  [1]{``#1''}%
\providecommand \bibnamefont  [1]{#1}%
\providecommand \bibfnamefont [1]{#1}%
\providecommand \citenamefont [1]{#1}%
\providecommand \href@noop [0]{\@secondoftwo}%
\providecommand \href [0]{\begingroup \@sanitize@url \@href}%
\providecommand \@href[1]{\@@startlink{#1}\@@href}%
\providecommand \@@href[1]{\endgroup#1\@@endlink}%
\providecommand \@sanitize@url [0]{\catcode `\\12\catcode `\$12\catcode
  `\&12\catcode `\#12\catcode `\^12\catcode `\_12\catcode `\%12\relax}%
\providecommand \@@startlink[1]{}%
\providecommand \@@endlink[0]{}%
\providecommand \url  [0]{\begingroup\@sanitize@url \@url }%
\providecommand \@url [1]{\endgroup\@href {#1}{\urlprefix }}%
\providecommand \urlprefix  [0]{URL }%
\providecommand \Eprint [0]{\href }%
\providecommand \doibase [0]{https://doi.org/}%
\providecommand \selectlanguage [0]{\@gobble}%
\providecommand \bibinfo  [0]{\@secondoftwo}%
\providecommand \bibfield  [0]{\@secondoftwo}%
\providecommand \translation [1]{[#1]}%
\providecommand \BibitemOpen [0]{}%
\providecommand \bibitemStop [0]{}%
\providecommand \bibitemNoStop [0]{.\EOS\space}%
\providecommand \EOS [0]{\spacefactor3000\relax}%
\providecommand \BibitemShut  [1]{\csname bibitem#1\endcsname}%
\let\auto@bib@innerbib\@empty
\bibitem [{\citenamefont {Slosar}\ \emph {et~al.}(2013)\citenamefont {Slosar},
  \citenamefont {Ir{\v s}i{\v c}}, \citenamefont {Kirkby}, \citenamefont
  {Bailey}, \citenamefont {Busca}, \citenamefont {Delubac}, \citenamefont
  {Rich}, \citenamefont {{\'Eric Aubourg}}, \citenamefont {Bautista},
  \citenamefont {Bhardwaj}, \citenamefont {Blomqvist}, \citenamefont {Bolton},
  \citenamefont {Bovy}, \citenamefont {{Joel Brownstein}}, \citenamefont
  {Carithers}, \citenamefont {Croft}, \citenamefont {Dawson}, \citenamefont
  {{Font-Ribera}}, \citenamefont {Goff}, \citenamefont {{Shirley Ho}},
  \citenamefont {Honscheid}, \citenamefont {Lee}, \citenamefont {Margala},
  \citenamefont {McDonald}, \citenamefont {Medolin}, \citenamefont {{Jordi
  Miralda-Escud\'e}}, \citenamefont {Myers}, \citenamefont {Nichol},
  \citenamefont {Noterdaeme}, \citenamefont {{Nathalie Palanque-Delabrouille}},
  \citenamefont {P{\^a}ris}, \citenamefont {Petitjean}, \citenamefont {Pieri},
  \citenamefont {Pi{\v s}kur}, \citenamefont {Roe}, \citenamefont {Ross},
  \citenamefont {Rossi}, \citenamefont {Schlegel}, \citenamefont {Schneider},
  \citenamefont {Suzuki}, \citenamefont {Sheldon}, \citenamefont {Seljak},
  \citenamefont {Viel}, \citenamefont {Weinberg},\ and\ \citenamefont
  {Y{\`e}che}}]{slosarMeasurementBaryonAcoustic2013}%
  \BibitemOpen
  \bibfield  {author} {\bibinfo {author} {\bibfnamefont {A.}~\bibnamefont
  {Slosar}}, \bibinfo {author} {\bibfnamefont {V.}~\bibnamefont {Ir{\v s}i{\v
  c}}}, \bibinfo {author} {\bibfnamefont {D.}~\bibnamefont {Kirkby}}, \bibinfo
  {author} {\bibfnamefont {S.}~\bibnamefont {Bailey}}, \bibinfo {author}
  {\bibfnamefont {N.~G.}\ \bibnamefont {Busca}}, \bibinfo {author}
  {\bibfnamefont {T.}~\bibnamefont {Delubac}}, \bibinfo {author} {\bibfnamefont
  {J.}~\bibnamefont {Rich}}, \bibinfo {author} {\bibnamefont {{\'Eric
  Aubourg}}}, \bibinfo {author} {\bibfnamefont {J.~E.}\ \bibnamefont
  {Bautista}}, \bibinfo {author} {\bibfnamefont {V.}~\bibnamefont {Bhardwaj}},
  \bibinfo {author} {\bibfnamefont {M.}~\bibnamefont {Blomqvist}}, \bibinfo
  {author} {\bibfnamefont {A.~S.}\ \bibnamefont {Bolton}}, \bibinfo {author}
  {\bibfnamefont {J.}~\bibnamefont {Bovy}}, \bibinfo {author} {\bibnamefont
  {{Joel Brownstein}}}, \bibinfo {author} {\bibfnamefont {B.}~\bibnamefont
  {Carithers}}, \bibinfo {author} {\bibfnamefont {R.~A.~C.}\ \bibnamefont
  {Croft}}, \bibinfo {author} {\bibfnamefont {K.~S.}\ \bibnamefont {Dawson}},
  \bibinfo {author} {\bibfnamefont {A.}~\bibnamefont {{Font-Ribera}}}, \bibinfo
  {author} {\bibfnamefont {J.-M.~L.}\ \bibnamefont {Goff}}, \bibinfo {author}
  {\bibnamefont {{Shirley Ho}}}, \bibinfo {author} {\bibfnamefont
  {K.}~\bibnamefont {Honscheid}}, \bibinfo {author} {\bibfnamefont {K.-G.}\
  \bibnamefont {Lee}}, \bibinfo {author} {\bibfnamefont {D.}~\bibnamefont
  {Margala}}, \bibinfo {author} {\bibfnamefont {P.}~\bibnamefont {McDonald}},
  \bibinfo {author} {\bibfnamefont {B.}~\bibnamefont {Medolin}}, \bibinfo
  {author} {\bibnamefont {{Jordi Miralda-Escud\'e}}}, \bibinfo {author}
  {\bibfnamefont {A.~D.}\ \bibnamefont {Myers}}, \bibinfo {author}
  {\bibfnamefont {R.~C.}\ \bibnamefont {Nichol}}, \bibinfo {author}
  {\bibfnamefont {P.}~\bibnamefont {Noterdaeme}}, \bibinfo {author}
  {\bibnamefont {{Nathalie Palanque-Delabrouille}}}, \bibinfo {author}
  {\bibfnamefont {I.}~\bibnamefont {P{\^a}ris}}, \bibinfo {author}
  {\bibfnamefont {P.}~\bibnamefont {Petitjean}}, \bibinfo {author}
  {\bibfnamefont {M.~M.}\ \bibnamefont {Pieri}}, \bibinfo {author}
  {\bibfnamefont {Y.}~\bibnamefont {Pi{\v s}kur}}, \bibinfo {author}
  {\bibfnamefont {N.~A.}\ \bibnamefont {Roe}}, \bibinfo {author} {\bibfnamefont
  {N.~P.}\ \bibnamefont {Ross}}, \bibinfo {author} {\bibfnamefont
  {G.}~\bibnamefont {Rossi}}, \bibinfo {author} {\bibfnamefont {D.~J.}\
  \bibnamefont {Schlegel}}, \bibinfo {author} {\bibfnamefont {D.~P.}\
  \bibnamefont {Schneider}}, \bibinfo {author} {\bibfnamefont {N.}~\bibnamefont
  {Suzuki}}, \bibinfo {author} {\bibfnamefont {E.~S.}\ \bibnamefont {Sheldon}},
  \bibinfo {author} {\bibfnamefont {U.}~\bibnamefont {Seljak}}, \bibinfo
  {author} {\bibfnamefont {M.}~\bibnamefont {Viel}}, \bibinfo {author}
  {\bibfnamefont {D.~H.}\ \bibnamefont {Weinberg}},\ and\ \bibinfo {author}
  {\bibfnamefont {C.}~\bibnamefont {Y{\`e}che}},\ }\bibfield  {title} {\bibinfo
  {title} {Measurement of baryon acoustic oscillations in the
  {{Lyman-$\alpha$}} forest fluctuations in {{BOSS}} data release 9},\ }\href
  {https://doi.org/10.1088/1475-7516/2013/04/026} {\bibfield  {journal}
  {\bibinfo  {journal} {\jcap}\ }\textbf {\bibinfo {volume} {2013}},\ \bibinfo
  {pages} {026} (\bibinfo {year} {2013})}\BibitemShut {NoStop}%
\bibitem [{\citenamefont {{Font-Ribera}}\ \emph {et~al.}(2018)\citenamefont
  {{Font-Ribera}}, \citenamefont {{McDonald}},\ and\ \citenamefont
  {{Slosar}}}]{fontriberaEstimate3DPowerLya2018}%
  \BibitemOpen
  \bibfield  {author} {\bibinfo {author} {\bibfnamefont {A.}~\bibnamefont
  {{Font-Ribera}}}, \bibinfo {author} {\bibfnamefont {P.}~\bibnamefont
  {{McDonald}}},\ and\ \bibinfo {author} {\bibfnamefont {A.}~\bibnamefont
  {{Slosar}}},\ }\bibfield  {title} {\bibinfo {title} {{How to estimate the 3D
  power spectrum of the Lyman-{\ensuremath{\alpha}} forest}},\ }\href
  {https://doi.org/10.1088/1475-7516/2018/01/003} {\bibfield  {journal}
  {\bibinfo  {journal} {\jcap}\ }\textbf {\bibinfo {volume} {2018}},\ \bibinfo
  {eid} {003} (\bibinfo {year} {2018})},\ \Eprint
  {https://arxiv.org/abs/1710.11036} {arXiv:1710.11036 [astro-ph.CO]}
  \BibitemShut {NoStop}%
\bibitem [{\citenamefont {{de Belsunce}}\ \emph {et~al.}(2024)\citenamefont
  {{de Belsunce}}, \citenamefont {{Philcox}}, \citenamefont
  {{Ir{\v{s}}i{\v{c}}}}, \citenamefont {{McDonald}}, \citenamefont {{Guy}},\
  and\ \citenamefont
  {{Palanque-Delabrouille}}}]{belsunce3dLymanAlphaPowerSpectru2024}%
  \BibitemOpen
  \bibfield  {author} {\bibinfo {author} {\bibfnamefont {R.}~\bibnamefont {{de
  Belsunce}}}, \bibinfo {author} {\bibfnamefont {O.~H.~E.}\ \bibnamefont
  {{Philcox}}}, \bibinfo {author} {\bibfnamefont {V.}~\bibnamefont
  {{Ir{\v{s}}i{\v{c}}}}}, \bibinfo {author} {\bibfnamefont {P.}~\bibnamefont
  {{McDonald}}}, \bibinfo {author} {\bibfnamefont {J.}~\bibnamefont {{Guy}}},\
  and\ \bibinfo {author} {\bibfnamefont {N.}~\bibnamefont
  {{Palanque-Delabrouille}}},\ }\bibfield  {title} {\bibinfo {title} {{The 3D
  Lyman-{\ensuremath{\alpha}} forest power spectrum from eBOSS DR16}},\ }\href
  {https://doi.org/10.1093/mnras/stae2035} {\bibfield  {journal} {\bibinfo
  {journal} {\mnras}\ }\textbf {\bibinfo {volume} {533}},\ \bibinfo {pages}
  {3756} (\bibinfo {year} {2024})},\ \Eprint {https://arxiv.org/abs/2403.08241}
  {arXiv:2403.08241 [astro-ph.CO]} \BibitemShut {NoStop}%
\bibitem [{\citenamefont {{Kara{\c{c}}ayl{\i}}}\ and\ \citenamefont
  {{Hirata}}(2025)}]{karacayliOptimal3dEstimator2025}%
  \BibitemOpen
  \bibfield  {author} {\bibinfo {author} {\bibfnamefont {N.~G.}\ \bibnamefont
  {{Kara{\c{c}}ayl{\i}}}}\ and\ \bibinfo {author} {\bibfnamefont {C.~M.}\
  \bibnamefont {{Hirata}}},\ }\bibfield  {title} {\bibinfo {title} {{Light in
  the dark forest. Part I. An efficient optimal estimator for 3D Lyman-alpha
  forest power spectrum}},\ }\href
  {https://doi.org/10.1088/1475-7516/2025/07/085} {\bibfield  {journal}
  {\bibinfo  {journal} {\jcap}\ }\textbf {\bibinfo {volume} {2025}},\ \bibinfo
  {eid} {085} (\bibinfo {year} {2025})},\ \Eprint
  {https://arxiv.org/abs/2503.15619} {arXiv:2503.15619 [astro-ph.CO]}
  \BibitemShut {NoStop}%
\bibitem [{\citenamefont {{Croft}}\ \emph {et~al.}(1998)\citenamefont
  {{Croft}}, \citenamefont {{Weinberg}}, \citenamefont {{Katz}},\ and\
  \citenamefont {{Hernquist}}}]{croftRecoveryPowerSpectrum1998}%
  \BibitemOpen
  \bibfield  {author} {\bibinfo {author} {\bibfnamefont {R.~A.~C.}\
  \bibnamefont {{Croft}}}, \bibinfo {author} {\bibfnamefont {D.~H.}\
  \bibnamefont {{Weinberg}}}, \bibinfo {author} {\bibfnamefont
  {N.}~\bibnamefont {{Katz}}},\ and\ \bibinfo {author} {\bibfnamefont
  {L.}~\bibnamefont {{Hernquist}}},\ }\bibfield  {title} {\bibinfo {title}
  {{Recovery of the Power Spectrum of Mass Fluctuations from Observations of
  the Ly{\ensuremath{\alpha}} Forest}},\ }\href
  {https://doi.org/10.1086/305289} {\bibfield  {journal} {\bibinfo  {journal}
  {\apj}\ }\textbf {\bibinfo {volume} {495}},\ \bibinfo {pages} {44} (\bibinfo
  {year} {1998})},\ \Eprint {https://arxiv.org/abs/astro-ph/9708018}
  {arXiv:astro-ph/9708018 [astro-ph]} \BibitemShut {NoStop}%
\bibitem [{\citenamefont {McDonald}\ \emph {et~al.}(2006)\citenamefont
  {McDonald}, \citenamefont {Seljak}, \citenamefont {Burles}, \citenamefont
  {Schlegel}, \citenamefont {Weinberg}, \citenamefont {Cen}, \citenamefont
  {Shih}, \citenamefont {Schaye}, \citenamefont {Schneider}, \citenamefont
  {Bahcall}, \citenamefont {Briggs}, \citenamefont {Brinkmann}, \citenamefont
  {Brunner}, \citenamefont {Fukugita}, \citenamefont {Gunn}, \citenamefont
  {Ivezi{\'c}}, \citenamefont {Kent}, \citenamefont {Lupton},\ and\
  \citenamefont {Berk}}]{mcdonaldLyUpalphaForest2006}%
  \BibitemOpen
  \bibfield  {author} {\bibinfo {author} {\bibfnamefont {P.}~\bibnamefont
  {McDonald}}, \bibinfo {author} {\bibfnamefont {U.}~\bibnamefont {Seljak}},
  \bibinfo {author} {\bibfnamefont {S.}~\bibnamefont {Burles}}, \bibinfo
  {author} {\bibfnamefont {D.~J.}\ \bibnamefont {Schlegel}}, \bibinfo {author}
  {\bibfnamefont {D.~H.}\ \bibnamefont {Weinberg}}, \bibinfo {author}
  {\bibfnamefont {R.}~\bibnamefont {Cen}}, \bibinfo {author} {\bibfnamefont
  {D.}~\bibnamefont {Shih}}, \bibinfo {author} {\bibfnamefont {J.}~\bibnamefont
  {Schaye}}, \bibinfo {author} {\bibfnamefont {D.~P.}\ \bibnamefont
  {Schneider}}, \bibinfo {author} {\bibfnamefont {N.~A.}\ \bibnamefont
  {Bahcall}}, \bibinfo {author} {\bibfnamefont {J.~W.}\ \bibnamefont {Briggs}},
  \bibinfo {author} {\bibfnamefont {J.}~\bibnamefont {Brinkmann}}, \bibinfo
  {author} {\bibfnamefont {R.~J.}\ \bibnamefont {Brunner}}, \bibinfo {author}
  {\bibfnamefont {M.}~\bibnamefont {Fukugita}}, \bibinfo {author}
  {\bibfnamefont {J.~E.}\ \bibnamefont {Gunn}}, \bibinfo {author}
  {\bibfnamefont {{\v Z}.}~\bibnamefont {Ivezi{\'c}}}, \bibinfo {author}
  {\bibfnamefont {S.}~\bibnamefont {Kent}}, \bibinfo {author} {\bibfnamefont
  {R.~H.}\ \bibnamefont {Lupton}},\ and\ \bibinfo {author} {\bibfnamefont
  {D.~E.~V.}\ \bibnamefont {Berk}},\ }\bibfield  {title} {\bibinfo {title} {The
  {{Ly}}alpha {{Forest Power Spectrum}} from the {{Sloan Digital Sky
  Survey}}},\ }\href {https://doi.org/10.1086/444361} {\bibfield  {journal}
  {\bibinfo  {journal} {\apjs}\ }\textbf {\bibinfo {volume} {163}},\ \bibinfo
  {pages} {80} (\bibinfo {year} {2006})}\BibitemShut {NoStop}%
\bibitem [{\citenamefont {{Palanque-Delabrouille}}\ \emph
  {et~al.}(2013)\citenamefont {{Palanque-Delabrouille}}, \citenamefont
  {{Y{\`e}che}}, \citenamefont {{Borde}}, \citenamefont {{Le Goff}},
  \citenamefont {{Rossi}}, \citenamefont {{Viel}}, \citenamefont {{Aubourg}},
  \citenamefont {{Bailey}}, \citenamefont {{Bautista}}, \citenamefont
  {{Blomqvist}}, \citenamefont {{Bolton}}, \citenamefont {{Bolton}},
  \citenamefont {{Busca}}, \citenamefont {{Carithers}}, \citenamefont
  {{Croft}}, \citenamefont {{Dawson}}, \citenamefont {{Delubac}}, \citenamefont
  {{Font-Ribera}}, \citenamefont {{Ho}}, \citenamefont {{Kirkby}},
  \citenamefont {{Lee}}, \citenamefont {{Margala}}, \citenamefont
  {{Miralda-Escud{\'e}}}, \citenamefont {{Muna}}, \citenamefont {{Myers}},
  \citenamefont {{Noterdaeme}}, \citenamefont {{P{\^a}ris}}, \citenamefont
  {{Petitjean}}, \citenamefont {{Pieri}}, \citenamefont {{Rich}}, \citenamefont
  {{Rollinde}}, \citenamefont {{Ross}}, \citenamefont {{Schlegel}},
  \citenamefont {{Schneider}}, \citenamefont {{Slosar}},\ and\ \citenamefont
  {{Weinberg}}}]{palanque-delabrouilleOnedimensionalLyalphaForest2013}%
  \BibitemOpen
  \bibfield  {author} {\bibinfo {author} {\bibfnamefont {N.}~\bibnamefont
  {{Palanque-Delabrouille}}}, \bibinfo {author} {\bibfnamefont
  {C.}~\bibnamefont {{Y{\`e}che}}}, \bibinfo {author} {\bibfnamefont
  {A.}~\bibnamefont {{Borde}}}, \bibinfo {author} {\bibfnamefont {J.-M.}\
  \bibnamefont {{Le Goff}}}, \bibinfo {author} {\bibfnamefont {G.}~\bibnamefont
  {{Rossi}}}, \bibinfo {author} {\bibfnamefont {M.}~\bibnamefont {{Viel}}},
  \bibinfo {author} {\bibfnamefont {{\'E}.}~\bibnamefont {{Aubourg}}}, \bibinfo
  {author} {\bibfnamefont {S.}~\bibnamefont {{Bailey}}}, \bibinfo {author}
  {\bibfnamefont {J.}~\bibnamefont {{Bautista}}}, \bibinfo {author}
  {\bibfnamefont {M.}~\bibnamefont {{Blomqvist}}}, \bibinfo {author}
  {\bibfnamefont {A.}~\bibnamefont {{Bolton}}}, \bibinfo {author}
  {\bibfnamefont {J.~S.}\ \bibnamefont {{Bolton}}}, \bibinfo {author}
  {\bibfnamefont {N.~G.}\ \bibnamefont {{Busca}}}, \bibinfo {author}
  {\bibfnamefont {B.}~\bibnamefont {{Carithers}}}, \bibinfo {author}
  {\bibfnamefont {R.~A.~C.}\ \bibnamefont {{Croft}}}, \bibinfo {author}
  {\bibfnamefont {K.~S.}\ \bibnamefont {{Dawson}}}, \bibinfo {author}
  {\bibfnamefont {T.}~\bibnamefont {{Delubac}}}, \bibinfo {author}
  {\bibfnamefont {A.}~\bibnamefont {{Font-Ribera}}}, \bibinfo {author}
  {\bibfnamefont {S.}~\bibnamefont {{Ho}}}, \bibinfo {author} {\bibfnamefont
  {D.}~\bibnamefont {{Kirkby}}}, \bibinfo {author} {\bibfnamefont {K.-G.}\
  \bibnamefont {{Lee}}}, \bibinfo {author} {\bibfnamefont {D.}~\bibnamefont
  {{Margala}}}, \bibinfo {author} {\bibfnamefont {J.}~\bibnamefont
  {{Miralda-Escud{\'e}}}}, \bibinfo {author} {\bibfnamefont {D.}~\bibnamefont
  {{Muna}}}, \bibinfo {author} {\bibfnamefont {A.~D.}\ \bibnamefont {{Myers}}},
  \bibinfo {author} {\bibfnamefont {P.}~\bibnamefont {{Noterdaeme}}}, \bibinfo
  {author} {\bibfnamefont {I.}~\bibnamefont {{P{\^a}ris}}}, \bibinfo {author}
  {\bibfnamefont {P.}~\bibnamefont {{Petitjean}}}, \bibinfo {author}
  {\bibfnamefont {M.~M.}\ \bibnamefont {{Pieri}}}, \bibinfo {author}
  {\bibfnamefont {J.}~\bibnamefont {{Rich}}}, \bibinfo {author} {\bibfnamefont
  {E.}~\bibnamefont {{Rollinde}}}, \bibinfo {author} {\bibfnamefont {N.~P.}\
  \bibnamefont {{Ross}}}, \bibinfo {author} {\bibfnamefont {D.~J.}\
  \bibnamefont {{Schlegel}}}, \bibinfo {author} {\bibfnamefont {D.~P.}\
  \bibnamefont {{Schneider}}}, \bibinfo {author} {\bibfnamefont
  {A.}~\bibnamefont {{Slosar}}},\ and\ \bibinfo {author} {\bibfnamefont
  {D.~H.}\ \bibnamefont {{Weinberg}}},\ }\bibfield  {title} {\bibinfo {title}
  {{The one-dimensional Ly{\ensuremath{\alpha}} forest power spectrum from
  BOSS}},\ }\href {https://doi.org/10.1051/0004-6361/201322130} {\bibfield
  {journal} {\bibinfo  {journal} {\aap}\ }\textbf {\bibinfo {volume} {559}},\
  \bibinfo {eid} {A85} (\bibinfo {year} {2013})},\ \Eprint
  {https://arxiv.org/abs/1306.5896} {arXiv:1306.5896 [astro-ph.CO]}
  \BibitemShut {NoStop}%
\bibitem [{\citenamefont {Chabanier}\ \emph {et~al.}(2019)\citenamefont
  {Chabanier}, \citenamefont {{Palanque-Delabrouille}}, \citenamefont
  {Y{\`e}che}, \citenamefont {Goff}, \citenamefont {Armengaud}, \citenamefont
  {Bautista}, \citenamefont {Blomqvist}, \citenamefont {Busca}, \citenamefont
  {Dawson}, \citenamefont {Etourneau}, \citenamefont {{Font-Ribera}},
  \citenamefont {Lee}, \citenamefont {des Bourboux}, \citenamefont {Pieri},
  \citenamefont {Rich}, \citenamefont {Rossi}, \citenamefont {Schneider},\ and\
  \citenamefont {Slosar}}]{chabanierOnedimensionalPowerSpectrum2019}%
  \BibitemOpen
  \bibfield  {author} {\bibinfo {author} {\bibfnamefont {S.}~\bibnamefont
  {Chabanier}}, \bibinfo {author} {\bibfnamefont {N.}~\bibnamefont
  {{Palanque-Delabrouille}}}, \bibinfo {author} {\bibfnamefont
  {C.}~\bibnamefont {Y{\`e}che}}, \bibinfo {author} {\bibfnamefont {J.-M.~L.}\
  \bibnamefont {Goff}}, \bibinfo {author} {\bibfnamefont {E.}~\bibnamefont
  {Armengaud}}, \bibinfo {author} {\bibfnamefont {J.}~\bibnamefont {Bautista}},
  \bibinfo {author} {\bibfnamefont {M.}~\bibnamefont {Blomqvist}}, \bibinfo
  {author} {\bibfnamefont {N.}~\bibnamefont {Busca}}, \bibinfo {author}
  {\bibfnamefont {K.}~\bibnamefont {Dawson}}, \bibinfo {author} {\bibfnamefont
  {T.}~\bibnamefont {Etourneau}}, \bibinfo {author} {\bibfnamefont
  {A.}~\bibnamefont {{Font-Ribera}}}, \bibinfo {author} {\bibfnamefont
  {Y.}~\bibnamefont {Lee}}, \bibinfo {author} {\bibfnamefont {H.~d.~M.}\
  \bibnamefont {des Bourboux}}, \bibinfo {author} {\bibfnamefont
  {M.}~\bibnamefont {Pieri}}, \bibinfo {author} {\bibfnamefont
  {J.}~\bibnamefont {Rich}}, \bibinfo {author} {\bibfnamefont {G.}~\bibnamefont
  {Rossi}}, \bibinfo {author} {\bibfnamefont {D.}~\bibnamefont {Schneider}},\
  and\ \bibinfo {author} {\bibfnamefont {A.}~\bibnamefont {Slosar}},\
  }\bibfield  {title} {\bibinfo {title} {The one-dimensional power spectrum
  from the {{SDSS DR14 Ly}}\ensuremath{\alpha} forests},\ }\href
  {https://doi.org/10.1088/1475-7516/2019/07/017} {\bibfield  {journal}
  {\bibinfo  {journal} {\jcap}\ }\textbf {\bibinfo {volume} {2019}},\ \bibinfo
  {pages} {017} (\bibinfo {year} {2019})}\BibitemShut {NoStop}%
\bibitem [{\citenamefont {{Kara{\c{c}}ayl{\i}}}\ \emph
  {et~al.}(2024)\citenamefont {{Kara{\c{c}}ayl{\i}}}, \citenamefont
  {{Martini}}, \citenamefont {{Guy}}, \citenamefont {{Ravoux}}, \citenamefont
  {{Karim}}, \citenamefont {{Armengaud}}, \citenamefont {{Walther}},
  \citenamefont {{Aguilar}}, \citenamefont {{Ahlen}}, \citenamefont {{Bailey}},
  \citenamefont {{Bautista}}, \citenamefont {{Beltran}}, \citenamefont
  {{Brooks}}, \citenamefont {{Cabayol-Garcia}}, \citenamefont {{Chabanier}},
  \citenamefont {{Chaussidon}}, \citenamefont {{Chaves-Montero}}, \citenamefont
  {{Dawson}}, \citenamefont {{de la Cruz}}, \citenamefont {{de la Macorra}},
  \citenamefont {{Doel}}, \citenamefont {{Font-Ribera}}, \citenamefont
  {{Forero-Romero}}, \citenamefont {{Gontcho}}, \citenamefont
  {{Gonzalez-Morales}}, \citenamefont {{Gordon}}, \citenamefont
  {{Herrera-Alcantar}}, \citenamefont {{Honscheid}}, \citenamefont
  {{Ir{\v{s}}i{\v{c}}}}, \citenamefont {{Ishak}}, \citenamefont {{Kehoe}},
  \citenamefont {{Kisner}}, \citenamefont {{Kremin}}, \citenamefont
  {{Landriau}}, \citenamefont {{Le Guillou}}, \citenamefont {{Levi}},
  \citenamefont {{Luki{\'c}}}, \citenamefont {{Meisner}}, \citenamefont
  {{Miquel}}, \citenamefont {{Moustakas}}, \citenamefont {{Mueller}},
  \citenamefont {{Mu{\~n}oz-Guti{\'e}rrez}}, \citenamefont {{Napolitano}},
  \citenamefont {{Nie}}, \citenamefont {{Niz}}, \citenamefont
  {{Palanque-Delabrouille}}, \citenamefont {{Percival}}, \citenamefont
  {{Pieri}}, \citenamefont {{Poppett}}, \citenamefont {{Prada}}, \citenamefont
  {{P{\'e}rez-R{\`a}fols}}, \citenamefont {{Ram{\'\i}rez-P{\'e}rez}},
  \citenamefont {{Rossi}}, \citenamefont {{Sanchez}}, \citenamefont {{Seo}},
  \citenamefont {{Sinigaglia}}, \citenamefont {{Tan}}, \citenamefont
  {{Tarl{\'e}}}, \citenamefont {{Wang}}, \citenamefont {{Weaver}},
  \citenamefont {{Y{\'e}che}},\ and\ \citenamefont
  {{Zhou}}}]{karacayliOptimal1dDesiEdr2023}%
  \BibitemOpen
  \bibfield  {author} {\bibinfo {author} {\bibfnamefont {N.~G.}\ \bibnamefont
  {{Kara{\c{c}}ayl{\i}}}}, \bibinfo {author} {\bibfnamefont {P.}~\bibnamefont
  {{Martini}}}, \bibinfo {author} {\bibfnamefont {J.}~\bibnamefont {{Guy}}},
  \bibinfo {author} {\bibfnamefont {C.}~\bibnamefont {{Ravoux}}}, \bibinfo
  {author} {\bibfnamefont {M.~L.~A.}\ \bibnamefont {{Karim}}}, \bibinfo
  {author} {\bibfnamefont {E.}~\bibnamefont {{Armengaud}}}, \bibinfo {author}
  {\bibfnamefont {M.}~\bibnamefont {{Walther}}}, \bibinfo {author}
  {\bibfnamefont {J.}~\bibnamefont {{Aguilar}}}, \bibinfo {author}
  {\bibfnamefont {S.}~\bibnamefont {{Ahlen}}}, \bibinfo {author} {\bibfnamefont
  {S.}~\bibnamefont {{Bailey}}}, \bibinfo {author} {\bibfnamefont
  {J.}~\bibnamefont {{Bautista}}}, \bibinfo {author} {\bibfnamefont {S.~F.}\
  \bibnamefont {{Beltran}}}, \bibinfo {author} {\bibfnamefont {D.}~\bibnamefont
  {{Brooks}}}, \bibinfo {author} {\bibfnamefont {L.}~\bibnamefont
  {{Cabayol-Garcia}}}, \bibinfo {author} {\bibfnamefont {S.}~\bibnamefont
  {{Chabanier}}}, \bibinfo {author} {\bibfnamefont {E.}~\bibnamefont
  {{Chaussidon}}}, \bibinfo {author} {\bibfnamefont {J.}~\bibnamefont
  {{Chaves-Montero}}}, \bibinfo {author} {\bibfnamefont {K.}~\bibnamefont
  {{Dawson}}}, \bibinfo {author} {\bibfnamefont {R.}~\bibnamefont {{de la
  Cruz}}}, \bibinfo {author} {\bibfnamefont {A.}~\bibnamefont {{de la
  Macorra}}}, \bibinfo {author} {\bibfnamefont {P.}~\bibnamefont {{Doel}}},
  \bibinfo {author} {\bibfnamefont {A.}~\bibnamefont {{Font-Ribera}}}, \bibinfo
  {author} {\bibfnamefont {J.~E.}\ \bibnamefont {{Forero-Romero}}}, \bibinfo
  {author} {\bibfnamefont {S.~G.~A.}\ \bibnamefont {{Gontcho}}}, \bibinfo
  {author} {\bibfnamefont {A.~X.}\ \bibnamefont {{Gonzalez-Morales}}}, \bibinfo
  {author} {\bibfnamefont {C.}~\bibnamefont {{Gordon}}}, \bibinfo {author}
  {\bibfnamefont {H.~K.}\ \bibnamefont {{Herrera-Alcantar}}}, \bibinfo {author}
  {\bibfnamefont {K.}~\bibnamefont {{Honscheid}}}, \bibinfo {author}
  {\bibfnamefont {V.}~\bibnamefont {{Ir{\v{s}}i{\v{c}}}}}, \bibinfo {author}
  {\bibfnamefont {M.}~\bibnamefont {{Ishak}}}, \bibinfo {author} {\bibfnamefont
  {R.}~\bibnamefont {{Kehoe}}}, \bibinfo {author} {\bibfnamefont
  {T.}~\bibnamefont {{Kisner}}}, \bibinfo {author} {\bibfnamefont
  {A.}~\bibnamefont {{Kremin}}}, \bibinfo {author} {\bibfnamefont
  {M.}~\bibnamefont {{Landriau}}}, \bibinfo {author} {\bibfnamefont
  {L.}~\bibnamefont {{Le Guillou}}}, \bibinfo {author} {\bibfnamefont {M.~E.}\
  \bibnamefont {{Levi}}}, \bibinfo {author} {\bibfnamefont {Z.}~\bibnamefont
  {{Luki{\'c}}}}, \bibinfo {author} {\bibfnamefont {A.}~\bibnamefont
  {{Meisner}}}, \bibinfo {author} {\bibfnamefont {R.}~\bibnamefont {{Miquel}}},
  \bibinfo {author} {\bibfnamefont {J.}~\bibnamefont {{Moustakas}}}, \bibinfo
  {author} {\bibfnamefont {E.}~\bibnamefont {{Mueller}}}, \bibinfo {author}
  {\bibfnamefont {A.}~\bibnamefont {{Mu{\~n}oz-Guti{\'e}rrez}}}, \bibinfo
  {author} {\bibfnamefont {L.}~\bibnamefont {{Napolitano}}}, \bibinfo {author}
  {\bibfnamefont {J.}~\bibnamefont {{Nie}}}, \bibinfo {author} {\bibfnamefont
  {G.}~\bibnamefont {{Niz}}}, \bibinfo {author} {\bibfnamefont
  {N.}~\bibnamefont {{Palanque-Delabrouille}}}, \bibinfo {author}
  {\bibfnamefont {W.~J.}\ \bibnamefont {{Percival}}}, \bibinfo {author}
  {\bibfnamefont {M.}~\bibnamefont {{Pieri}}}, \bibinfo {author} {\bibfnamefont
  {C.}~\bibnamefont {{Poppett}}}, \bibinfo {author} {\bibfnamefont
  {F.}~\bibnamefont {{Prada}}}, \bibinfo {author} {\bibfnamefont
  {I.}~\bibnamefont {{P{\'e}rez-R{\`a}fols}}}, \bibinfo {author} {\bibfnamefont
  {C.}~\bibnamefont {{Ram{\'\i}rez-P{\'e}rez}}}, \bibinfo {author}
  {\bibfnamefont {G.}~\bibnamefont {{Rossi}}}, \bibinfo {author} {\bibfnamefont
  {E.}~\bibnamefont {{Sanchez}}}, \bibinfo {author} {\bibfnamefont
  {H.}~\bibnamefont {{Seo}}}, \bibinfo {author} {\bibfnamefont
  {F.}~\bibnamefont {{Sinigaglia}}}, \bibinfo {author} {\bibfnamefont
  {T.}~\bibnamefont {{Tan}}}, \bibinfo {author} {\bibfnamefont
  {G.}~\bibnamefont {{Tarl{\'e}}}}, \bibinfo {author} {\bibfnamefont
  {B.}~\bibnamefont {{Wang}}}, \bibinfo {author} {\bibfnamefont {B.~A.}\
  \bibnamefont {{Weaver}}}, \bibinfo {author} {\bibfnamefont {C.}~\bibnamefont
  {{Y{\'e}che}}},\ and\ \bibinfo {author} {\bibfnamefont {Z.}~\bibnamefont
  {{Zhou}}},\ }\bibfield  {title} {\bibinfo {title} {{Optimal 1D
  Ly{\ensuremath{\alpha}} Forest Power Spectrum Estimation - III. DESI early
  data}},\ }\bibfield  {journal} {\bibinfo  {journal} {\mnras}\ }\href
  {https://doi.org/10.1093/mnras/stae171} {10.1093/mnras/stae171} (\bibinfo
  {year} {2024}),\ \Eprint {https://arxiv.org/abs/2306.06316} {arXiv:2306.06316
  [astro-ph.CO]} \BibitemShut {NoStop}%
\bibitem [{\citenamefont {{Ravoux}}\ \emph {et~al.}(2023)\citenamefont
  {{Ravoux}}, \citenamefont {{Abdul Karim}}, \citenamefont {{Armengaud}},
  \citenamefont {{Walther}}, \citenamefont {{Kara{\c{c}}ayl{\i}}},
  \citenamefont {{Martini}}, \citenamefont {{Guy}}, \citenamefont {{Aguilar}},
  \citenamefont {{Ahlen}}, \citenamefont {{Bailey}}, \citenamefont
  {{Bautista}}, \citenamefont {{Beltran}}, \citenamefont {{Brooks}},
  \citenamefont {{Cabayol-Garcia}}, \citenamefont {{Chabanier}}, \citenamefont
  {{Chaussidon}}, \citenamefont {{Chaves-Montero}}, \citenamefont {{Dawson}},
  \citenamefont {{de la Cruz}}, \citenamefont {{de la Macorra}}, \citenamefont
  {{Doel}}, \citenamefont {{Fanning}}, \citenamefont {{Font-Ribera}},
  \citenamefont {{Forero-Romero}}, \citenamefont {{Gontcho A Gontcho}},
  \citenamefont {{Gonzalez-Morales}}, \citenamefont {{Gordon}}, \citenamefont
  {{Herrera-Alcantar}}, \citenamefont {{Honscheid}}, \citenamefont
  {{Ir{\v{s}}i{\v{c}}}}, \citenamefont {{Ishak}}, \citenamefont {{Kehoe}},
  \citenamefont {{Kisner}}, \citenamefont {{Kremin}}, \citenamefont
  {{Landriau}}, \citenamefont {{Le Guillou}}, \citenamefont {{Levi}},
  \citenamefont {{Luki{\'c}}}, \citenamefont {{Magneville}}, \citenamefont
  {{Meisner}}, \citenamefont {{Miquel}}, \citenamefont {{Moustakas}},
  \citenamefont {{Mueller}}, \citenamefont {{Mu{\~n}oz-Guti{\'e}rrez}},
  \citenamefont {{Napolitano}}, \citenamefont {{Nie}}, \citenamefont {{Niz}},
  \citenamefont {{Palanque-Delabrouille}}, \citenamefont {{Percival}},
  \citenamefont {{P{\'e}rez-R{\`a}fols}}, \citenamefont {{Pieri}},
  \citenamefont {{Poppett}}, \citenamefont {{Prada}}, \citenamefont
  {{Ram{\'\i}rez P{\'e}rez}}, \citenamefont {{Rossi}}, \citenamefont
  {{Sanchez}}, \citenamefont {{Schlegel}}, \citenamefont {{Schubnell}},
  \citenamefont {{Seo}}, \citenamefont {{Sinigaglia}}, \citenamefont {{Tan}},
  \citenamefont {{Tarl{\'e}}}, \citenamefont {{Wang}}, \citenamefont
  {{Weaver}}, \citenamefont {{Y{\`e}che}},\ and\ \citenamefont
  {{Zhou}}}]{ravouxFFTP1dEDR2023}%
  \BibitemOpen
  \bibfield  {author} {\bibinfo {author} {\bibfnamefont {C.}~\bibnamefont
  {{Ravoux}}}, \bibinfo {author} {\bibfnamefont {M.~L.}\ \bibnamefont {{Abdul
  Karim}}}, \bibinfo {author} {\bibfnamefont {E.}~\bibnamefont {{Armengaud}}},
  \bibinfo {author} {\bibfnamefont {M.}~\bibnamefont {{Walther}}}, \bibinfo
  {author} {\bibfnamefont {N.~G.}\ \bibnamefont {{Kara{\c{c}}ayl{\i}}}},
  \bibinfo {author} {\bibfnamefont {P.}~\bibnamefont {{Martini}}}, \bibinfo
  {author} {\bibfnamefont {J.}~\bibnamefont {{Guy}}}, \bibinfo {author}
  {\bibfnamefont {J.~N.}\ \bibnamefont {{Aguilar}}}, \bibinfo {author}
  {\bibfnamefont {S.}~\bibnamefont {{Ahlen}}}, \bibinfo {author} {\bibfnamefont
  {S.}~\bibnamefont {{Bailey}}}, \bibinfo {author} {\bibfnamefont
  {J.}~\bibnamefont {{Bautista}}}, \bibinfo {author} {\bibfnamefont {S.~F.}\
  \bibnamefont {{Beltran}}}, \bibinfo {author} {\bibfnamefont {D.}~\bibnamefont
  {{Brooks}}}, \bibinfo {author} {\bibfnamefont {L.}~\bibnamefont
  {{Cabayol-Garcia}}}, \bibinfo {author} {\bibfnamefont {S.}~\bibnamefont
  {{Chabanier}}}, \bibinfo {author} {\bibfnamefont {E.}~\bibnamefont
  {{Chaussidon}}}, \bibinfo {author} {\bibfnamefont {J.}~\bibnamefont
  {{Chaves-Montero}}}, \bibinfo {author} {\bibfnamefont {K.}~\bibnamefont
  {{Dawson}}}, \bibinfo {author} {\bibfnamefont {R.}~\bibnamefont {{de la
  Cruz}}}, \bibinfo {author} {\bibfnamefont {A.}~\bibnamefont {{de la
  Macorra}}}, \bibinfo {author} {\bibfnamefont {P.}~\bibnamefont {{Doel}}},
  \bibinfo {author} {\bibfnamefont {K.}~\bibnamefont {{Fanning}}}, \bibinfo
  {author} {\bibfnamefont {A.}~\bibnamefont {{Font-Ribera}}}, \bibinfo {author}
  {\bibfnamefont {J.}~\bibnamefont {{Forero-Romero}}}, \bibinfo {author}
  {\bibfnamefont {S.}~\bibnamefont {{Gontcho A Gontcho}}}, \bibinfo {author}
  {\bibfnamefont {A.~X.}\ \bibnamefont {{Gonzalez-Morales}}}, \bibinfo {author}
  {\bibfnamefont {C.}~\bibnamefont {{Gordon}}}, \bibinfo {author}
  {\bibfnamefont {H.~K.}\ \bibnamefont {{Herrera-Alcantar}}}, \bibinfo {author}
  {\bibfnamefont {K.}~\bibnamefont {{Honscheid}}}, \bibinfo {author}
  {\bibfnamefont {V.}~\bibnamefont {{Ir{\v{s}}i{\v{c}}}}}, \bibinfo {author}
  {\bibfnamefont {M.}~\bibnamefont {{Ishak}}}, \bibinfo {author} {\bibfnamefont
  {R.}~\bibnamefont {{Kehoe}}}, \bibinfo {author} {\bibfnamefont
  {T.}~\bibnamefont {{Kisner}}}, \bibinfo {author} {\bibfnamefont
  {A.}~\bibnamefont {{Kremin}}}, \bibinfo {author} {\bibfnamefont
  {M.}~\bibnamefont {{Landriau}}}, \bibinfo {author} {\bibfnamefont
  {L.}~\bibnamefont {{Le Guillou}}}, \bibinfo {author} {\bibfnamefont
  {M.}~\bibnamefont {{Levi}}}, \bibinfo {author} {\bibfnamefont
  {Z.}~\bibnamefont {{Luki{\'c}}}}, \bibinfo {author} {\bibfnamefont
  {C.}~\bibnamefont {{Magneville}}}, \bibinfo {author} {\bibfnamefont
  {A.}~\bibnamefont {{Meisner}}}, \bibinfo {author} {\bibfnamefont
  {R.}~\bibnamefont {{Miquel}}}, \bibinfo {author} {\bibfnamefont
  {J.}~\bibnamefont {{Moustakas}}}, \bibinfo {author} {\bibfnamefont {E.-M.}\
  \bibnamefont {{Mueller}}}, \bibinfo {author} {\bibfnamefont {A.}~\bibnamefont
  {{Mu{\~n}oz-Guti{\'e}rrez}}}, \bibinfo {author} {\bibfnamefont
  {L.}~\bibnamefont {{Napolitano}}}, \bibinfo {author} {\bibfnamefont
  {J.}~\bibnamefont {{Nie}}}, \bibinfo {author} {\bibfnamefont
  {G.}~\bibnamefont {{Niz}}}, \bibinfo {author} {\bibfnamefont
  {N.}~\bibnamefont {{Palanque-Delabrouille}}}, \bibinfo {author}
  {\bibfnamefont {W.}~\bibnamefont {{Percival}}}, \bibinfo {author}
  {\bibfnamefont {I.}~\bibnamefont {{P{\'e}rez-R{\`a}fols}}}, \bibinfo {author}
  {\bibfnamefont {M.}~\bibnamefont {{Pieri}}}, \bibinfo {author} {\bibfnamefont
  {C.}~\bibnamefont {{Poppett}}}, \bibinfo {author} {\bibfnamefont
  {F.}~\bibnamefont {{Prada}}}, \bibinfo {author} {\bibfnamefont
  {C.}~\bibnamefont {{Ram{\'\i}rez P{\'e}rez}}}, \bibinfo {author}
  {\bibfnamefont {G.}~\bibnamefont {{Rossi}}}, \bibinfo {author} {\bibfnamefont
  {E.}~\bibnamefont {{Sanchez}}}, \bibinfo {author} {\bibfnamefont
  {D.}~\bibnamefont {{Schlegel}}}, \bibinfo {author} {\bibfnamefont
  {M.}~\bibnamefont {{Schubnell}}}, \bibinfo {author} {\bibfnamefont {H.-J.}\
  \bibnamefont {{Seo}}}, \bibinfo {author} {\bibfnamefont {F.}~\bibnamefont
  {{Sinigaglia}}}, \bibinfo {author} {\bibfnamefont {T.}~\bibnamefont {{Tan}}},
  \bibinfo {author} {\bibfnamefont {G.}~\bibnamefont {{Tarl{\'e}}}}, \bibinfo
  {author} {\bibfnamefont {B.}~\bibnamefont {{Wang}}}, \bibinfo {author}
  {\bibfnamefont {B.}~\bibnamefont {{Weaver}}}, \bibinfo {author}
  {\bibfnamefont {C.}~\bibnamefont {{Y{\`e}che}}},\ and\ \bibinfo {author}
  {\bibfnamefont {Z.}~\bibnamefont {{Zhou}}},\ }\bibfield  {title} {\bibinfo
  {title} {{The Dark Energy Spectroscopic Instrument: one-dimensional power
  spectrum from first Ly {\ensuremath{\alpha}} forest samples with Fast Fourier
  Transform}},\ }\href {https://doi.org/10.1093/mnras/stad3008} {\bibfield
  {journal} {\bibinfo  {journal} {\mnras}\ }\textbf {\bibinfo {volume} {526}},\
  \bibinfo {pages} {5118} (\bibinfo {year} {2023})},\ \Eprint
  {https://arxiv.org/abs/2306.06311} {arXiv:2306.06311 [astro-ph.CO]}
  \BibitemShut {NoStop}%
\bibitem [{\citenamefont {{Kara{\c{c}}ayl{\i}}}\ \emph
  {et~al.}(2025)\citenamefont {{Kara{\c{c}}ayl{\i}}}, \citenamefont
  {{Martini}}, \citenamefont {{Aguilar}}, \citenamefont {{Ahlen}},
  \citenamefont {{Armengaud}}, \citenamefont {{Bailey}}, \citenamefont
  {{Bault}}, \citenamefont {{Bianchi}}, \citenamefont {{Brodzeller}},
  \citenamefont {{Brooks}}, \citenamefont {{Chaves-Montero}}, \citenamefont
  {{Claybaugh}}, \citenamefont {{Cuceu}}, \citenamefont {{de la Macorra}},
  \citenamefont {{Dey}}, \citenamefont {{Dey}}, \citenamefont {{Doel}},
  \citenamefont {{Ferraro}}, \citenamefont {{Font-Ribera}}, \citenamefont
  {{Forero-Romero}}, \citenamefont {{Gazta{\~n}aga}}, \citenamefont
  {{Gontcho}}, \citenamefont {{Gutierrez}}, \citenamefont {{Guy}},
  \citenamefont {{Hahn}}, \citenamefont {{Herrera-Alcantar}}, \citenamefont
  {{Honscheid}}, \citenamefont {{Ishak}}, \citenamefont {{Kehoe}},
  \citenamefont {{Kirkby}}, \citenamefont {{Kremin}}, \citenamefont
  {{Landriau}}, \citenamefont {{Le Goff}}, \citenamefont {{Le Guillou}},
  \citenamefont {{Levi}}, \citenamefont {{Manera}}, \citenamefont {{Meisner}},
  \citenamefont {{Miquel}}, \citenamefont {{Montero-Camacho}}, \citenamefont
  {{Nadathur}}, \citenamefont {{Niz}}, \citenamefont {{Palanque-Delabrouille}},
  \citenamefont {{Pan}}, \citenamefont {{Percival}}, \citenamefont {{Pieri}},
  \citenamefont {{Prada}}, \citenamefont {{P{\'e}rez-R{\`a}fols}},
  \citenamefont {{Ravoux}}, \citenamefont {{Rossi}}, \citenamefont {{Sanchez}},
  \citenamefont {{Saulder}}, \citenamefont {{Schlegel}}, \citenamefont
  {{Schubnell}}, \citenamefont {{Seo}}, \citenamefont {{Siudek}}, \citenamefont
  {{Sprayberry}}, \citenamefont {{Tan}}, \citenamefont {{Tang}}, \citenamefont
  {{Tarl{\'e}}}, \citenamefont {{Walther}}, \citenamefont {{Weaver}},
  \citenamefont {{Yu}}, \citenamefont {{Zhou}},\ and\ \citenamefont
  {{Zou}}}]{karacayliQmleP1dDesiDr12024}%
  \BibitemOpen
  \bibfield  {author} {\bibinfo {author} {\bibfnamefont {N.~G.}\ \bibnamefont
  {{Kara{\c{c}}ayl{\i}}}}, \bibinfo {author} {\bibfnamefont {P.}~\bibnamefont
  {{Martini}}}, \bibinfo {author} {\bibfnamefont {J.}~\bibnamefont
  {{Aguilar}}}, \bibinfo {author} {\bibfnamefont {S.}~\bibnamefont {{Ahlen}}},
  \bibinfo {author} {\bibfnamefont {E.}~\bibnamefont {{Armengaud}}}, \bibinfo
  {author} {\bibfnamefont {S.}~\bibnamefont {{Bailey}}}, \bibinfo {author}
  {\bibfnamefont {A.}~\bibnamefont {{Bault}}}, \bibinfo {author} {\bibfnamefont
  {D.}~\bibnamefont {{Bianchi}}}, \bibinfo {author} {\bibfnamefont
  {A.}~\bibnamefont {{Brodzeller}}}, \bibinfo {author} {\bibfnamefont
  {D.}~\bibnamefont {{Brooks}}}, \bibinfo {author} {\bibfnamefont
  {J.}~\bibnamefont {{Chaves-Montero}}}, \bibinfo {author} {\bibfnamefont
  {T.}~\bibnamefont {{Claybaugh}}}, \bibinfo {author} {\bibfnamefont
  {A.}~\bibnamefont {{Cuceu}}}, \bibinfo {author} {\bibfnamefont
  {A.}~\bibnamefont {{de la Macorra}}}, \bibinfo {author} {\bibfnamefont
  {A.}~\bibnamefont {{Dey}}}, \bibinfo {author} {\bibfnamefont
  {B.}~\bibnamefont {{Dey}}}, \bibinfo {author} {\bibfnamefont
  {P.}~\bibnamefont {{Doel}}}, \bibinfo {author} {\bibfnamefont
  {S.}~\bibnamefont {{Ferraro}}}, \bibinfo {author} {\bibfnamefont
  {A.}~\bibnamefont {{Font-Ribera}}}, \bibinfo {author} {\bibfnamefont {J.~E.}\
  \bibnamefont {{Forero-Romero}}}, \bibinfo {author} {\bibfnamefont
  {E.}~\bibnamefont {{Gazta{\~n}aga}}}, \bibinfo {author} {\bibfnamefont
  {S.~G.~A.}\ \bibnamefont {{Gontcho}}}, \bibinfo {author} {\bibfnamefont
  {G.}~\bibnamefont {{Gutierrez}}}, \bibinfo {author} {\bibfnamefont
  {J.}~\bibnamefont {{Guy}}}, \bibinfo {author} {\bibfnamefont
  {C.}~\bibnamefont {{Hahn}}}, \bibinfo {author} {\bibfnamefont {H.~K.}\
  \bibnamefont {{Herrera-Alcantar}}}, \bibinfo {author} {\bibfnamefont
  {K.}~\bibnamefont {{Honscheid}}}, \bibinfo {author} {\bibfnamefont
  {M.}~\bibnamefont {{Ishak}}}, \bibinfo {author} {\bibfnamefont
  {R.}~\bibnamefont {{Kehoe}}}, \bibinfo {author} {\bibfnamefont
  {D.}~\bibnamefont {{Kirkby}}}, \bibinfo {author} {\bibfnamefont
  {A.}~\bibnamefont {{Kremin}}}, \bibinfo {author} {\bibfnamefont
  {M.}~\bibnamefont {{Landriau}}}, \bibinfo {author} {\bibfnamefont {J.~M.}\
  \bibnamefont {{Le Goff}}}, \bibinfo {author} {\bibfnamefont {L.}~\bibnamefont
  {{Le Guillou}}}, \bibinfo {author} {\bibfnamefont {M.~E.}\ \bibnamefont
  {{Levi}}}, \bibinfo {author} {\bibfnamefont {M.}~\bibnamefont {{Manera}}},
  \bibinfo {author} {\bibfnamefont {A.}~\bibnamefont {{Meisner}}}, \bibinfo
  {author} {\bibfnamefont {R.}~\bibnamefont {{Miquel}}}, \bibinfo {author}
  {\bibfnamefont {P.}~\bibnamefont {{Montero-Camacho}}}, \bibinfo {author}
  {\bibfnamefont {S.}~\bibnamefont {{Nadathur}}}, \bibinfo {author}
  {\bibfnamefont {G.}~\bibnamefont {{Niz}}}, \bibinfo {author} {\bibfnamefont
  {N.}~\bibnamefont {{Palanque-Delabrouille}}}, \bibinfo {author}
  {\bibfnamefont {Z.}~\bibnamefont {{Pan}}}, \bibinfo {author} {\bibfnamefont
  {W.~J.}\ \bibnamefont {{Percival}}}, \bibinfo {author} {\bibfnamefont
  {M.~M.}\ \bibnamefont {{Pieri}}}, \bibinfo {author} {\bibfnamefont
  {F.}~\bibnamefont {{Prada}}}, \bibinfo {author} {\bibfnamefont
  {I.}~\bibnamefont {{P{\'e}rez-R{\`a}fols}}}, \bibinfo {author} {\bibfnamefont
  {C.}~\bibnamefont {{Ravoux}}}, \bibinfo {author} {\bibfnamefont
  {G.}~\bibnamefont {{Rossi}}}, \bibinfo {author} {\bibfnamefont
  {E.}~\bibnamefont {{Sanchez}}}, \bibinfo {author} {\bibfnamefont
  {C.}~\bibnamefont {{Saulder}}}, \bibinfo {author} {\bibfnamefont
  {D.}~\bibnamefont {{Schlegel}}}, \bibinfo {author} {\bibfnamefont
  {M.}~\bibnamefont {{Schubnell}}}, \bibinfo {author} {\bibfnamefont
  {H.}~\bibnamefont {{Seo}}}, \bibinfo {author} {\bibfnamefont
  {M.}~\bibnamefont {{Siudek}}}, \bibinfo {author} {\bibfnamefont
  {D.}~\bibnamefont {{Sprayberry}}}, \bibinfo {author} {\bibfnamefont
  {T.}~\bibnamefont {{Tan}}}, \bibinfo {author} {\bibfnamefont {J.-J.}\
  \bibnamefont {{Tang}}}, \bibinfo {author} {\bibfnamefont {G.}~\bibnamefont
  {{Tarl{\'e}}}}, \bibinfo {author} {\bibfnamefont {M.}~\bibnamefont
  {{Walther}}}, \bibinfo {author} {\bibfnamefont {B.~A.}\ \bibnamefont
  {{Weaver}}}, \bibinfo {author} {\bibfnamefont {J.}~\bibnamefont {{Yu}}},
  \bibinfo {author} {\bibfnamefont {R.}~\bibnamefont {{Zhou}}},\ and\ \bibinfo
  {author} {\bibfnamefont {H.}~\bibnamefont {{Zou}}},\ }\bibfield  {title}
  {\bibinfo {title} {{DESI DR1 Ly{\ensuremath{\alpha}} 1D power spectrum: the
  optimal estimator measurement}},\ }\href
  {https://doi.org/10.1088/1475-7516/2025/10/004} {\bibfield  {journal}
  {\bibinfo  {journal} {\jcap}\ }\textbf {\bibinfo {volume} {2025}},\ \bibinfo
  {eid} {004} (\bibinfo {year} {2025})},\ \Eprint
  {https://arxiv.org/abs/2505.07974} {arXiv:2505.07974 [astro-ph.CO]}
  \BibitemShut {NoStop}%
\bibitem [{\citenamefont {{Ravoux}}\ \emph {et~al.}(2025)\citenamefont
  {{Ravoux}}, \citenamefont {{Abdul-Karim}}, \citenamefont {{Le Goff}},
  \citenamefont {{Armengaud}}, \citenamefont {{Aguilar}}, \citenamefont
  {{Ahlen}}, \citenamefont {{Bailey}}, \citenamefont {{Bianchi}}, \citenamefont
  {{Brodzeller}}, \citenamefont {{Brooks}}, \citenamefont {{Chaves-Montero}},
  \citenamefont {{Claybaugh}}, \citenamefont {{Cuceu}}, \citenamefont {{de
  Belsunce}}, \citenamefont {{de la Macorra}}, \citenamefont {{Dey}},
  \citenamefont {{Ding}}, \citenamefont {{Doel}}, \citenamefont {{Ferraro}},
  \citenamefont {{Font-Ribera}}, \citenamefont {{Forero-Romero}}, \citenamefont
  {{Gazta{\~n}aga}}, \citenamefont {{G{\"o}ksel Kara{\c{c}}ayl{\i}}},
  \citenamefont {{Gontcho}}, \citenamefont {{Gutierrez}}, \citenamefont
  {{Guy}}, \citenamefont {{Herrera-Alcantar}}, \citenamefont {{Ishak}},
  \citenamefont {{Kehoe}}, \citenamefont {{Kirkby}}, \citenamefont {{Kisner}},
  \citenamefont {{Kremin}}, \citenamefont {{Landriau}}, \citenamefont {{Le
  Guillou}}, \citenamefont {{Levi}}, \citenamefont {{Manera}}, \citenamefont
  {{Martini}}, \citenamefont {{Meisner}}, \citenamefont {{Miquel}},
  \citenamefont {{Montero-Camacho}}, \citenamefont {{Mu{\~n}oz-Guti{\'e}rrez}},
  \citenamefont {{Nadathur}}, \citenamefont {{Niz}}, \citenamefont
  {{Palanque-Delabrouille}}, \citenamefont {{Pan}}, \citenamefont {{Percival}},
  \citenamefont {{P{\'e}rez-R{\`a}fols}}, \citenamefont {{Pieri}},
  \citenamefont {{Prada}}, \citenamefont {{Rossi}}, \citenamefont {{Sanchez}},
  \citenamefont {{Saulder}}, \citenamefont {{Schlegel}}, \citenamefont
  {{Schubnell}}, \citenamefont {{Seo}}, \citenamefont {{Silber}}, \citenamefont
  {{Siudek}}, \citenamefont {{Sprayberry}}, \citenamefont {{Tan}},
  \citenamefont {{Tang}}, \citenamefont {{Tarl{\'e}}}, \citenamefont
  {{Walther}}, \citenamefont {{Weaver}}, \citenamefont {{Y{\`e}che}},
  \citenamefont {{Yu}}, \citenamefont {{Zhou}},\ and\ \citenamefont
  {{Zou}}}]{ravouxFFTP1dDesiDr12024}%
  \BibitemOpen
  \bibfield  {author} {\bibinfo {author} {\bibfnamefont {C.}~\bibnamefont
  {{Ravoux}}}, \bibinfo {author} {\bibfnamefont {M.-L.}\ \bibnamefont
  {{Abdul-Karim}}}, \bibinfo {author} {\bibfnamefont {J.-M.}\ \bibnamefont {{Le
  Goff}}}, \bibinfo {author} {\bibfnamefont {E.}~\bibnamefont {{Armengaud}}},
  \bibinfo {author} {\bibfnamefont {J.~N.}\ \bibnamefont {{Aguilar}}}, \bibinfo
  {author} {\bibfnamefont {S.}~\bibnamefont {{Ahlen}}}, \bibinfo {author}
  {\bibfnamefont {S.}~\bibnamefont {{Bailey}}}, \bibinfo {author}
  {\bibfnamefont {D.}~\bibnamefont {{Bianchi}}}, \bibinfo {author}
  {\bibfnamefont {A.}~\bibnamefont {{Brodzeller}}}, \bibinfo {author}
  {\bibfnamefont {D.}~\bibnamefont {{Brooks}}}, \bibinfo {author}
  {\bibfnamefont {J.}~\bibnamefont {{Chaves-Montero}}}, \bibinfo {author}
  {\bibfnamefont {T.}~\bibnamefont {{Claybaugh}}}, \bibinfo {author}
  {\bibfnamefont {A.}~\bibnamefont {{Cuceu}}}, \bibinfo {author} {\bibfnamefont
  {R.}~\bibnamefont {{de Belsunce}}}, \bibinfo {author} {\bibfnamefont
  {A.}~\bibnamefont {{de la Macorra}}}, \bibinfo {author} {\bibfnamefont
  {A.}~\bibnamefont {{Dey}}}, \bibinfo {author} {\bibfnamefont
  {Z.}~\bibnamefont {{Ding}}}, \bibinfo {author} {\bibfnamefont
  {P.}~\bibnamefont {{Doel}}}, \bibinfo {author} {\bibfnamefont
  {S.}~\bibnamefont {{Ferraro}}}, \bibinfo {author} {\bibfnamefont
  {A.}~\bibnamefont {{Font-Ribera}}}, \bibinfo {author} {\bibfnamefont {J.~E.}\
  \bibnamefont {{Forero-Romero}}}, \bibinfo {author} {\bibfnamefont
  {E.}~\bibnamefont {{Gazta{\~n}aga}}}, \bibinfo {author} {\bibfnamefont
  {N.}~\bibnamefont {{G{\"o}ksel Kara{\c{c}}ayl{\i}}}}, \bibinfo {author}
  {\bibfnamefont {S.~G.~A.}\ \bibnamefont {{Gontcho}}}, \bibinfo {author}
  {\bibfnamefont {G.}~\bibnamefont {{Gutierrez}}}, \bibinfo {author}
  {\bibfnamefont {J.}~\bibnamefont {{Guy}}}, \bibinfo {author} {\bibfnamefont
  {H.~K.}\ \bibnamefont {{Herrera-Alcantar}}}, \bibinfo {author} {\bibfnamefont
  {M.}~\bibnamefont {{Ishak}}}, \bibinfo {author} {\bibfnamefont
  {R.}~\bibnamefont {{Kehoe}}}, \bibinfo {author} {\bibfnamefont
  {D.}~\bibnamefont {{Kirkby}}}, \bibinfo {author} {\bibfnamefont
  {T.}~\bibnamefont {{Kisner}}}, \bibinfo {author} {\bibfnamefont
  {A.}~\bibnamefont {{Kremin}}}, \bibinfo {author} {\bibfnamefont
  {M.}~\bibnamefont {{Landriau}}}, \bibinfo {author} {\bibfnamefont
  {L.}~\bibnamefont {{Le Guillou}}}, \bibinfo {author} {\bibfnamefont {M.~E.}\
  \bibnamefont {{Levi}}}, \bibinfo {author} {\bibfnamefont {M.}~\bibnamefont
  {{Manera}}}, \bibinfo {author} {\bibfnamefont {P.}~\bibnamefont {{Martini}}},
  \bibinfo {author} {\bibfnamefont {A.}~\bibnamefont {{Meisner}}}, \bibinfo
  {author} {\bibfnamefont {R.}~\bibnamefont {{Miquel}}}, \bibinfo {author}
  {\bibfnamefont {P.}~\bibnamefont {{Montero-Camacho}}}, \bibinfo {author}
  {\bibfnamefont {A.}~\bibnamefont {{Mu{\~n}oz-Guti{\'e}rrez}}}, \bibinfo
  {author} {\bibfnamefont {S.}~\bibnamefont {{Nadathur}}}, \bibinfo {author}
  {\bibfnamefont {G.}~\bibnamefont {{Niz}}}, \bibinfo {author} {\bibfnamefont
  {N.}~\bibnamefont {{Palanque-Delabrouille}}}, \bibinfo {author}
  {\bibfnamefont {Z.}~\bibnamefont {{Pan}}}, \bibinfo {author} {\bibfnamefont
  {W.~J.}\ \bibnamefont {{Percival}}}, \bibinfo {author} {\bibfnamefont
  {I.}~\bibnamefont {{P{\'e}rez-R{\`a}fols}}}, \bibinfo {author} {\bibfnamefont
  {M.~M.}\ \bibnamefont {{Pieri}}}, \bibinfo {author} {\bibfnamefont
  {F.}~\bibnamefont {{Prada}}}, \bibinfo {author} {\bibfnamefont
  {G.}~\bibnamefont {{Rossi}}}, \bibinfo {author} {\bibfnamefont
  {E.}~\bibnamefont {{Sanchez}}}, \bibinfo {author} {\bibfnamefont
  {C.}~\bibnamefont {{Saulder}}}, \bibinfo {author} {\bibfnamefont
  {D.}~\bibnamefont {{Schlegel}}}, \bibinfo {author} {\bibfnamefont
  {M.}~\bibnamefont {{Schubnell}}}, \bibinfo {author} {\bibfnamefont {H.-J.}\
  \bibnamefont {{Seo}}}, \bibinfo {author} {\bibfnamefont {J.~H.}\ \bibnamefont
  {{Silber}}}, \bibinfo {author} {\bibfnamefont {M.}~\bibnamefont {{Siudek}}},
  \bibinfo {author} {\bibfnamefont {D.}~\bibnamefont {{Sprayberry}}}, \bibinfo
  {author} {\bibfnamefont {T.}~\bibnamefont {{Tan}}}, \bibinfo {author}
  {\bibfnamefont {J.-J.}\ \bibnamefont {{Tang}}}, \bibinfo {author}
  {\bibfnamefont {G.}~\bibnamefont {{Tarl{\'e}}}}, \bibinfo {author}
  {\bibfnamefont {M.}~\bibnamefont {{Walther}}}, \bibinfo {author}
  {\bibfnamefont {B.~A.}\ \bibnamefont {{Weaver}}}, \bibinfo {author}
  {\bibfnamefont {C.}~\bibnamefont {{Y{\`e}che}}}, \bibinfo {author}
  {\bibfnamefont {J.}~\bibnamefont {{Yu}}}, \bibinfo {author} {\bibfnamefont
  {R.}~\bibnamefont {{Zhou}}},\ and\ \bibinfo {author} {\bibfnamefont
  {H.}~\bibnamefont {{Zou}}},\ }\bibfield  {title} {\bibinfo {title} {{DESI DR1
  Ly{\ensuremath{\alpha}} 1D power spectrum: The Fast Fourier Transform
  estimator measurement}},\ }\href {https://doi.org/10.48550/arXiv.2505.09493}
  {\bibfield  {journal} {\bibinfo  {journal} {arXiv e-prints}\ ,\ \bibinfo
  {eid} {arXiv:2505.09493}} (\bibinfo {year} {2025})},\ \Eprint
  {https://arxiv.org/abs/2505.09493} {arXiv:2505.09493 [astro-ph.CO]}
  \BibitemShut {NoStop}%
\bibitem [{\citenamefont {{Bautista}}\ \emph {et~al.}(2017)\citenamefont
  {{Bautista}}, \citenamefont {{Busca}}, \citenamefont {{Guy}}, \citenamefont
  {{Rich}}, \citenamefont {{Blomqvist}}, \citenamefont {{du Mas des Bourboux}},
  \citenamefont {{Pieri}}, \citenamefont {{Font-Ribera}}, \citenamefont
  {{Bailey}}, \citenamefont {{Delubac}}, \citenamefont {{Kirkby}},
  \citenamefont {{Le Goff}}, \citenamefont {{Margala}}, \citenamefont
  {{Slosar}}, \citenamefont {{Vazquez}}, \citenamefont {{Brownstein}},
  \citenamefont {{Dawson}}, \citenamefont {{Eisenstein}}, \citenamefont
  {{Miralda-Escud{\'e}}}, \citenamefont {{Noterdaeme}}, \citenamefont
  {{Palanque-Delabrouille}}, \citenamefont {{P{\^a}ris}}, \citenamefont
  {{Petitjean}}, \citenamefont {{Ross}}, \citenamefont {{Schneider}},
  \citenamefont {{Weinberg}},\ and\ \citenamefont
  {{Y{\`e}che}}}]{bautistajuliane.MeasurementBaryonAcoustic2017}%
  \BibitemOpen
  \bibfield  {author} {\bibinfo {author} {\bibfnamefont {J.~E.}\ \bibnamefont
  {{Bautista}}}, \bibinfo {author} {\bibfnamefont {N.~G.}\ \bibnamefont
  {{Busca}}}, \bibinfo {author} {\bibfnamefont {J.}~\bibnamefont {{Guy}}},
  \bibinfo {author} {\bibfnamefont {J.}~\bibnamefont {{Rich}}}, \bibinfo
  {author} {\bibfnamefont {M.}~\bibnamefont {{Blomqvist}}}, \bibinfo {author}
  {\bibfnamefont {H.}~\bibnamefont {{du Mas des Bourboux}}}, \bibinfo {author}
  {\bibfnamefont {M.~M.}\ \bibnamefont {{Pieri}}}, \bibinfo {author}
  {\bibfnamefont {A.}~\bibnamefont {{Font-Ribera}}}, \bibinfo {author}
  {\bibfnamefont {S.}~\bibnamefont {{Bailey}}}, \bibinfo {author}
  {\bibfnamefont {T.}~\bibnamefont {{Delubac}}}, \bibinfo {author}
  {\bibfnamefont {D.}~\bibnamefont {{Kirkby}}}, \bibinfo {author}
  {\bibfnamefont {J.-M.}\ \bibnamefont {{Le Goff}}}, \bibinfo {author}
  {\bibfnamefont {D.}~\bibnamefont {{Margala}}}, \bibinfo {author}
  {\bibfnamefont {A.}~\bibnamefont {{Slosar}}}, \bibinfo {author}
  {\bibfnamefont {J.~A.}\ \bibnamefont {{Vazquez}}}, \bibinfo {author}
  {\bibfnamefont {J.~R.}\ \bibnamefont {{Brownstein}}}, \bibinfo {author}
  {\bibfnamefont {K.~S.}\ \bibnamefont {{Dawson}}}, \bibinfo {author}
  {\bibfnamefont {D.~J.}\ \bibnamefont {{Eisenstein}}}, \bibinfo {author}
  {\bibfnamefont {J.}~\bibnamefont {{Miralda-Escud{\'e}}}}, \bibinfo {author}
  {\bibfnamefont {P.}~\bibnamefont {{Noterdaeme}}}, \bibinfo {author}
  {\bibfnamefont {N.}~\bibnamefont {{Palanque-Delabrouille}}}, \bibinfo
  {author} {\bibfnamefont {I.}~\bibnamefont {{P{\^a}ris}}}, \bibinfo {author}
  {\bibfnamefont {P.}~\bibnamefont {{Petitjean}}}, \bibinfo {author}
  {\bibfnamefont {N.~P.}\ \bibnamefont {{Ross}}}, \bibinfo {author}
  {\bibfnamefont {D.~P.}\ \bibnamefont {{Schneider}}}, \bibinfo {author}
  {\bibfnamefont {D.~H.}\ \bibnamefont {{Weinberg}}},\ and\ \bibinfo {author}
  {\bibfnamefont {C.}~\bibnamefont {{Y{\`e}che}}},\ }\bibfield  {title}
  {\bibinfo {title} {{Measurement of baryon acoustic oscillation correlations
  at z = 2.3 with SDSS DR12 Ly{\ensuremath{\alpha}}-Forests}},\ }\href
  {https://doi.org/10.1051/0004-6361/201730533} {\bibfield  {journal} {\bibinfo
   {journal} {\aap}\ }\textbf {\bibinfo {volume} {603}},\ \bibinfo {eid} {A12}
  (\bibinfo {year} {2017})},\ \Eprint {https://arxiv.org/abs/1702.00176}
  {arXiv:1702.00176 [astro-ph.CO]} \BibitemShut {NoStop}%
\bibitem [{\citenamefont {{du Mas des Bourboux}}\ \emph
  {et~al.}(2020)\citenamefont {{du Mas des Bourboux}}, \citenamefont {{Rich}},
  \citenamefont {{Font-Ribera}}, \citenamefont {{de Sainte Agathe}},
  \citenamefont {{Farr}}, \citenamefont {{Etourneau}}, \citenamefont {{Le
  Goff}}, \citenamefont {{Cuceu}}, \citenamefont {{Balland}}, \citenamefont
  {{Bautista}}, \citenamefont {{Blomqvist}}, \citenamefont {{Brinkmann}},
  \citenamefont {{Brownstein}}, \citenamefont {{Chabanier}}, \citenamefont
  {{Chaussidon}}, \citenamefont {{Dawson}}, \citenamefont
  {{Gonz{\'a}lez-Morales}}, \citenamefont {{Guy}}, \citenamefont {{Lyke}},
  \citenamefont {{de la Macorra}}, \citenamefont {{Mueller}}, \citenamefont
  {{Myers}}, \citenamefont {{Nitschelm}}, \citenamefont {{Mu{\~n}oz
  Guti{\'e}rrez}}, \citenamefont {{Palanque-Delabrouille}}, \citenamefont
  {{Parker}}, \citenamefont {{Percival}}, \citenamefont
  {{P{\'e}rez-R{\`a}fols}}, \citenamefont {{Petitjean}}, \citenamefont
  {{Pieri}}, \citenamefont {{Ravoux}}, \citenamefont {{Rossi}}, \citenamefont
  {{Schneider}}, \citenamefont {{Seo}}, \citenamefont {{Slosar}}, \citenamefont
  {{Stermer}}, \citenamefont {{Vivek}}, \citenamefont {{Y{\`e}che}},\ and\
  \citenamefont {{Youles}}}]{bourbouxCompletedSDSSIVExtended2020}%
  \BibitemOpen
  \bibfield  {author} {\bibinfo {author} {\bibfnamefont {H.}~\bibnamefont {{du
  Mas des Bourboux}}}, \bibinfo {author} {\bibfnamefont {J.}~\bibnamefont
  {{Rich}}}, \bibinfo {author} {\bibfnamefont {A.}~\bibnamefont
  {{Font-Ribera}}}, \bibinfo {author} {\bibfnamefont {V.}~\bibnamefont {{de
  Sainte Agathe}}}, \bibinfo {author} {\bibfnamefont {J.}~\bibnamefont
  {{Farr}}}, \bibinfo {author} {\bibfnamefont {T.}~\bibnamefont {{Etourneau}}},
  \bibinfo {author} {\bibfnamefont {J.-M.}\ \bibnamefont {{Le Goff}}}, \bibinfo
  {author} {\bibfnamefont {A.}~\bibnamefont {{Cuceu}}}, \bibinfo {author}
  {\bibfnamefont {C.}~\bibnamefont {{Balland}}}, \bibinfo {author}
  {\bibfnamefont {J.~E.}\ \bibnamefont {{Bautista}}}, \bibinfo {author}
  {\bibfnamefont {M.}~\bibnamefont {{Blomqvist}}}, \bibinfo {author}
  {\bibfnamefont {J.}~\bibnamefont {{Brinkmann}}}, \bibinfo {author}
  {\bibfnamefont {J.~R.}\ \bibnamefont {{Brownstein}}}, \bibinfo {author}
  {\bibfnamefont {S.}~\bibnamefont {{Chabanier}}}, \bibinfo {author}
  {\bibfnamefont {E.}~\bibnamefont {{Chaussidon}}}, \bibinfo {author}
  {\bibfnamefont {K.}~\bibnamefont {{Dawson}}}, \bibinfo {author}
  {\bibfnamefont {A.~X.}\ \bibnamefont {{Gonz{\'a}lez-Morales}}}, \bibinfo
  {author} {\bibfnamefont {J.}~\bibnamefont {{Guy}}}, \bibinfo {author}
  {\bibfnamefont {B.~W.}\ \bibnamefont {{Lyke}}}, \bibinfo {author}
  {\bibfnamefont {A.}~\bibnamefont {{de la Macorra}}}, \bibinfo {author}
  {\bibfnamefont {E.-M.}\ \bibnamefont {{Mueller}}}, \bibinfo {author}
  {\bibfnamefont {A.~D.}\ \bibnamefont {{Myers}}}, \bibinfo {author}
  {\bibfnamefont {C.}~\bibnamefont {{Nitschelm}}}, \bibinfo {author}
  {\bibfnamefont {A.}~\bibnamefont {{Mu{\~n}oz Guti{\'e}rrez}}}, \bibinfo
  {author} {\bibfnamefont {N.}~\bibnamefont {{Palanque-Delabrouille}}},
  \bibinfo {author} {\bibfnamefont {J.}~\bibnamefont {{Parker}}}, \bibinfo
  {author} {\bibfnamefont {W.~J.}\ \bibnamefont {{Percival}}}, \bibinfo
  {author} {\bibfnamefont {I.}~\bibnamefont {{P{\'e}rez-R{\`a}fols}}}, \bibinfo
  {author} {\bibfnamefont {P.}~\bibnamefont {{Petitjean}}}, \bibinfo {author}
  {\bibfnamefont {M.~M.}\ \bibnamefont {{Pieri}}}, \bibinfo {author}
  {\bibfnamefont {C.}~\bibnamefont {{Ravoux}}}, \bibinfo {author}
  {\bibfnamefont {G.}~\bibnamefont {{Rossi}}}, \bibinfo {author} {\bibfnamefont
  {D.~P.}\ \bibnamefont {{Schneider}}}, \bibinfo {author} {\bibfnamefont
  {H.-J.}\ \bibnamefont {{Seo}}}, \bibinfo {author} {\bibfnamefont
  {A.}~\bibnamefont {{Slosar}}}, \bibinfo {author} {\bibfnamefont
  {J.}~\bibnamefont {{Stermer}}}, \bibinfo {author} {\bibfnamefont
  {M.}~\bibnamefont {{Vivek}}}, \bibinfo {author} {\bibfnamefont
  {C.}~\bibnamefont {{Y{\`e}che}}},\ and\ \bibinfo {author} {\bibfnamefont
  {S.}~\bibnamefont {{Youles}}},\ }\bibfield  {title} {\bibinfo {title} {{The
  Completed SDSS-IV Extended Baryon Oscillation Spectroscopic Survey: Baryon
  Acoustic Oscillations with Ly{\ensuremath{\alpha}} Forests}},\ }\href
  {https://doi.org/10.3847/1538-4357/abb085} {\bibfield  {journal} {\bibinfo
  {journal} {\apj}\ }\textbf {\bibinfo {volume} {901}},\ \bibinfo {eid} {153}
  (\bibinfo {year} {2020})},\ \Eprint {https://arxiv.org/abs/2007.08995}
  {arXiv:2007.08995 [astro-ph.CO]} \BibitemShut {NoStop}%
\bibitem [{\citenamefont {{Adame}}\ \emph {et~al.}(2025)\citenamefont
  {{Adame}}, \citenamefont {{Aguilar}}, \citenamefont {{Ahlen}}, \citenamefont
  {{Alam}}, \citenamefont {{Alexander}}, \citenamefont {{Alvarez}},
  \citenamefont {{Alves}}, \citenamefont {{Anand}}, \citenamefont {{Andrade}},
  \citenamefont {{Armengaud}}, \citenamefont {{Avila}}, \citenamefont
  {{Aviles}}, \citenamefont {{Awan}}, \citenamefont {{Bailey}}, \citenamefont
  {{Baltay}}, \citenamefont {{Bault}}, \citenamefont {{Bautista}},
  \citenamefont {{Behera}}, \citenamefont {{BenZvi}}, \citenamefont
  {{Beutler}}, \citenamefont {{Bianchi}}, \citenamefont {{Blake}},
  \citenamefont {{Blum}}, \citenamefont {{Brieden}}, \citenamefont
  {{Brodzeller}}, \citenamefont {{Brooks}}, \citenamefont {{Buckley-Geer}},
  \citenamefont {{Burtin}}, \citenamefont {{Calderon}}, \citenamefont
  {{Canning}}, \citenamefont {{Carnero Rosell}}, \citenamefont {{Cereskaite}},
  \citenamefont {{Cervantes-Cota}}, \citenamefont {{Chabanier}}, \citenamefont
  {{Chaussidon}}, \citenamefont {{Chaves-Montero}}, \citenamefont {{Chen}},
  \citenamefont {{Chen}}, \citenamefont {{Claybaugh}}, \citenamefont {{Cole}},
  \citenamefont {{Cuceu}}, \citenamefont {{Davis}}, \citenamefont {{Dawson}},
  \citenamefont {{de la Cruz}}, \citenamefont {{de la Macorra}}, \citenamefont
  {{de Mattia}}, \citenamefont {{Deiosso}}, \citenamefont {{Dey}},
  \citenamefont {{Dey}}, \citenamefont {{Ding}}, \citenamefont {{Ding}},
  \citenamefont {{Doel}}, \citenamefont {{Edelstein}}, \citenamefont
  {{Eftekharzadeh}}, \citenamefont {{Eisenstein}}, \citenamefont {{Elliott}},
  \citenamefont {{Fagrelius}}, \citenamefont {{Fanning}}, \citenamefont
  {{Ferraro}}, \citenamefont {{Ereza}}, \citenamefont {{Findlay}},
  \citenamefont {{Flaugher}}, \citenamefont {{Font-Ribera}}, \citenamefont
  {{Forero-S{\'a}nchez}}, \citenamefont {{Forero-Romero}}, \citenamefont
  {{Garcia-Quintero}}, \citenamefont {{Gazta{\~n}aga}}, \citenamefont
  {{Gil-Mar{\'\i}n}}, \citenamefont {{Gontcho}}, \citenamefont
  {{Gonzalez-Morales}}, \citenamefont {{Gonzalez-Perez}}, \citenamefont
  {{Gordon}}, \citenamefont {{Green}}, \citenamefont {{Gruen}}, \citenamefont
  {{Gsponer}}, \citenamefont {{Gutierrez}}, \citenamefont {{Guy}},
  \citenamefont {{Hadzhiyska}}, \citenamefont {{Hahn}}, \citenamefont
  {{Hanif}}, \citenamefont {{Herrera-Alcantar}}, \citenamefont {{Honscheid}},
  \citenamefont {{Howlett}}, \citenamefont {{Huterer}}, \citenamefont
  {{Ir{\v{s}}i{\v{c}}}}, \citenamefont {{Ishak}}, \citenamefont {{Juneau}},
  \citenamefont {{Kara{\c{c}}ayl{\i}}}, \citenamefont {{Kehoe}}, \citenamefont
  {{Kent}}, \citenamefont {{Kirkby}}, \citenamefont {{Kremin}}, \citenamefont
  {{Krolewski}}, \citenamefont {{Lai}}, \citenamefont {{Lan}}, \citenamefont
  {{Landriau}}, \citenamefont {{Lang}}, \citenamefont {{Lasker}}, \citenamefont
  {{Le Goff}}, \citenamefont {{Le Guillou}}, \citenamefont {{Leauthaud}},
  \citenamefont {{Levi}}, \citenamefont {{Li}}, \citenamefont {{Linder}},
  \citenamefont {{Lodha}}, \citenamefont {{Magneville}}, \citenamefont
  {{Manera}}, \citenamefont {{Margala}}, \citenamefont {{Martini}},
  \citenamefont {{Maus}}, \citenamefont {{McDonald}}, \citenamefont
  {{Medina-Varela}}, \citenamefont {{Meisner}}, \citenamefont
  {{Mena-Fern{\'a}ndez}}, \citenamefont {{Miquel}}, \citenamefont {{Moon}},
  \citenamefont {{Moore}}, \citenamefont {{Moustakas}}, \citenamefont
  {{Mueller}}, \citenamefont {{Mu{\~n}oz-Guti{\'e}rrez}}, \citenamefont
  {{Myers}}, \citenamefont {{Nadathur}}, \citenamefont {{Napolitano}},
  \citenamefont {{Neveux}}, \citenamefont {{Newman}}, \citenamefont {{Nguyen}},
  \citenamefont {{Nie}}, \citenamefont {{Niz}}, \citenamefont {{Noriega}},
  \citenamefont {{Padmanabhan}}, \citenamefont {{Paillas}}, \citenamefont
  {{Palanque-Delabrouille}}, \citenamefont {{Pan}}, \citenamefont {{Penmetsa}},
  \citenamefont {{Percival}}, \citenamefont {{Pieri}}, \citenamefont {{Pinon}},
  \citenamefont {{Poppett}}, \citenamefont {{Porredon}}, \citenamefont
  {{Prada}}, \citenamefont {{P{\'e}rez-Fern{\'a}ndez}}, \citenamefont
  {{P{\'e}rez-R{\`a}fols}}, \citenamefont {{Rabinowitz}}, \citenamefont
  {{Raichoor}}, \citenamefont {{Ram{\'\i}rez-P{\'e}rez}}, \citenamefont
  {{Ramirez-Solano}}, \citenamefont {{Rashkovetskyi}}, \citenamefont
  {{Ravoux}}, \citenamefont {{Rezaie}}, \citenamefont {{Rich}}, \citenamefont
  {{Rocher}}, \citenamefont {{Rockosi}}, \citenamefont {{Roe}}, \citenamefont
  {{Rosado-Marin}}, \citenamefont {{Ross}}, \citenamefont {{Rossi}},
  \citenamefont {{Ruggeri}}, \citenamefont {{Ruhlmann-Kleider}}, \citenamefont
  {{Samushia}}, \citenamefont {{Sanchez}}, \citenamefont {{Saulder}},
  \citenamefont {{Schlafly}}, \citenamefont {{Schlegel}}, \citenamefont
  {{Schubnell}}, \citenamefont {{Seo}}, \citenamefont {{Sharples}},
  \citenamefont {{Silber}}, \citenamefont {{Sinigaglia}}, \citenamefont
  {{Slosar}}, \citenamefont {{Smith}}, \citenamefont {{Sprayberry}},
  \citenamefont {{Tan}}, \citenamefont {{Tarl{\'e}}}, \citenamefont {{Trusov}},
  \citenamefont {{Vaisakh}}, \citenamefont {{Valcin}}, \citenamefont
  {{Valdes}}, \citenamefont {{Vargas-Maga{\~n}a}}, \citenamefont {{Verde}},
  \citenamefont {{Walther}}, \citenamefont {{Wang}}, \citenamefont {{Wang}},
  \citenamefont {{Weaver}}, \citenamefont {{Weaverdyck}}, \citenamefont
  {{Wechsler}}, \citenamefont {{Weinberg}}, \citenamefont {{White}},
  \citenamefont {{Yu}}, \citenamefont {{Yu}}, \citenamefont {{Yuan}},
  \citenamefont {{Y{\`e}che}}, \citenamefont {{Zaborowski}}, \citenamefont
  {{Zarrouk}}, \citenamefont {{Zhang}}, \citenamefont {{Zhao}}, \citenamefont
  {{Zhao}}, \citenamefont {{Zhou}}, \citenamefont {{Zou}},\ and\ \citenamefont
  {{DESI Collaboration}}}]{desiKp6BaoLya2024}%
  \BibitemOpen
  \bibfield  {author} {\bibinfo {author} {\bibfnamefont {A.~G.}\ \bibnamefont
  {{Adame}}}, \bibinfo {author} {\bibfnamefont {J.}~\bibnamefont {{Aguilar}}},
  \bibinfo {author} {\bibfnamefont {S.}~\bibnamefont {{Ahlen}}}, \bibinfo
  {author} {\bibfnamefont {S.}~\bibnamefont {{Alam}}}, \bibinfo {author}
  {\bibfnamefont {D.~M.}\ \bibnamefont {{Alexander}}}, \bibinfo {author}
  {\bibfnamefont {M.}~\bibnamefont {{Alvarez}}}, \bibinfo {author}
  {\bibfnamefont {O.}~\bibnamefont {{Alves}}}, \bibinfo {author} {\bibfnamefont
  {A.}~\bibnamefont {{Anand}}}, \bibinfo {author} {\bibfnamefont
  {U.}~\bibnamefont {{Andrade}}}, \bibinfo {author} {\bibfnamefont
  {E.}~\bibnamefont {{Armengaud}}}, \bibinfo {author} {\bibfnamefont
  {S.}~\bibnamefont {{Avila}}}, \bibinfo {author} {\bibfnamefont
  {A.}~\bibnamefont {{Aviles}}}, \bibinfo {author} {\bibfnamefont
  {H.}~\bibnamefont {{Awan}}}, \bibinfo {author} {\bibfnamefont
  {S.}~\bibnamefont {{Bailey}}}, \bibinfo {author} {\bibfnamefont
  {C.}~\bibnamefont {{Baltay}}}, \bibinfo {author} {\bibfnamefont
  {A.}~\bibnamefont {{Bault}}}, \bibinfo {author} {\bibfnamefont
  {J.}~\bibnamefont {{Bautista}}}, \bibinfo {author} {\bibfnamefont
  {J.}~\bibnamefont {{Behera}}}, \bibinfo {author} {\bibfnamefont
  {S.}~\bibnamefont {{BenZvi}}}, \bibinfo {author} {\bibfnamefont
  {F.}~\bibnamefont {{Beutler}}}, \bibinfo {author} {\bibfnamefont
  {D.}~\bibnamefont {{Bianchi}}}, \bibinfo {author} {\bibfnamefont
  {C.}~\bibnamefont {{Blake}}}, \bibinfo {author} {\bibfnamefont
  {R.}~\bibnamefont {{Blum}}}, \bibinfo {author} {\bibfnamefont
  {S.}~\bibnamefont {{Brieden}}}, \bibinfo {author} {\bibfnamefont
  {A.}~\bibnamefont {{Brodzeller}}}, \bibinfo {author} {\bibfnamefont
  {D.}~\bibnamefont {{Brooks}}}, \bibinfo {author} {\bibfnamefont
  {E.}~\bibnamefont {{Buckley-Geer}}}, \bibinfo {author} {\bibfnamefont
  {E.}~\bibnamefont {{Burtin}}}, \bibinfo {author} {\bibfnamefont
  {R.}~\bibnamefont {{Calderon}}}, \bibinfo {author} {\bibfnamefont
  {R.}~\bibnamefont {{Canning}}}, \bibinfo {author} {\bibfnamefont
  {A.}~\bibnamefont {{Carnero Rosell}}}, \bibinfo {author} {\bibfnamefont
  {R.}~\bibnamefont {{Cereskaite}}}, \bibinfo {author} {\bibfnamefont {J.~L.}\
  \bibnamefont {{Cervantes-Cota}}}, \bibinfo {author} {\bibfnamefont
  {S.}~\bibnamefont {{Chabanier}}}, \bibinfo {author} {\bibfnamefont
  {E.}~\bibnamefont {{Chaussidon}}}, \bibinfo {author} {\bibfnamefont
  {J.}~\bibnamefont {{Chaves-Montero}}}, \bibinfo {author} {\bibfnamefont
  {S.}~\bibnamefont {{Chen}}}, \bibinfo {author} {\bibfnamefont
  {X.}~\bibnamefont {{Chen}}}, \bibinfo {author} {\bibfnamefont
  {T.}~\bibnamefont {{Claybaugh}}}, \bibinfo {author} {\bibfnamefont
  {S.}~\bibnamefont {{Cole}}}, \bibinfo {author} {\bibfnamefont
  {A.}~\bibnamefont {{Cuceu}}}, \bibinfo {author} {\bibfnamefont {T.~M.}\
  \bibnamefont {{Davis}}}, \bibinfo {author} {\bibfnamefont {K.}~\bibnamefont
  {{Dawson}}}, \bibinfo {author} {\bibfnamefont {R.}~\bibnamefont {{de la
  Cruz}}}, \bibinfo {author} {\bibfnamefont {A.}~\bibnamefont {{de la
  Macorra}}}, \bibinfo {author} {\bibfnamefont {A.}~\bibnamefont {{de
  Mattia}}}, \bibinfo {author} {\bibfnamefont {N.}~\bibnamefont {{Deiosso}}},
  \bibinfo {author} {\bibfnamefont {A.}~\bibnamefont {{Dey}}}, \bibinfo
  {author} {\bibfnamefont {B.}~\bibnamefont {{Dey}}}, \bibinfo {author}
  {\bibfnamefont {J.}~\bibnamefont {{Ding}}}, \bibinfo {author} {\bibfnamefont
  {Z.}~\bibnamefont {{Ding}}}, \bibinfo {author} {\bibfnamefont
  {P.}~\bibnamefont {{Doel}}}, \bibinfo {author} {\bibfnamefont
  {J.}~\bibnamefont {{Edelstein}}}, \bibinfo {author} {\bibfnamefont
  {S.}~\bibnamefont {{Eftekharzadeh}}}, \bibinfo {author} {\bibfnamefont
  {D.~J.}\ \bibnamefont {{Eisenstein}}}, \bibinfo {author} {\bibfnamefont
  {A.}~\bibnamefont {{Elliott}}}, \bibinfo {author} {\bibfnamefont
  {P.}~\bibnamefont {{Fagrelius}}}, \bibinfo {author} {\bibfnamefont
  {K.}~\bibnamefont {{Fanning}}}, \bibinfo {author} {\bibfnamefont
  {S.}~\bibnamefont {{Ferraro}}}, \bibinfo {author} {\bibfnamefont
  {J.}~\bibnamefont {{Ereza}}}, \bibinfo {author} {\bibfnamefont
  {N.}~\bibnamefont {{Findlay}}}, \bibinfo {author} {\bibfnamefont
  {B.}~\bibnamefont {{Flaugher}}}, \bibinfo {author} {\bibfnamefont
  {A.}~\bibnamefont {{Font-Ribera}}}, \bibinfo {author} {\bibfnamefont
  {D.}~\bibnamefont {{Forero-S{\'a}nchez}}}, \bibinfo {author} {\bibfnamefont
  {J.~E.}\ \bibnamefont {{Forero-Romero}}}, \bibinfo {author} {\bibfnamefont
  {C.}~\bibnamefont {{Garcia-Quintero}}}, \bibinfo {author} {\bibfnamefont
  {E.}~\bibnamefont {{Gazta{\~n}aga}}}, \bibinfo {author} {\bibfnamefont
  {H.}~\bibnamefont {{Gil-Mar{\'\i}n}}}, \bibinfo {author} {\bibfnamefont
  {S.~G.~A.}\ \bibnamefont {{Gontcho}}}, \bibinfo {author} {\bibfnamefont
  {A.~X.}\ \bibnamefont {{Gonzalez-Morales}}}, \bibinfo {author} {\bibfnamefont
  {V.}~\bibnamefont {{Gonzalez-Perez}}}, \bibinfo {author} {\bibfnamefont
  {C.}~\bibnamefont {{Gordon}}}, \bibinfo {author} {\bibfnamefont
  {D.}~\bibnamefont {{Green}}}, \bibinfo {author} {\bibfnamefont
  {D.}~\bibnamefont {{Gruen}}}, \bibinfo {author} {\bibfnamefont
  {R.}~\bibnamefont {{Gsponer}}}, \bibinfo {author} {\bibfnamefont
  {G.}~\bibnamefont {{Gutierrez}}}, \bibinfo {author} {\bibfnamefont
  {J.}~\bibnamefont {{Guy}}}, \bibinfo {author} {\bibfnamefont
  {B.}~\bibnamefont {{Hadzhiyska}}}, \bibinfo {author} {\bibfnamefont
  {C.}~\bibnamefont {{Hahn}}}, \bibinfo {author} {\bibfnamefont {M.~M.~S.}\
  \bibnamefont {{Hanif}}}, \bibinfo {author} {\bibfnamefont {H.~K.}\
  \bibnamefont {{Herrera-Alcantar}}}, \bibinfo {author} {\bibfnamefont
  {K.}~\bibnamefont {{Honscheid}}}, \bibinfo {author} {\bibfnamefont
  {C.}~\bibnamefont {{Howlett}}}, \bibinfo {author} {\bibfnamefont
  {D.}~\bibnamefont {{Huterer}}}, \bibinfo {author} {\bibfnamefont
  {V.}~\bibnamefont {{Ir{\v{s}}i{\v{c}}}}}, \bibinfo {author} {\bibfnamefont
  {M.}~\bibnamefont {{Ishak}}}, \bibinfo {author} {\bibfnamefont
  {S.}~\bibnamefont {{Juneau}}}, \bibinfo {author} {\bibfnamefont {N.~G.}\
  \bibnamefont {{Kara{\c{c}}ayl{\i}}}}, \bibinfo {author} {\bibfnamefont
  {R.}~\bibnamefont {{Kehoe}}}, \bibinfo {author} {\bibfnamefont
  {S.}~\bibnamefont {{Kent}}}, \bibinfo {author} {\bibfnamefont
  {D.}~\bibnamefont {{Kirkby}}}, \bibinfo {author} {\bibfnamefont
  {A.}~\bibnamefont {{Kremin}}}, \bibinfo {author} {\bibfnamefont
  {A.}~\bibnamefont {{Krolewski}}}, \bibinfo {author} {\bibfnamefont
  {Y.}~\bibnamefont {{Lai}}}, \bibinfo {author} {\bibfnamefont {T.~W.}\
  \bibnamefont {{Lan}}}, \bibinfo {author} {\bibfnamefont {M.}~\bibnamefont
  {{Landriau}}}, \bibinfo {author} {\bibfnamefont {D.}~\bibnamefont {{Lang}}},
  \bibinfo {author} {\bibfnamefont {J.}~\bibnamefont {{Lasker}}}, \bibinfo
  {author} {\bibfnamefont {J.~M.}\ \bibnamefont {{Le Goff}}}, \bibinfo {author}
  {\bibfnamefont {L.}~\bibnamefont {{Le Guillou}}}, \bibinfo {author}
  {\bibfnamefont {A.}~\bibnamefont {{Leauthaud}}}, \bibinfo {author}
  {\bibfnamefont {M.~E.}\ \bibnamefont {{Levi}}}, \bibinfo {author}
  {\bibfnamefont {T.~S.}\ \bibnamefont {{Li}}}, \bibinfo {author}
  {\bibfnamefont {E.}~\bibnamefont {{Linder}}}, \bibinfo {author}
  {\bibfnamefont {K.}~\bibnamefont {{Lodha}}}, \bibinfo {author} {\bibfnamefont
  {C.}~\bibnamefont {{Magneville}}}, \bibinfo {author} {\bibfnamefont
  {M.}~\bibnamefont {{Manera}}}, \bibinfo {author} {\bibfnamefont
  {D.}~\bibnamefont {{Margala}}}, \bibinfo {author} {\bibfnamefont
  {P.}~\bibnamefont {{Martini}}}, \bibinfo {author} {\bibfnamefont
  {M.}~\bibnamefont {{Maus}}}, \bibinfo {author} {\bibfnamefont
  {P.}~\bibnamefont {{McDonald}}}, \bibinfo {author} {\bibfnamefont
  {L.}~\bibnamefont {{Medina-Varela}}}, \bibinfo {author} {\bibfnamefont
  {A.}~\bibnamefont {{Meisner}}}, \bibinfo {author} {\bibfnamefont
  {J.}~\bibnamefont {{Mena-Fern{\'a}ndez}}}, \bibinfo {author} {\bibfnamefont
  {R.}~\bibnamefont {{Miquel}}}, \bibinfo {author} {\bibfnamefont
  {J.}~\bibnamefont {{Moon}}}, \bibinfo {author} {\bibfnamefont
  {S.}~\bibnamefont {{Moore}}}, \bibinfo {author} {\bibfnamefont
  {J.}~\bibnamefont {{Moustakas}}}, \bibinfo {author} {\bibfnamefont
  {E.}~\bibnamefont {{Mueller}}}, \bibinfo {author} {\bibfnamefont
  {A.}~\bibnamefont {{Mu{\~n}oz-Guti{\'e}rrez}}}, \bibinfo {author}
  {\bibfnamefont {A.~D.}\ \bibnamefont {{Myers}}}, \bibinfo {author}
  {\bibfnamefont {S.}~\bibnamefont {{Nadathur}}}, \bibinfo {author}
  {\bibfnamefont {L.}~\bibnamefont {{Napolitano}}}, \bibinfo {author}
  {\bibfnamefont {R.}~\bibnamefont {{Neveux}}}, \bibinfo {author}
  {\bibfnamefont {J.~A.}\ \bibnamefont {{Newman}}}, \bibinfo {author}
  {\bibfnamefont {N.~M.}\ \bibnamefont {{Nguyen}}}, \bibinfo {author}
  {\bibfnamefont {J.}~\bibnamefont {{Nie}}}, \bibinfo {author} {\bibfnamefont
  {G.}~\bibnamefont {{Niz}}}, \bibinfo {author} {\bibfnamefont {H.~E.}\
  \bibnamefont {{Noriega}}}, \bibinfo {author} {\bibfnamefont {N.}~\bibnamefont
  {{Padmanabhan}}}, \bibinfo {author} {\bibfnamefont {E.}~\bibnamefont
  {{Paillas}}}, \bibinfo {author} {\bibfnamefont {N.}~\bibnamefont
  {{Palanque-Delabrouille}}}, \bibinfo {author} {\bibfnamefont
  {J.}~\bibnamefont {{Pan}}}, \bibinfo {author} {\bibfnamefont
  {S.}~\bibnamefont {{Penmetsa}}}, \bibinfo {author} {\bibfnamefont {W.~J.}\
  \bibnamefont {{Percival}}}, \bibinfo {author} {\bibfnamefont {M.~M.}\
  \bibnamefont {{Pieri}}}, \bibinfo {author} {\bibfnamefont {M.}~\bibnamefont
  {{Pinon}}}, \bibinfo {author} {\bibfnamefont {C.}~\bibnamefont {{Poppett}}},
  \bibinfo {author} {\bibfnamefont {A.}~\bibnamefont {{Porredon}}}, \bibinfo
  {author} {\bibfnamefont {F.}~\bibnamefont {{Prada}}}, \bibinfo {author}
  {\bibfnamefont {A.}~\bibnamefont {{P{\'e}rez-Fern{\'a}ndez}}}, \bibinfo
  {author} {\bibfnamefont {I.}~\bibnamefont {{P{\'e}rez-R{\`a}fols}}}, \bibinfo
  {author} {\bibfnamefont {D.}~\bibnamefont {{Rabinowitz}}}, \bibinfo {author}
  {\bibfnamefont {A.}~\bibnamefont {{Raichoor}}}, \bibinfo {author}
  {\bibfnamefont {C.}~\bibnamefont {{Ram{\'\i}rez-P{\'e}rez}}}, \bibinfo
  {author} {\bibfnamefont {S.}~\bibnamefont {{Ramirez-Solano}}}, \bibinfo
  {author} {\bibfnamefont {M.}~\bibnamefont {{Rashkovetskyi}}}, \bibinfo
  {author} {\bibfnamefont {C.}~\bibnamefont {{Ravoux}}}, \bibinfo {author}
  {\bibfnamefont {M.}~\bibnamefont {{Rezaie}}}, \bibinfo {author}
  {\bibfnamefont {J.}~\bibnamefont {{Rich}}}, \bibinfo {author} {\bibfnamefont
  {A.}~\bibnamefont {{Rocher}}}, \bibinfo {author} {\bibfnamefont
  {C.}~\bibnamefont {{Rockosi}}}, \bibinfo {author} {\bibfnamefont {N.~A.}\
  \bibnamefont {{Roe}}}, \bibinfo {author} {\bibfnamefont {A.}~\bibnamefont
  {{Rosado-Marin}}}, \bibinfo {author} {\bibfnamefont {A.~J.}\ \bibnamefont
  {{Ross}}}, \bibinfo {author} {\bibfnamefont {G.}~\bibnamefont {{Rossi}}},
  \bibinfo {author} {\bibfnamefont {R.}~\bibnamefont {{Ruggeri}}}, \bibinfo
  {author} {\bibfnamefont {V.}~\bibnamefont {{Ruhlmann-Kleider}}}, \bibinfo
  {author} {\bibfnamefont {L.}~\bibnamefont {{Samushia}}}, \bibinfo {author}
  {\bibfnamefont {E.}~\bibnamefont {{Sanchez}}}, \bibinfo {author}
  {\bibfnamefont {C.}~\bibnamefont {{Saulder}}}, \bibinfo {author}
  {\bibfnamefont {E.~F.}\ \bibnamefont {{Schlafly}}}, \bibinfo {author}
  {\bibfnamefont {D.}~\bibnamefont {{Schlegel}}}, \bibinfo {author}
  {\bibfnamefont {M.}~\bibnamefont {{Schubnell}}}, \bibinfo {author}
  {\bibfnamefont {H.}~\bibnamefont {{Seo}}}, \bibinfo {author} {\bibfnamefont
  {R.}~\bibnamefont {{Sharples}}}, \bibinfo {author} {\bibfnamefont
  {J.}~\bibnamefont {{Silber}}}, \bibinfo {author} {\bibfnamefont
  {F.}~\bibnamefont {{Sinigaglia}}}, \bibinfo {author} {\bibfnamefont
  {A.}~\bibnamefont {{Slosar}}}, \bibinfo {author} {\bibfnamefont
  {A.}~\bibnamefont {{Smith}}}, \bibinfo {author} {\bibfnamefont
  {D.}~\bibnamefont {{Sprayberry}}}, \bibinfo {author} {\bibfnamefont
  {T.}~\bibnamefont {{Tan}}}, \bibinfo {author} {\bibfnamefont
  {G.}~\bibnamefont {{Tarl{\'e}}}}, \bibinfo {author} {\bibfnamefont
  {S.}~\bibnamefont {{Trusov}}}, \bibinfo {author} {\bibfnamefont
  {R.}~\bibnamefont {{Vaisakh}}}, \bibinfo {author} {\bibfnamefont
  {D.}~\bibnamefont {{Valcin}}}, \bibinfo {author} {\bibfnamefont
  {F.}~\bibnamefont {{Valdes}}}, \bibinfo {author} {\bibfnamefont
  {M.}~\bibnamefont {{Vargas-Maga{\~n}a}}}, \bibinfo {author} {\bibfnamefont
  {L.}~\bibnamefont {{Verde}}}, \bibinfo {author} {\bibfnamefont
  {M.}~\bibnamefont {{Walther}}}, \bibinfo {author} {\bibfnamefont
  {B.}~\bibnamefont {{Wang}}}, \bibinfo {author} {\bibfnamefont {M.~S.}\
  \bibnamefont {{Wang}}}, \bibinfo {author} {\bibfnamefont {B.~A.}\
  \bibnamefont {{Weaver}}}, \bibinfo {author} {\bibfnamefont {N.}~\bibnamefont
  {{Weaverdyck}}}, \bibinfo {author} {\bibfnamefont {R.~H.}\ \bibnamefont
  {{Wechsler}}}, \bibinfo {author} {\bibfnamefont {D.~H.}\ \bibnamefont
  {{Weinberg}}}, \bibinfo {author} {\bibfnamefont {M.}~\bibnamefont {{White}}},
  \bibinfo {author} {\bibfnamefont {J.}~\bibnamefont {{Yu}}}, \bibinfo {author}
  {\bibfnamefont {Y.}~\bibnamefont {{Yu}}}, \bibinfo {author} {\bibfnamefont
  {S.}~\bibnamefont {{Yuan}}}, \bibinfo {author} {\bibfnamefont
  {C.}~\bibnamefont {{Y{\`e}che}}}, \bibinfo {author} {\bibfnamefont {E.~A.}\
  \bibnamefont {{Zaborowski}}}, \bibinfo {author} {\bibfnamefont
  {P.}~\bibnamefont {{Zarrouk}}}, \bibinfo {author} {\bibfnamefont
  {H.}~\bibnamefont {{Zhang}}}, \bibinfo {author} {\bibfnamefont
  {C.}~\bibnamefont {{Zhao}}}, \bibinfo {author} {\bibfnamefont
  {R.}~\bibnamefont {{Zhao}}}, \bibinfo {author} {\bibfnamefont
  {R.}~\bibnamefont {{Zhou}}}, \bibinfo {author} {\bibfnamefont
  {H.}~\bibnamefont {{Zou}}},\ and\ \bibinfo {author} {\bibnamefont {{DESI
  Collaboration}}},\ }\bibfield  {title} {\bibinfo {title} {{DESI 2024 IV:
  Baryon Acoustic Oscillations from the Lyman alpha forest}},\ }\href
  {https://doi.org/10.1088/1475-7516/2025/01/124} {\bibfield  {journal}
  {\bibinfo  {journal} {\jcap}\ }\textbf {\bibinfo {volume} {2025}},\ \bibinfo
  {eid} {124} (\bibinfo {year} {2025})},\ \Eprint
  {https://arxiv.org/abs/2404.03001} {arXiv:2404.03001 [astro-ph.CO]}
  \BibitemShut {NoStop}%
\bibitem [{\citenamefont {{DESI Collaboration}}\ \emph
  {et~al.}(2025)\citenamefont {{DESI Collaboration}}, \citenamefont
  {{Abdul-Karim}}, \citenamefont {{Aguilar}}, \citenamefont {{Ahlen}},
  \citenamefont {{Allende Prieto}}, \citenamefont {{Alves}}, \citenamefont
  {{Anand}}, \citenamefont {{Andrade}}, \citenamefont {{Armengaud}},
  \citenamefont {{Aviles}}, \citenamefont {{Bailey}}, \citenamefont {{Bault}},
  \citenamefont {{BenZvi}}, \citenamefont {{Bianchi}}, \citenamefont {{Blake}},
  \citenamefont {{Brodzeller}}, \citenamefont {{Brooks}}, \citenamefont
  {{Buckley-Geer}}, \citenamefont {{Burtin}}, \citenamefont {{Calderon}},
  \citenamefont {{Canning}}, \citenamefont {{Carnero Rosell}}, \citenamefont
  {{Carrilho}}, \citenamefont {{Casas}}, \citenamefont {{Castander}},
  \citenamefont {{Cereskaite}}, \citenamefont {{Charles}}, \citenamefont
  {{Chaussidon}}, \citenamefont {{Chaves-Montero}}, \citenamefont {{Chebat}},
  \citenamefont {{Claybaugh}}, \citenamefont {{Cole}}, \citenamefont
  {{Cooper}}, \citenamefont {{Cuceu}}, \citenamefont {{Dawson}}, \citenamefont
  {{de Belsunce}}, \citenamefont {{de la Macorra}}, \citenamefont {{de
  Mattia}}, \citenamefont {{Deiosso}}, \citenamefont {{Della Costa}},
  \citenamefont {{Dey}}, \citenamefont {{Dey}}, \citenamefont {{Ding}},
  \citenamefont {{Doel}}, \citenamefont {{Edelstein}}, \citenamefont
  {{Eisenstein}}, \citenamefont {{Elbers}}, \citenamefont {{Fagrelius}},
  \citenamefont {{Fanning}}, \citenamefont {{Ferraro}}, \citenamefont
  {{Font-Ribera}}, \citenamefont {{Forero-Romero}}, \citenamefont
  {{Garcia-Quintero}}, \citenamefont {{Garrison}}, \citenamefont
  {{Gazta\~{n}aga}}, \citenamefont {{Gil-Mar\'in}}, \citenamefont {{Gontcho}},
  \citenamefont {{Gonzalez-Morales}}, \citenamefont {{Gordon}}, \citenamefont
  {{Green}}, \citenamefont {{Gutierrez}}, \citenamefont {{Guy}}, \citenamefont
  {{Hahn}}, \citenamefont {{Herbold}}, \citenamefont {{Herrera-Alcantar}},
  \citenamefont {{Ho}}, \citenamefont {{Honscheid}}, \citenamefont {{Howlett}},
  \citenamefont {{Huterer}}, \citenamefont {{Ishak}}, \citenamefont {{Juneau}},
  \citenamefont {{Kara{\c{c}}ayl{\i}}}, \citenamefont {{Kehoe}}, \citenamefont
  {{Kent}}, \citenamefont {{Kirkby}}, \citenamefont {{Kisner}}, \citenamefont
  {{Kitaura}}, \citenamefont {{Koposov}}, \citenamefont {{Kremin}},
  \citenamefont {{Lahav}}, \citenamefont {{Lamman}}, \citenamefont
  {{Landriau}}, \citenamefont {{Lang}}, \citenamefont {{Lasker}}, \citenamefont
  {{Le Goff}}, \citenamefont {{Le Guillou}}, \citenamefont {{Leauthaud}},
  \citenamefont {{Levi}}, \citenamefont {{Li}}, \citenamefont {{Li}},
  \citenamefont {{Lodha}}, \citenamefont {{Lokken}}, \citenamefont
  {{Magneville}}, \citenamefont {{Manera}}, \citenamefont {{Martini}},
  \citenamefont {{Matthewson}}, \citenamefont {{McDonald}}, \citenamefont
  {{Meisner}}, \citenamefont {{Mena-Fern\'andez}}, \citenamefont {{Miquel}},
  \citenamefont {{Moustakas}}, \citenamefont {{Mu\~{n}oz Santos}},
  \citenamefont {{Mu\~{n}oz-Guti\'errez}}, \citenamefont {{Myers}},
  \citenamefont {{Newman}}, \citenamefont {{Niz}}, \citenamefont {{Noriega}},
  \citenamefont {{Paillas}}, \citenamefont {{Palanque-Delabrouille}},
  \citenamefont {{Pan}}, \citenamefont {{Percival}}, \citenamefont
  {{P\'erez-R\`afols}}, \citenamefont {{Pieri}}, \citenamefont {{Poppett}},
  \citenamefont {{Prada}}, \citenamefont {{Rabinowitz}}, \citenamefont
  {{Raichoor}}, \citenamefont {{Ram\'irez-P\'erez}}, \citenamefont
  {{Rashkovetskyi}}, \citenamefont {{Ravoux}}, \citenamefont {{Rich}},
  \citenamefont {{Rockosi}}, \citenamefont {{Ross}}, \citenamefont {{Rossi}},
  \citenamefont {{Ruhlmann-Kleider}}, \citenamefont {{Sanchez}}, \citenamefont
  {{Sanders}}, \citenamefont {{Satyavolu}}, \citenamefont {{Schlegel}},
  \citenamefont {{Schubnell}}, \citenamefont {{Seo}}, \citenamefont
  {{Shafieloo}}, \citenamefont {{Sharples}}, \citenamefont {{Silber}},
  \citenamefont {{Sinigaglia}}, \citenamefont {{Sprayberry}}, \citenamefont
  {{Tan}}, \citenamefont {{Tarl\'e}}, \citenamefont {{Taylor}}, \citenamefont
  {{Turner}}, \citenamefont {{Valdes}}, \citenamefont {{Vargas-Maga\~{n}a}},
  \citenamefont {{Walther}}, \citenamefont {{Weaver}}, \citenamefont
  {{Wolfson}}, \citenamefont {{Y\`eche}}, \citenamefont {{Zarrouk}},
  \citenamefont {{Zhou}},\ and\ \citenamefont {{Zou}}}]{desiY3LyaBAO2025}%
  \BibitemOpen
  \bibfield  {author} {\bibinfo {author} {\bibnamefont {{DESI Collaboration}}},
  \bibinfo {author} {\bibfnamefont {M.}~\bibnamefont {{Abdul-Karim}}}, \bibinfo
  {author} {\bibfnamefont {J.}~\bibnamefont {{Aguilar}}}, \bibinfo {author}
  {\bibfnamefont {S.}~\bibnamefont {{Ahlen}}}, \bibinfo {author} {\bibfnamefont
  {C.}~\bibnamefont {{Allende Prieto}}}, \bibinfo {author} {\bibfnamefont
  {O.}~\bibnamefont {{Alves}}}, \bibinfo {author} {\bibfnamefont
  {A.}~\bibnamefont {{Anand}}}, \bibinfo {author} {\bibfnamefont
  {U.}~\bibnamefont {{Andrade}}}, \bibinfo {author} {\bibfnamefont
  {E.}~\bibnamefont {{Armengaud}}}, \bibinfo {author} {\bibfnamefont
  {A.}~\bibnamefont {{Aviles}}}, \bibinfo {author} {\bibfnamefont
  {S.}~\bibnamefont {{Bailey}}}, \bibinfo {author} {\bibfnamefont
  {A.}~\bibnamefont {{Bault}}}, \bibinfo {author} {\bibfnamefont
  {S.}~\bibnamefont {{BenZvi}}}, \bibinfo {author} {\bibfnamefont
  {D.}~\bibnamefont {{Bianchi}}}, \bibinfo {author} {\bibfnamefont
  {C.}~\bibnamefont {{Blake}}}, \bibinfo {author} {\bibfnamefont
  {A.}~\bibnamefont {{Brodzeller}}}, \bibinfo {author} {\bibfnamefont
  {D.}~\bibnamefont {{Brooks}}}, \bibinfo {author} {\bibfnamefont
  {E.}~\bibnamefont {{Buckley-Geer}}}, \bibinfo {author} {\bibfnamefont
  {E.}~\bibnamefont {{Burtin}}}, \bibinfo {author} {\bibfnamefont
  {R.}~\bibnamefont {{Calderon}}}, \bibinfo {author} {\bibfnamefont
  {R.}~\bibnamefont {{Canning}}}, \bibinfo {author} {\bibfnamefont
  {A.}~\bibnamefont {{Carnero Rosell}}}, \bibinfo {author} {\bibfnamefont
  {P.}~\bibnamefont {{Carrilho}}}, \bibinfo {author} {\bibfnamefont
  {L.}~\bibnamefont {{Casas}}}, \bibinfo {author} {\bibfnamefont {F.~J.}\
  \bibnamefont {{Castander}}}, \bibinfo {author} {\bibfnamefont
  {R.}~\bibnamefont {{Cereskaite}}}, \bibinfo {author} {\bibfnamefont
  {M.}~\bibnamefont {{Charles}}}, \bibinfo {author} {\bibfnamefont
  {E.}~\bibnamefont {{Chaussidon}}}, \bibinfo {author} {\bibfnamefont
  {J.}~\bibnamefont {{Chaves-Montero}}}, \bibinfo {author} {\bibfnamefont
  {D.}~\bibnamefont {{Chebat}}}, \bibinfo {author} {\bibfnamefont
  {T.}~\bibnamefont {{Claybaugh}}}, \bibinfo {author} {\bibfnamefont
  {S.}~\bibnamefont {{Cole}}}, \bibinfo {author} {\bibfnamefont {A.~P.}\
  \bibnamefont {{Cooper}}}, \bibinfo {author} {\bibfnamefont {A.}~\bibnamefont
  {{Cuceu}}}, \bibinfo {author} {\bibfnamefont {K.~S.}\ \bibnamefont
  {{Dawson}}}, \bibinfo {author} {\bibfnamefont {R.}~\bibnamefont {{de
  Belsunce}}}, \bibinfo {author} {\bibfnamefont {A.}~\bibnamefont {{de la
  Macorra}}}, \bibinfo {author} {\bibfnamefont {A.}~\bibnamefont {{de
  Mattia}}}, \bibinfo {author} {\bibfnamefont {N.}~\bibnamefont {{Deiosso}}},
  \bibinfo {author} {\bibfnamefont {J.}~\bibnamefont {{Della Costa}}}, \bibinfo
  {author} {\bibfnamefont {A.}~\bibnamefont {{Dey}}}, \bibinfo {author}
  {\bibfnamefont {B.}~\bibnamefont {{Dey}}}, \bibinfo {author} {\bibfnamefont
  {Z.}~\bibnamefont {{Ding}}}, \bibinfo {author} {\bibfnamefont
  {P.}~\bibnamefont {{Doel}}}, \bibinfo {author} {\bibfnamefont
  {J.}~\bibnamefont {{Edelstein}}}, \bibinfo {author} {\bibfnamefont {D.~J.}\
  \bibnamefont {{Eisenstein}}}, \bibinfo {author} {\bibfnamefont
  {W.}~\bibnamefont {{Elbers}}}, \bibinfo {author} {\bibfnamefont
  {P.}~\bibnamefont {{Fagrelius}}}, \bibinfo {author} {\bibfnamefont
  {K.}~\bibnamefont {{Fanning}}}, \bibinfo {author} {\bibfnamefont
  {S.}~\bibnamefont {{Ferraro}}}, \bibinfo {author} {\bibfnamefont
  {A.}~\bibnamefont {{Font-Ribera}}}, \bibinfo {author} {\bibfnamefont {J.~E.}\
  \bibnamefont {{Forero-Romero}}}, \bibinfo {author} {\bibfnamefont
  {C.}~\bibnamefont {{Garcia-Quintero}}}, \bibinfo {author} {\bibfnamefont
  {L.~H.}\ \bibnamefont {{Garrison}}}, \bibinfo {author} {\bibfnamefont
  {E.}~\bibnamefont {{Gazta\~{n}aga}}}, \bibinfo {author} {\bibfnamefont
  {H.}~\bibnamefont {{Gil-Mar\'in}}}, \bibinfo {author} {\bibfnamefont
  {S.~G.~A.}\ \bibnamefont {{Gontcho}}}, \bibinfo {author} {\bibfnamefont
  {A.~X.}\ \bibnamefont {{Gonzalez-Morales}}}, \bibinfo {author} {\bibfnamefont
  {C.}~\bibnamefont {{Gordon}}}, \bibinfo {author} {\bibfnamefont
  {D.}~\bibnamefont {{Green}}}, \bibinfo {author} {\bibfnamefont
  {G.}~\bibnamefont {{Gutierrez}}}, \bibinfo {author} {\bibfnamefont
  {J.}~\bibnamefont {{Guy}}}, \bibinfo {author} {\bibfnamefont
  {C.}~\bibnamefont {{Hahn}}}, \bibinfo {author} {\bibfnamefont
  {M.}~\bibnamefont {{Herbold}}}, \bibinfo {author} {\bibfnamefont {H.~K.}\
  \bibnamefont {{Herrera-Alcantar}}}, \bibinfo {author} {\bibfnamefont {M.~F.}\
  \bibnamefont {{Ho}}}, \bibinfo {author} {\bibfnamefont {K.}~\bibnamefont
  {{Honscheid}}}, \bibinfo {author} {\bibfnamefont {C.}~\bibnamefont
  {{Howlett}}}, \bibinfo {author} {\bibfnamefont {D.}~\bibnamefont
  {{Huterer}}}, \bibinfo {author} {\bibfnamefont {M.}~\bibnamefont {{Ishak}}},
  \bibinfo {author} {\bibfnamefont {S.}~\bibnamefont {{Juneau}}}, \bibinfo
  {author} {\bibfnamefont {N.~G.}\ \bibnamefont {{Kara{\c{c}}ayl{\i}}}},
  \bibinfo {author} {\bibfnamefont {R.}~\bibnamefont {{Kehoe}}}, \bibinfo
  {author} {\bibfnamefont {S.}~\bibnamefont {{Kent}}}, \bibinfo {author}
  {\bibfnamefont {D.}~\bibnamefont {{Kirkby}}}, \bibinfo {author}
  {\bibfnamefont {T.}~\bibnamefont {{Kisner}}}, \bibinfo {author}
  {\bibfnamefont {F.~S.}\ \bibnamefont {{Kitaura}}}, \bibinfo {author}
  {\bibfnamefont {S.~E.}\ \bibnamefont {{Koposov}}}, \bibinfo {author}
  {\bibfnamefont {A.}~\bibnamefont {{Kremin}}}, \bibinfo {author}
  {\bibfnamefont {O.}~\bibnamefont {{Lahav}}}, \bibinfo {author} {\bibfnamefont
  {C.}~\bibnamefont {{Lamman}}}, \bibinfo {author} {\bibfnamefont
  {M.}~\bibnamefont {{Landriau}}}, \bibinfo {author} {\bibfnamefont
  {D.}~\bibnamefont {{Lang}}}, \bibinfo {author} {\bibfnamefont
  {J.}~\bibnamefont {{Lasker}}}, \bibinfo {author} {\bibfnamefont {J.~M.}\
  \bibnamefont {{Le Goff}}}, \bibinfo {author} {\bibfnamefont {L.}~\bibnamefont
  {{Le Guillou}}}, \bibinfo {author} {\bibfnamefont {A.}~\bibnamefont
  {{Leauthaud}}}, \bibinfo {author} {\bibfnamefont {M.~E.}\ \bibnamefont
  {{Levi}}}, \bibinfo {author} {\bibfnamefont {Q.}~\bibnamefont {{Li}}},
  \bibinfo {author} {\bibfnamefont {T.~S.}\ \bibnamefont {{Li}}}, \bibinfo
  {author} {\bibfnamefont {K.}~\bibnamefont {{Lodha}}}, \bibinfo {author}
  {\bibfnamefont {M.}~\bibnamefont {{Lokken}}}, \bibinfo {author}
  {\bibfnamefont {C.}~\bibnamefont {{Magneville}}}, \bibinfo {author}
  {\bibfnamefont {M.}~\bibnamefont {{Manera}}}, \bibinfo {author}
  {\bibfnamefont {P.}~\bibnamefont {{Martini}}}, \bibinfo {author}
  {\bibfnamefont {W.~L.}\ \bibnamefont {{Matthewson}}}, \bibinfo {author}
  {\bibfnamefont {P.}~\bibnamefont {{McDonald}}}, \bibinfo {author}
  {\bibfnamefont {A.}~\bibnamefont {{Meisner}}}, \bibinfo {author}
  {\bibfnamefont {J.}~\bibnamefont {{Mena-Fern\'andez}}}, \bibinfo {author}
  {\bibfnamefont {R.}~\bibnamefont {{Miquel}}}, \bibinfo {author}
  {\bibfnamefont {J.}~\bibnamefont {{Moustakas}}}, \bibinfo {author}
  {\bibfnamefont {D.}~\bibnamefont {{Mu\~{n}oz Santos}}}, \bibinfo {author}
  {\bibfnamefont {A.}~\bibnamefont {{Mu\~{n}oz-Guti\'errez}}}, \bibinfo
  {author} {\bibfnamefont {A.~D.}\ \bibnamefont {{Myers}}}, \bibinfo {author}
  {\bibfnamefont {J.~A.}\ \bibnamefont {{Newman}}}, \bibinfo {author}
  {\bibfnamefont {G.}~\bibnamefont {{Niz}}}, \bibinfo {author} {\bibfnamefont
  {H.~E.}\ \bibnamefont {{Noriega}}}, \bibinfo {author} {\bibfnamefont
  {E.}~\bibnamefont {{Paillas}}}, \bibinfo {author} {\bibfnamefont
  {N.}~\bibnamefont {{Palanque-Delabrouille}}}, \bibinfo {author}
  {\bibfnamefont {J.}~\bibnamefont {{Pan}}}, \bibinfo {author} {\bibfnamefont
  {W.~J.}\ \bibnamefont {{Percival}}}, \bibinfo {author} {\bibfnamefont
  {I.}~\bibnamefont {{P\'erez-R\`afols}}}, \bibinfo {author} {\bibfnamefont
  {M.~M.}\ \bibnamefont {{Pieri}}}, \bibinfo {author} {\bibfnamefont
  {C.}~\bibnamefont {{Poppett}}}, \bibinfo {author} {\bibfnamefont
  {F.}~\bibnamefont {{Prada}}}, \bibinfo {author} {\bibfnamefont
  {D.}~\bibnamefont {{Rabinowitz}}}, \bibinfo {author} {\bibfnamefont
  {A.}~\bibnamefont {{Raichoor}}}, \bibinfo {author} {\bibfnamefont
  {C.}~\bibnamefont {{Ram\'irez-P\'erez}}}, \bibinfo {author} {\bibfnamefont
  {M.}~\bibnamefont {{Rashkovetskyi}}}, \bibinfo {author} {\bibfnamefont
  {C.}~\bibnamefont {{Ravoux}}}, \bibinfo {author} {\bibfnamefont
  {J.}~\bibnamefont {{Rich}}}, \bibinfo {author} {\bibfnamefont
  {C.}~\bibnamefont {{Rockosi}}}, \bibinfo {author} {\bibfnamefont {A.~J.}\
  \bibnamefont {{Ross}}}, \bibinfo {author} {\bibfnamefont {G.}~\bibnamefont
  {{Rossi}}}, \bibinfo {author} {\bibfnamefont {V.}~\bibnamefont
  {{Ruhlmann-Kleider}}}, \bibinfo {author} {\bibfnamefont {E.}~\bibnamefont
  {{Sanchez}}}, \bibinfo {author} {\bibfnamefont {N.}~\bibnamefont
  {{Sanders}}}, \bibinfo {author} {\bibfnamefont {S.}~\bibnamefont
  {{Satyavolu}}}, \bibinfo {author} {\bibfnamefont {D.}~\bibnamefont
  {{Schlegel}}}, \bibinfo {author} {\bibfnamefont {M.}~\bibnamefont
  {{Schubnell}}}, \bibinfo {author} {\bibfnamefont {H.}~\bibnamefont {{Seo}}},
  \bibinfo {author} {\bibfnamefont {A.}~\bibnamefont {{Shafieloo}}}, \bibinfo
  {author} {\bibfnamefont {R.}~\bibnamefont {{Sharples}}}, \bibinfo {author}
  {\bibfnamefont {J.}~\bibnamefont {{Silber}}}, \bibinfo {author}
  {\bibfnamefont {F.}~\bibnamefont {{Sinigaglia}}}, \bibinfo {author}
  {\bibfnamefont {D.}~\bibnamefont {{Sprayberry}}}, \bibinfo {author}
  {\bibfnamefont {T.}~\bibnamefont {{Tan}}}, \bibinfo {author} {\bibfnamefont
  {G.}~\bibnamefont {{Tarl\'e}}}, \bibinfo {author} {\bibfnamefont
  {P.}~\bibnamefont {{Taylor}}}, \bibinfo {author} {\bibfnamefont
  {W.}~\bibnamefont {{Turner}}}, \bibinfo {author} {\bibfnamefont
  {F.}~\bibnamefont {{Valdes}}}, \bibinfo {author} {\bibfnamefont
  {M.}~\bibnamefont {{Vargas-Maga\~{n}a}}}, \bibinfo {author} {\bibfnamefont
  {M.}~\bibnamefont {{Walther}}}, \bibinfo {author} {\bibfnamefont {B.~A.}\
  \bibnamefont {{Weaver}}}, \bibinfo {author} {\bibfnamefont {M.}~\bibnamefont
  {{Wolfson}}}, \bibinfo {author} {\bibfnamefont {C.}~\bibnamefont
  {{Y\`eche}}}, \bibinfo {author} {\bibfnamefont {P.}~\bibnamefont
  {{Zarrouk}}}, \bibinfo {author} {\bibfnamefont {R.}~\bibnamefont {{Zhou}}},\
  and\ \bibinfo {author} {\bibfnamefont {H.}~\bibnamefont {{Zou}}},\ }\bibfield
   {title} {\bibinfo {title} {{DESI DR2 Results I: Baryon Acoustic Oscillations
  from the Lyman Alpha Forest}},\ }\href
  {https://doi.org/10.48550/arXiv.2503.14739} {\bibfield  {journal} {\bibinfo
  {journal} {arXiv e-prints}\ ,\ \bibinfo {eid} {arXiv:2503.14739}} (\bibinfo
  {year} {2025})},\ \Eprint {https://arxiv.org/abs/2503.14739}
  {arXiv:2503.14739 [astro-ph.CO]} \BibitemShut {NoStop}%
\bibitem [{\citenamefont {{Abdul Karim}}\ \emph {et~al.}(2024)\citenamefont
  {{Abdul Karim}}, \citenamefont {{Armengaud}}, \citenamefont {{Mention}},
  \citenamefont {{Chabanier}}, \citenamefont {{Ravoux}},\ and\ \citenamefont
  {{Luki{\'c}}}}]{abdulkarimMeasurementofPcross2024}%
  \BibitemOpen
  \bibfield  {author} {\bibinfo {author} {\bibfnamefont {M.~L.}\ \bibnamefont
  {{Abdul Karim}}}, \bibinfo {author} {\bibfnamefont {E.}~\bibnamefont
  {{Armengaud}}}, \bibinfo {author} {\bibfnamefont {G.}~\bibnamefont
  {{Mention}}}, \bibinfo {author} {\bibfnamefont {S.}~\bibnamefont
  {{Chabanier}}}, \bibinfo {author} {\bibfnamefont {C.}~\bibnamefont
  {{Ravoux}}},\ and\ \bibinfo {author} {\bibfnamefont {Z.}~\bibnamefont
  {{Luki{\'c}}}},\ }\bibfield  {title} {\bibinfo {title} {{Measurement of the
  small-scale 3D Lyman-{\ensuremath{\alpha}} forest power spectrum}},\ }\href
  {https://doi.org/10.1088/1475-7516/2024/05/088} {\bibfield  {journal}
  {\bibinfo  {journal} {\jcap}\ }\textbf {\bibinfo {volume} {2024}},\ \bibinfo
  {eid} {088} (\bibinfo {year} {2024})},\ \Eprint
  {https://arxiv.org/abs/2310.09116} {arXiv:2310.09116 [astro-ph.CO]}
  \BibitemShut {NoStop}%
\bibitem [{\citenamefont {Croft}\ \emph {et~al.}(1999)\citenamefont {Croft},
  \citenamefont {Weinberg}, \citenamefont {Pettini}, \citenamefont
  {Hernquist},\ and\ \citenamefont {Katz}}]{croftPowerSpectrumMass1999}%
  \BibitemOpen
  \bibfield  {author} {\bibinfo {author} {\bibfnamefont {R.~A.~C.}\
  \bibnamefont {Croft}}, \bibinfo {author} {\bibfnamefont {D.~H.}\ \bibnamefont
  {Weinberg}}, \bibinfo {author} {\bibfnamefont {M.}~\bibnamefont {Pettini}},
  \bibinfo {author} {\bibfnamefont {L.}~\bibnamefont {Hernquist}},\ and\
  \bibinfo {author} {\bibfnamefont {N.}~\bibnamefont {Katz}},\ }\bibfield
  {title} {\bibinfo {title} {The {{Power Spectrum}} of {{Mass Fluctuations
  Measured}} from the {{Ly$\alpha$ Forest}} at {{Redshift}} z = 2.5},\ }\href
  {https://doi.org/10.1086/307438} {\bibfield  {journal} {\bibinfo  {journal}
  {\apj}\ }\textbf {\bibinfo {volume} {520}},\ \bibinfo {pages} {1} (\bibinfo
  {year} {1999})}\BibitemShut {NoStop}%
\bibitem [{\citenamefont {{Chabanier}}\ \emph {et~al.}(2019)\citenamefont
  {{Chabanier}}, \citenamefont {{Millea}},\ and\ \citenamefont
  {{Palanque-Delabrouille}}}]{chabanierMatterPowerSpectrum2019}%
  \BibitemOpen
  \bibfield  {author} {\bibinfo {author} {\bibfnamefont {S.}~\bibnamefont
  {{Chabanier}}}, \bibinfo {author} {\bibfnamefont {M.}~\bibnamefont
  {{Millea}}},\ and\ \bibinfo {author} {\bibfnamefont {N.}~\bibnamefont
  {{Palanque-Delabrouille}}},\ }\bibfield  {title} {\bibinfo {title} {{Matter
  power spectrum: from Ly {\ensuremath{\alpha}} forest to CMB scales}},\ }\href
  {https://doi.org/10.1093/mnras/stz2310} {\bibfield  {journal} {\bibinfo
  {journal} {\mnras}\ }\textbf {\bibinfo {volume} {489}},\ \bibinfo {pages}
  {2247} (\bibinfo {year} {2019})},\ \Eprint {https://arxiv.org/abs/1905.08103}
  {arXiv:1905.08103 [astro-ph.CO]} \BibitemShut {NoStop}%
\bibitem [{\citenamefont {{Tegmark}}\ and\ \citenamefont
  {{Zaldarriaga}}(2002)}]{tegmarkMatterPowerReconstruction2002}%
  \BibitemOpen
  \bibfield  {author} {\bibinfo {author} {\bibfnamefont {M.}~\bibnamefont
  {{Tegmark}}}\ and\ \bibinfo {author} {\bibfnamefont {M.}~\bibnamefont
  {{Zaldarriaga}}},\ }\bibfield  {title} {\bibinfo {title} {{Separating the
  early universe from the late universe: Cosmological parameter estimation
  beyond the black box}},\ }\href {https://doi.org/10.1103/PhysRevD.66.103508}
  {\bibfield  {journal} {\bibinfo  {journal} {\prd}\ }\textbf {\bibinfo
  {volume} {66}},\ \bibinfo {eid} {103508} (\bibinfo {year} {2002})},\ \Eprint
  {https://arxiv.org/abs/astro-ph/0207047} {arXiv:astro-ph/0207047 [astro-ph]}
  \BibitemShut {NoStop}%
\bibitem [{\citenamefont {{Levi}}\ \emph {et~al.}(2013)\citenamefont {{Levi}},
  \citenamefont {{Bebek}}, \citenamefont {{Beers}}, \citenamefont {{Blum}},
  \citenamefont {{Cahn}}, \citenamefont {{Eisenstein}}, \citenamefont
  {{Flaugher}}, \citenamefont {{Honscheid}}, \citenamefont {{Kron}},
  \citenamefont {{Lahav}}, \citenamefont {{McDonald}}, \citenamefont {{Roe}},
  \citenamefont {{Schlegel}},\ and\ \citenamefont {{representing the DESI
  collaboration}}}]{leviDESIExperimentWhitepaper2013}%
  \BibitemOpen
  \bibfield  {author} {\bibinfo {author} {\bibfnamefont {M.}~\bibnamefont
  {{Levi}}}, \bibinfo {author} {\bibfnamefont {C.}~\bibnamefont {{Bebek}}},
  \bibinfo {author} {\bibfnamefont {T.}~\bibnamefont {{Beers}}}, \bibinfo
  {author} {\bibfnamefont {R.}~\bibnamefont {{Blum}}}, \bibinfo {author}
  {\bibfnamefont {R.}~\bibnamefont {{Cahn}}}, \bibinfo {author} {\bibfnamefont
  {D.}~\bibnamefont {{Eisenstein}}}, \bibinfo {author} {\bibfnamefont
  {B.}~\bibnamefont {{Flaugher}}}, \bibinfo {author} {\bibfnamefont
  {K.}~\bibnamefont {{Honscheid}}}, \bibinfo {author} {\bibfnamefont
  {R.}~\bibnamefont {{Kron}}}, \bibinfo {author} {\bibfnamefont
  {O.}~\bibnamefont {{Lahav}}}, \bibinfo {author} {\bibfnamefont
  {P.}~\bibnamefont {{McDonald}}}, \bibinfo {author} {\bibfnamefont
  {N.}~\bibnamefont {{Roe}}}, \bibinfo {author} {\bibfnamefont
  {D.}~\bibnamefont {{Schlegel}}},\ and\ \bibinfo {author} {\bibnamefont
  {{representing the DESI collaboration}}},\ }\bibfield  {title} {\bibinfo
  {title} {{The DESI Experiment, a whitepaper for Snowmass 2013}},\ }\href
  {https://doi.org/10.48550/arXiv.1308.0847} {\bibfield  {journal} {\bibinfo
  {journal} {arXiv e-prints}\ ,\ \bibinfo {eid} {arXiv:1308.0847}} (\bibinfo
  {year} {2013})},\ \Eprint {https://arxiv.org/abs/1308.0847} {arXiv:1308.0847
  [astro-ph.CO]} \BibitemShut {NoStop}%
\bibitem [{\citenamefont {Baydin}\ \emph {et~al.}(2018)\citenamefont {Baydin},
  \citenamefont {Pearlmutter}, \citenamefont {Radul},\ and\ \citenamefont
  {Siskind}}]{baydin_automatic_2018}%
  \BibitemOpen
  \bibfield  {author} {\bibinfo {author} {\bibfnamefont {A.~G.}\ \bibnamefont
  {Baydin}}, \bibinfo {author} {\bibfnamefont {B.~A.}\ \bibnamefont
  {Pearlmutter}}, \bibinfo {author} {\bibfnamefont {A.~A.}\ \bibnamefont
  {Radul}},\ and\ \bibinfo {author} {\bibfnamefont {J.~M.}\ \bibnamefont
  {Siskind}},\ }\href {https://doi.org/10.48550/arXiv.1502.05767} {\bibinfo
  {title} {Automatic differentiation in machine learning: a survey}} (\bibinfo
  {year} {2018})\BibitemShut {NoStop}%
\bibitem [{\citenamefont {{Duane}}\ \emph {et~al.}(1987)\citenamefont
  {{Duane}}, \citenamefont {{Kennedy}}, \citenamefont {{Pendleton}},\ and\
  \citenamefont {{Roweth}}}]{duaneHybridMonteCarlo1987}%
  \BibitemOpen
  \bibfield  {author} {\bibinfo {author} {\bibfnamefont {S.}~\bibnamefont
  {{Duane}}}, \bibinfo {author} {\bibfnamefont {A.~D.}\ \bibnamefont
  {{Kennedy}}}, \bibinfo {author} {\bibfnamefont {B.~J.}\ \bibnamefont
  {{Pendleton}}},\ and\ \bibinfo {author} {\bibfnamefont {D.}~\bibnamefont
  {{Roweth}}},\ }\bibfield  {title} {\bibinfo {title} {{Hybrid Monte Carlo}},\
  }\href {https://doi.org/10.1016/0370-2693(87)91197-X} {\bibfield  {journal}
  {\bibinfo  {journal} {Physics Letters B}\ }\textbf {\bibinfo {volume}
  {195}},\ \bibinfo {pages} {216} (\bibinfo {year} {1987})}\BibitemShut
  {NoStop}%
\bibitem [{\citenamefont {Betancourt}(2017)}]{betancourt_conceptual_2017}%
  \BibitemOpen
  \bibfield  {author} {\bibinfo {author} {\bibfnamefont {M.}~\bibnamefont
  {Betancourt}},\ }\href {https://doi.org/10.48550/arXiv.1701.02434} {\bibinfo
  {title} {A {Conceptual} {Introduction} to {Hamiltonian} {Monte} {Carlo}}}
  (\bibinfo {year} {2017})\BibitemShut {NoStop}%
\bibitem [{\citenamefont {{Arinyo-i-Prats}}\ \emph {et~al.}(2015)\citenamefont
  {{Arinyo-i-Prats}}, \citenamefont {{Miralda-Escud{\'e}}}, \citenamefont
  {{Viel}},\ and\ \citenamefont {{Cen}}}]{arinyoNonLinearPowerLya2015}%
  \BibitemOpen
  \bibfield  {author} {\bibinfo {author} {\bibfnamefont {A.}~\bibnamefont
  {{Arinyo-i-Prats}}}, \bibinfo {author} {\bibfnamefont {J.}~\bibnamefont
  {{Miralda-Escud{\'e}}}}, \bibinfo {author} {\bibfnamefont {M.}~\bibnamefont
  {{Viel}}},\ and\ \bibinfo {author} {\bibfnamefont {R.}~\bibnamefont
  {{Cen}}},\ }\bibfield  {title} {\bibinfo {title} {{The non-linear power
  spectrum of the Lyman alpha forest}},\ }\href
  {https://doi.org/10.1088/1475-7516/2015/12/017} {\bibfield  {journal}
  {\bibinfo  {journal} {\jcap}\ }\textbf {\bibinfo {volume} {2015}},\ \bibinfo
  {pages} {017} (\bibinfo {year} {2015})},\ \Eprint
  {https://arxiv.org/abs/1506.04519} {arXiv:1506.04519 [astro-ph.CO]}
  \BibitemShut {NoStop}%
\bibitem [{\citenamefont {{McDonald}}(2003)}]{mcdonaldPredictingLyaPower2003}%
  \BibitemOpen
  \bibfield  {author} {\bibinfo {author} {\bibfnamefont {P.}~\bibnamefont
  {{McDonald}}},\ }\bibfield  {title} {\bibinfo {title} {{Toward a Measurement
  of the Cosmological Geometry at z \raisebox{-0.5ex}\textasciitilde 2:
  Predicting Ly{\ensuremath{\alpha}} Forest Correlation in Three Dimensions and
  the Potential of Future Data Sets}},\ }\href {https://doi.org/10.1086/345945}
  {\bibfield  {journal} {\bibinfo  {journal} {\apj}\ }\textbf {\bibinfo
  {volume} {585}},\ \bibinfo {pages} {34} (\bibinfo {year} {2003})},\ \Eprint
  {https://arxiv.org/abs/astro-ph/0108064} {arXiv:astro-ph/0108064 [astro-ph]}
  \BibitemShut {NoStop}%
\bibitem [{\citenamefont {{Lewis}}\ \emph {et~al.}(2000)\citenamefont
  {{Lewis}}, \citenamefont {{Challinor}},\ and\ \citenamefont
  {{Lasenby}}}]{lewisCAMB2000}%
  \BibitemOpen
  \bibfield  {author} {\bibinfo {author} {\bibfnamefont {A.}~\bibnamefont
  {{Lewis}}}, \bibinfo {author} {\bibfnamefont {A.}~\bibnamefont
  {{Challinor}}},\ and\ \bibinfo {author} {\bibfnamefont {A.}~\bibnamefont
  {{Lasenby}}},\ }\bibfield  {title} {\bibinfo {title} {{Efficient Computation
  of Cosmic Microwave Background Anisotropies in Closed
  Friedmann-Robertson-Walker Models}},\ }\href {https://doi.org/10.1086/309179}
  {\bibfield  {journal} {\bibinfo  {journal} {\apj}\ }\textbf {\bibinfo
  {volume} {538}},\ \bibinfo {pages} {473} (\bibinfo {year} {2000})},\ \Eprint
  {https://arxiv.org/abs/astro-ph/9911177} {arXiv:astro-ph/9911177 [astro-ph]}
  \BibitemShut {NoStop}%
\bibitem [{\citenamefont {{Planck Collaboration}}\ \emph
  {et~al.}(2020)\citenamefont {{Planck Collaboration}}, \citenamefont
  {{Aghanim}}, \citenamefont {{Akrami}}, \citenamefont {{Ashdown}},
  \citenamefont {{Aumont}}, \citenamefont {{Baccigalupi}}, \citenamefont
  {{Ballardini}}, \citenamefont {{Banday}}, \citenamefont {{Barreiro}},
  \citenamefont {{Bartolo}}, \citenamefont {{Basak}}, \citenamefont {{Battye}},
  \citenamefont {{Benabed}}, \citenamefont {{Bernard}}, \citenamefont
  {{Bersanelli}}, \citenamefont {{Bielewicz}}, \citenamefont {{Bock}},
  \citenamefont {{Bond}}, \citenamefont {{Borrill}}, \citenamefont {{Bouchet}},
  \citenamefont {{Boulanger}}, \citenamefont {{Bucher}}, \citenamefont
  {{Burigana}}, \citenamefont {{Butler}}, \citenamefont {{Calabrese}},
  \citenamefont {{Cardoso}}, \citenamefont {{Carron}}, \citenamefont
  {{Challinor}}, \citenamefont {{Chiang}}, \citenamefont {{Chluba}},
  \citenamefont {{Colombo}}, \citenamefont {{Combet}}, \citenamefont
  {{Contreras}}, \citenamefont {{Crill}}, \citenamefont {{Cuttaia}},
  \citenamefont {{de Bernardis}}, \citenamefont {{de Zotti}}, \citenamefont
  {{Delabrouille}}, \citenamefont {{Delouis}}, \citenamefont {{Di Valentino}},
  \citenamefont {{Diego}}, \citenamefont {{Dor{\'e}}}, \citenamefont
  {{Douspis}}, \citenamefont {{Ducout}}, \citenamefont {{Dupac}}, \citenamefont
  {{Dusini}}, \citenamefont {{Efstathiou}}, \citenamefont {{Elsner}},
  \citenamefont {{En{\ss}lin}}, \citenamefont {{Eriksen}}, \citenamefont
  {{Fantaye}}, \citenamefont {{Farhang}}, \citenamefont {{Fergusson}},
  \citenamefont {{Fernandez-Cobos}}, \citenamefont {{Finelli}}, \citenamefont
  {{Forastieri}}, \citenamefont {{Frailis}}, \citenamefont {{Fraisse}},
  \citenamefont {{Franceschi}}, \citenamefont {{Frolov}}, \citenamefont
  {{Galeotta}}, \citenamefont {{Galli}}, \citenamefont {{Ganga}}, \citenamefont
  {{G{\'e}nova-Santos}}, \citenamefont {{Gerbino}}, \citenamefont {{Ghosh}},
  \citenamefont {{Gonz{\'a}lez-Nuevo}}, \citenamefont {{G{\'o}rski}},
  \citenamefont {{Gratton}}, \citenamefont {{Gruppuso}}, \citenamefont
  {{Gudmundsson}}, \citenamefont {{Hamann}}, \citenamefont {{Handley}},
  \citenamefont {{Hansen}}, \citenamefont {{Herranz}}, \citenamefont
  {{Hildebrandt}}, \citenamefont {{Hivon}}, \citenamefont {{Huang}},
  \citenamefont {{Jaffe}}, \citenamefont {{Jones}}, \citenamefont {{Karakci}},
  \citenamefont {{Keih{\"a}nen}}, \citenamefont {{Keskitalo}}, \citenamefont
  {{Kiiveri}}, \citenamefont {{Kim}}, \citenamefont {{Kisner}}, \citenamefont
  {{Knox}}, \citenamefont {{Krachmalnicoff}}, \citenamefont {{Kunz}},
  \citenamefont {{Kurki-Suonio}}, \citenamefont {{Lagache}}, \citenamefont
  {{Lamarre}}, \citenamefont {{Lasenby}}, \citenamefont {{Lattanzi}},
  \citenamefont {{Lawrence}}, \citenamefont {{Le Jeune}}, \citenamefont
  {{Lemos}}, \citenamefont {{Lesgourgues}}, \citenamefont {{Levrier}},
  \citenamefont {{Lewis}}, \citenamefont {{Liguori}}, \citenamefont {{Lilje}},
  \citenamefont {{Lilley}}, \citenamefont {{Lindholm}}, \citenamefont
  {{L{\'o}pez-Caniego}}, \citenamefont {{Lubin}}, \citenamefont {{Ma}},
  \citenamefont {{Mac{\'\i}as-P{\'e}rez}}, \citenamefont {{Maggio}},
  \citenamefont {{Maino}}, \citenamefont {{Mandolesi}}, \citenamefont
  {{Mangilli}}, \citenamefont {{Marcos-Caballero}}, \citenamefont {{Maris}},
  \citenamefont {{Martin}}, \citenamefont {{Martinelli}}, \citenamefont
  {{Mart{\'\i}nez-Gonz{\'a}lez}}, \citenamefont {{Matarrese}}, \citenamefont
  {{Mauri}}, \citenamefont {{McEwen}}, \citenamefont {{Meinhold}},
  \citenamefont {{Melchiorri}}, \citenamefont {{Mennella}}, \citenamefont
  {{Migliaccio}}, \citenamefont {{Millea}}, \citenamefont {{Mitra}},
  \citenamefont {{Miville-Desch{\^e}nes}}, \citenamefont {{Molinari}},
  \citenamefont {{Montier}}, \citenamefont {{Morgante}}, \citenamefont
  {{Moss}}, \citenamefont {{Natoli}}, \citenamefont {{N{\o}rgaard-Nielsen}},
  \citenamefont {{Pagano}}, \citenamefont {{Paoletti}}, \citenamefont
  {{Partridge}}, \citenamefont {{Patanchon}}, \citenamefont {{Peiris}},
  \citenamefont {{Perrotta}}, \citenamefont {{Pettorino}}, \citenamefont
  {{Piacentini}}, \citenamefont {{Polastri}}, \citenamefont {{Polenta}},
  \citenamefont {{Puget}}, \citenamefont {{Rachen}}, \citenamefont
  {{Reinecke}}, \citenamefont {{Remazeilles}}, \citenamefont {{Renzi}},
  \citenamefont {{Rocha}}, \citenamefont {{Rosset}}, \citenamefont {{Roudier}},
  \citenamefont {{Rubi{\~n}o-Mart{\'\i}n}}, \citenamefont {{Ruiz-Granados}},
  \citenamefont {{Salvati}}, \citenamefont {{Sandri}}, \citenamefont
  {{Savelainen}}, \citenamefont {{Scott}}, \citenamefont {{Shellard}},
  \citenamefont {{Sirignano}}, \citenamefont {{Sirri}}, \citenamefont
  {{Spencer}}, \citenamefont {{Sunyaev}}, \citenamefont {{Suur-Uski}},
  \citenamefont {{Tauber}}, \citenamefont {{Tavagnacco}}, \citenamefont
  {{Tenti}}, \citenamefont {{Toffolatti}}, \citenamefont {{Tomasi}},
  \citenamefont {{Trombetti}}, \citenamefont {{Valenziano}}, \citenamefont
  {{Valiviita}}, \citenamefont {{Van Tent}}, \citenamefont {{Vibert}},
  \citenamefont {{Vielva}}, \citenamefont {{Villa}}, \citenamefont
  {{Vittorio}}, \citenamefont {{Wandelt}}, \citenamefont {{Wehus}},
  \citenamefont {{White}}, \citenamefont {{White}}, \citenamefont {{Zacchei}},\
  and\ \citenamefont {{Zonca}}}]{collaborationPlanck2018Results2020}%
  \BibitemOpen
  \bibfield  {author} {\bibinfo {author} {\bibnamefont {{Planck
  Collaboration}}}, \bibinfo {author} {\bibfnamefont {N.}~\bibnamefont
  {{Aghanim}}}, \bibinfo {author} {\bibfnamefont {Y.}~\bibnamefont {{Akrami}}},
  \bibinfo {author} {\bibfnamefont {M.}~\bibnamefont {{Ashdown}}}, \bibinfo
  {author} {\bibfnamefont {J.}~\bibnamefont {{Aumont}}}, \bibinfo {author}
  {\bibfnamefont {C.}~\bibnamefont {{Baccigalupi}}}, \bibinfo {author}
  {\bibfnamefont {M.}~\bibnamefont {{Ballardini}}}, \bibinfo {author}
  {\bibfnamefont {A.~J.}\ \bibnamefont {{Banday}}}, \bibinfo {author}
  {\bibfnamefont {R.~B.}\ \bibnamefont {{Barreiro}}}, \bibinfo {author}
  {\bibfnamefont {N.}~\bibnamefont {{Bartolo}}}, \bibinfo {author}
  {\bibfnamefont {S.}~\bibnamefont {{Basak}}}, \bibinfo {author} {\bibfnamefont
  {R.}~\bibnamefont {{Battye}}}, \bibinfo {author} {\bibfnamefont
  {K.}~\bibnamefont {{Benabed}}}, \bibinfo {author} {\bibfnamefont {J.~P.}\
  \bibnamefont {{Bernard}}}, \bibinfo {author} {\bibfnamefont {M.}~\bibnamefont
  {{Bersanelli}}}, \bibinfo {author} {\bibfnamefont {P.}~\bibnamefont
  {{Bielewicz}}}, \bibinfo {author} {\bibfnamefont {J.~J.}\ \bibnamefont
  {{Bock}}}, \bibinfo {author} {\bibfnamefont {J.~R.}\ \bibnamefont {{Bond}}},
  \bibinfo {author} {\bibfnamefont {J.}~\bibnamefont {{Borrill}}}, \bibinfo
  {author} {\bibfnamefont {F.~R.}\ \bibnamefont {{Bouchet}}}, \bibinfo {author}
  {\bibfnamefont {F.}~\bibnamefont {{Boulanger}}}, \bibinfo {author}
  {\bibfnamefont {M.}~\bibnamefont {{Bucher}}}, \bibinfo {author}
  {\bibfnamefont {C.}~\bibnamefont {{Burigana}}}, \bibinfo {author}
  {\bibfnamefont {R.~C.}\ \bibnamefont {{Butler}}}, \bibinfo {author}
  {\bibfnamefont {E.}~\bibnamefont {{Calabrese}}}, \bibinfo {author}
  {\bibfnamefont {J.~F.}\ \bibnamefont {{Cardoso}}}, \bibinfo {author}
  {\bibfnamefont {J.}~\bibnamefont {{Carron}}}, \bibinfo {author}
  {\bibfnamefont {A.}~\bibnamefont {{Challinor}}}, \bibinfo {author}
  {\bibfnamefont {H.~C.}\ \bibnamefont {{Chiang}}}, \bibinfo {author}
  {\bibfnamefont {J.}~\bibnamefont {{Chluba}}}, \bibinfo {author}
  {\bibfnamefont {L.~P.~L.}\ \bibnamefont {{Colombo}}}, \bibinfo {author}
  {\bibfnamefont {C.}~\bibnamefont {{Combet}}}, \bibinfo {author}
  {\bibfnamefont {D.}~\bibnamefont {{Contreras}}}, \bibinfo {author}
  {\bibfnamefont {B.~P.}\ \bibnamefont {{Crill}}}, \bibinfo {author}
  {\bibfnamefont {F.}~\bibnamefont {{Cuttaia}}}, \bibinfo {author}
  {\bibfnamefont {P.}~\bibnamefont {{de Bernardis}}}, \bibinfo {author}
  {\bibfnamefont {G.}~\bibnamefont {{de Zotti}}}, \bibinfo {author}
  {\bibfnamefont {J.}~\bibnamefont {{Delabrouille}}}, \bibinfo {author}
  {\bibfnamefont {J.~M.}\ \bibnamefont {{Delouis}}}, \bibinfo {author}
  {\bibfnamefont {E.}~\bibnamefont {{Di Valentino}}}, \bibinfo {author}
  {\bibfnamefont {J.~M.}\ \bibnamefont {{Diego}}}, \bibinfo {author}
  {\bibfnamefont {O.}~\bibnamefont {{Dor{\'e}}}}, \bibinfo {author}
  {\bibfnamefont {M.}~\bibnamefont {{Douspis}}}, \bibinfo {author}
  {\bibfnamefont {A.}~\bibnamefont {{Ducout}}}, \bibinfo {author}
  {\bibfnamefont {X.}~\bibnamefont {{Dupac}}}, \bibinfo {author} {\bibfnamefont
  {S.}~\bibnamefont {{Dusini}}}, \bibinfo {author} {\bibfnamefont
  {G.}~\bibnamefont {{Efstathiou}}}, \bibinfo {author} {\bibfnamefont
  {F.}~\bibnamefont {{Elsner}}}, \bibinfo {author} {\bibfnamefont {T.~A.}\
  \bibnamefont {{En{\ss}lin}}}, \bibinfo {author} {\bibfnamefont {H.~K.}\
  \bibnamefont {{Eriksen}}}, \bibinfo {author} {\bibfnamefont {Y.}~\bibnamefont
  {{Fantaye}}}, \bibinfo {author} {\bibfnamefont {M.}~\bibnamefont
  {{Farhang}}}, \bibinfo {author} {\bibfnamefont {J.}~\bibnamefont
  {{Fergusson}}}, \bibinfo {author} {\bibfnamefont {R.}~\bibnamefont
  {{Fernandez-Cobos}}}, \bibinfo {author} {\bibfnamefont {F.}~\bibnamefont
  {{Finelli}}}, \bibinfo {author} {\bibfnamefont {F.}~\bibnamefont
  {{Forastieri}}}, \bibinfo {author} {\bibfnamefont {M.}~\bibnamefont
  {{Frailis}}}, \bibinfo {author} {\bibfnamefont {A.~A.}\ \bibnamefont
  {{Fraisse}}}, \bibinfo {author} {\bibfnamefont {E.}~\bibnamefont
  {{Franceschi}}}, \bibinfo {author} {\bibfnamefont {A.}~\bibnamefont
  {{Frolov}}}, \bibinfo {author} {\bibfnamefont {S.}~\bibnamefont
  {{Galeotta}}}, \bibinfo {author} {\bibfnamefont {S.}~\bibnamefont {{Galli}}},
  \bibinfo {author} {\bibfnamefont {K.}~\bibnamefont {{Ganga}}}, \bibinfo
  {author} {\bibfnamefont {R.~T.}\ \bibnamefont {{G{\'e}nova-Santos}}},
  \bibinfo {author} {\bibfnamefont {M.}~\bibnamefont {{Gerbino}}}, \bibinfo
  {author} {\bibfnamefont {T.}~\bibnamefont {{Ghosh}}}, \bibinfo {author}
  {\bibfnamefont {J.}~\bibnamefont {{Gonz{\'a}lez-Nuevo}}}, \bibinfo {author}
  {\bibfnamefont {K.~M.}\ \bibnamefont {{G{\'o}rski}}}, \bibinfo {author}
  {\bibfnamefont {S.}~\bibnamefont {{Gratton}}}, \bibinfo {author}
  {\bibfnamefont {A.}~\bibnamefont {{Gruppuso}}}, \bibinfo {author}
  {\bibfnamefont {J.~E.}\ \bibnamefont {{Gudmundsson}}}, \bibinfo {author}
  {\bibfnamefont {J.}~\bibnamefont {{Hamann}}}, \bibinfo {author}
  {\bibfnamefont {W.}~\bibnamefont {{Handley}}}, \bibinfo {author}
  {\bibfnamefont {F.~K.}\ \bibnamefont {{Hansen}}}, \bibinfo {author}
  {\bibfnamefont {D.}~\bibnamefont {{Herranz}}}, \bibinfo {author}
  {\bibfnamefont {S.~R.}\ \bibnamefont {{Hildebrandt}}}, \bibinfo {author}
  {\bibfnamefont {E.}~\bibnamefont {{Hivon}}}, \bibinfo {author} {\bibfnamefont
  {Z.}~\bibnamefont {{Huang}}}, \bibinfo {author} {\bibfnamefont {A.~H.}\
  \bibnamefont {{Jaffe}}}, \bibinfo {author} {\bibfnamefont {W.~C.}\
  \bibnamefont {{Jones}}}, \bibinfo {author} {\bibfnamefont {A.}~\bibnamefont
  {{Karakci}}}, \bibinfo {author} {\bibfnamefont {E.}~\bibnamefont
  {{Keih{\"a}nen}}}, \bibinfo {author} {\bibfnamefont {R.}~\bibnamefont
  {{Keskitalo}}}, \bibinfo {author} {\bibfnamefont {K.}~\bibnamefont
  {{Kiiveri}}}, \bibinfo {author} {\bibfnamefont {J.}~\bibnamefont {{Kim}}},
  \bibinfo {author} {\bibfnamefont {T.~S.}\ \bibnamefont {{Kisner}}}, \bibinfo
  {author} {\bibfnamefont {L.}~\bibnamefont {{Knox}}}, \bibinfo {author}
  {\bibfnamefont {N.}~\bibnamefont {{Krachmalnicoff}}}, \bibinfo {author}
  {\bibfnamefont {M.}~\bibnamefont {{Kunz}}}, \bibinfo {author} {\bibfnamefont
  {H.}~\bibnamefont {{Kurki-Suonio}}}, \bibinfo {author} {\bibfnamefont
  {G.}~\bibnamefont {{Lagache}}}, \bibinfo {author} {\bibfnamefont {J.~M.}\
  \bibnamefont {{Lamarre}}}, \bibinfo {author} {\bibfnamefont {A.}~\bibnamefont
  {{Lasenby}}}, \bibinfo {author} {\bibfnamefont {M.}~\bibnamefont
  {{Lattanzi}}}, \bibinfo {author} {\bibfnamefont {C.~R.}\ \bibnamefont
  {{Lawrence}}}, \bibinfo {author} {\bibfnamefont {M.}~\bibnamefont {{Le
  Jeune}}}, \bibinfo {author} {\bibfnamefont {P.}~\bibnamefont {{Lemos}}},
  \bibinfo {author} {\bibfnamefont {J.}~\bibnamefont {{Lesgourgues}}}, \bibinfo
  {author} {\bibfnamefont {F.}~\bibnamefont {{Levrier}}}, \bibinfo {author}
  {\bibfnamefont {A.}~\bibnamefont {{Lewis}}}, \bibinfo {author} {\bibfnamefont
  {M.}~\bibnamefont {{Liguori}}}, \bibinfo {author} {\bibfnamefont {P.~B.}\
  \bibnamefont {{Lilje}}}, \bibinfo {author} {\bibfnamefont {M.}~\bibnamefont
  {{Lilley}}}, \bibinfo {author} {\bibfnamefont {V.}~\bibnamefont
  {{Lindholm}}}, \bibinfo {author} {\bibfnamefont {M.}~\bibnamefont
  {{L{\'o}pez-Caniego}}}, \bibinfo {author} {\bibfnamefont {P.~M.}\
  \bibnamefont {{Lubin}}}, \bibinfo {author} {\bibfnamefont {Y.~Z.}\
  \bibnamefont {{Ma}}}, \bibinfo {author} {\bibfnamefont {J.~F.}\ \bibnamefont
  {{Mac{\'\i}as-P{\'e}rez}}}, \bibinfo {author} {\bibfnamefont
  {G.}~\bibnamefont {{Maggio}}}, \bibinfo {author} {\bibfnamefont
  {D.}~\bibnamefont {{Maino}}}, \bibinfo {author} {\bibfnamefont
  {N.}~\bibnamefont {{Mandolesi}}}, \bibinfo {author} {\bibfnamefont
  {A.}~\bibnamefont {{Mangilli}}}, \bibinfo {author} {\bibfnamefont
  {A.}~\bibnamefont {{Marcos-Caballero}}}, \bibinfo {author} {\bibfnamefont
  {M.}~\bibnamefont {{Maris}}}, \bibinfo {author} {\bibfnamefont {P.~G.}\
  \bibnamefont {{Martin}}}, \bibinfo {author} {\bibfnamefont {M.}~\bibnamefont
  {{Martinelli}}}, \bibinfo {author} {\bibfnamefont {E.}~\bibnamefont
  {{Mart{\'\i}nez-Gonz{\'a}lez}}}, \bibinfo {author} {\bibfnamefont
  {S.}~\bibnamefont {{Matarrese}}}, \bibinfo {author} {\bibfnamefont
  {N.}~\bibnamefont {{Mauri}}}, \bibinfo {author} {\bibfnamefont {J.~D.}\
  \bibnamefont {{McEwen}}}, \bibinfo {author} {\bibfnamefont {P.~R.}\
  \bibnamefont {{Meinhold}}}, \bibinfo {author} {\bibfnamefont
  {A.}~\bibnamefont {{Melchiorri}}}, \bibinfo {author} {\bibfnamefont
  {A.}~\bibnamefont {{Mennella}}}, \bibinfo {author} {\bibfnamefont
  {M.}~\bibnamefont {{Migliaccio}}}, \bibinfo {author} {\bibfnamefont
  {M.}~\bibnamefont {{Millea}}}, \bibinfo {author} {\bibfnamefont
  {S.}~\bibnamefont {{Mitra}}}, \bibinfo {author} {\bibfnamefont {M.~A.}\
  \bibnamefont {{Miville-Desch{\^e}nes}}}, \bibinfo {author} {\bibfnamefont
  {D.}~\bibnamefont {{Molinari}}}, \bibinfo {author} {\bibfnamefont
  {L.}~\bibnamefont {{Montier}}}, \bibinfo {author} {\bibfnamefont
  {G.}~\bibnamefont {{Morgante}}}, \bibinfo {author} {\bibfnamefont
  {A.}~\bibnamefont {{Moss}}}, \bibinfo {author} {\bibfnamefont
  {P.}~\bibnamefont {{Natoli}}}, \bibinfo {author} {\bibfnamefont {H.~U.}\
  \bibnamefont {{N{\o}rgaard-Nielsen}}}, \bibinfo {author} {\bibfnamefont
  {L.}~\bibnamefont {{Pagano}}}, \bibinfo {author} {\bibfnamefont
  {D.}~\bibnamefont {{Paoletti}}}, \bibinfo {author} {\bibfnamefont
  {B.}~\bibnamefont {{Partridge}}}, \bibinfo {author} {\bibfnamefont
  {G.}~\bibnamefont {{Patanchon}}}, \bibinfo {author} {\bibfnamefont {H.~V.}\
  \bibnamefont {{Peiris}}}, \bibinfo {author} {\bibfnamefont {F.}~\bibnamefont
  {{Perrotta}}}, \bibinfo {author} {\bibfnamefont {V.}~\bibnamefont
  {{Pettorino}}}, \bibinfo {author} {\bibfnamefont {F.}~\bibnamefont
  {{Piacentini}}}, \bibinfo {author} {\bibfnamefont {L.}~\bibnamefont
  {{Polastri}}}, \bibinfo {author} {\bibfnamefont {G.}~\bibnamefont
  {{Polenta}}}, \bibinfo {author} {\bibfnamefont {J.~L.}\ \bibnamefont
  {{Puget}}}, \bibinfo {author} {\bibfnamefont {J.~P.}\ \bibnamefont
  {{Rachen}}}, \bibinfo {author} {\bibfnamefont {M.}~\bibnamefont
  {{Reinecke}}}, \bibinfo {author} {\bibfnamefont {M.}~\bibnamefont
  {{Remazeilles}}}, \bibinfo {author} {\bibfnamefont {A.}~\bibnamefont
  {{Renzi}}}, \bibinfo {author} {\bibfnamefont {G.}~\bibnamefont {{Rocha}}},
  \bibinfo {author} {\bibfnamefont {C.}~\bibnamefont {{Rosset}}}, \bibinfo
  {author} {\bibfnamefont {G.}~\bibnamefont {{Roudier}}}, \bibinfo {author}
  {\bibfnamefont {J.~A.}\ \bibnamefont {{Rubi{\~n}o-Mart{\'\i}n}}}, \bibinfo
  {author} {\bibfnamefont {B.}~\bibnamefont {{Ruiz-Granados}}}, \bibinfo
  {author} {\bibfnamefont {L.}~\bibnamefont {{Salvati}}}, \bibinfo {author}
  {\bibfnamefont {M.}~\bibnamefont {{Sandri}}}, \bibinfo {author}
  {\bibfnamefont {M.}~\bibnamefont {{Savelainen}}}, \bibinfo {author}
  {\bibfnamefont {D.}~\bibnamefont {{Scott}}}, \bibinfo {author} {\bibfnamefont
  {E.~P.~S.}\ \bibnamefont {{Shellard}}}, \bibinfo {author} {\bibfnamefont
  {C.}~\bibnamefont {{Sirignano}}}, \bibinfo {author} {\bibfnamefont
  {G.}~\bibnamefont {{Sirri}}}, \bibinfo {author} {\bibfnamefont {L.~D.}\
  \bibnamefont {{Spencer}}}, \bibinfo {author} {\bibfnamefont {R.}~\bibnamefont
  {{Sunyaev}}}, \bibinfo {author} {\bibfnamefont {A.~S.}\ \bibnamefont
  {{Suur-Uski}}}, \bibinfo {author} {\bibfnamefont {J.~A.}\ \bibnamefont
  {{Tauber}}}, \bibinfo {author} {\bibfnamefont {D.}~\bibnamefont
  {{Tavagnacco}}}, \bibinfo {author} {\bibfnamefont {M.}~\bibnamefont
  {{Tenti}}}, \bibinfo {author} {\bibfnamefont {L.}~\bibnamefont
  {{Toffolatti}}}, \bibinfo {author} {\bibfnamefont {M.}~\bibnamefont
  {{Tomasi}}}, \bibinfo {author} {\bibfnamefont {T.}~\bibnamefont
  {{Trombetti}}}, \bibinfo {author} {\bibfnamefont {L.}~\bibnamefont
  {{Valenziano}}}, \bibinfo {author} {\bibfnamefont {J.}~\bibnamefont
  {{Valiviita}}}, \bibinfo {author} {\bibfnamefont {B.}~\bibnamefont {{Van
  Tent}}}, \bibinfo {author} {\bibfnamefont {L.}~\bibnamefont {{Vibert}}},
  \bibinfo {author} {\bibfnamefont {P.}~\bibnamefont {{Vielva}}}, \bibinfo
  {author} {\bibfnamefont {F.}~\bibnamefont {{Villa}}}, \bibinfo {author}
  {\bibfnamefont {N.}~\bibnamefont {{Vittorio}}}, \bibinfo {author}
  {\bibfnamefont {B.~D.}\ \bibnamefont {{Wandelt}}}, \bibinfo {author}
  {\bibfnamefont {I.~K.}\ \bibnamefont {{Wehus}}}, \bibinfo {author}
  {\bibfnamefont {M.}~\bibnamefont {{White}}}, \bibinfo {author} {\bibfnamefont
  {S.~D.~M.}\ \bibnamefont {{White}}}, \bibinfo {author} {\bibfnamefont
  {A.}~\bibnamefont {{Zacchei}}},\ and\ \bibinfo {author} {\bibfnamefont
  {A.}~\bibnamefont {{Zonca}}},\ }\bibfield  {title} {\bibinfo {title} {{Planck
  2018 results. VI. Cosmological parameters}},\ }\href
  {https://doi.org/10.1051/0004-6361/201833910} {\bibfield  {journal} {\bibinfo
   {journal} {\aap}\ }\textbf {\bibinfo {volume} {641}},\ \bibinfo {eid} {A6}
  (\bibinfo {year} {2020})},\ \Eprint {https://arxiv.org/abs/1807.06209}
  {arXiv:1807.06209 [astro-ph.CO]} \BibitemShut {NoStop}%
\bibitem [{\citenamefont {Gnedin}\ and\ \citenamefont
  {Hui}(1998)}]{gnedinProbingUniverseLya1998}%
  \BibitemOpen
  \bibfield  {author} {\bibinfo {author} {\bibfnamefont {N.~Y.}\ \bibnamefont
  {Gnedin}}\ and\ \bibinfo {author} {\bibfnamefont {L.}~\bibnamefont {Hui}},\
  }\bibfield  {title} {\bibinfo {title} {Probing the {{Universe}} with the
  {{Ly$\alpha$}} forest \textemdash{} {{I}}. {{Hydrodynamics}} of the
  low-density intergalactic medium},\ }\href
  {https://doi.org/10.1046/j.1365-8711.1998.01249.x} {\bibfield  {journal}
  {\bibinfo  {journal} {Monthly Notices of the Royal Astronomical Society}\
  }\textbf {\bibinfo {volume} {296}},\ \bibinfo {pages} {44} (\bibinfo {year}
  {1998})}\BibitemShut {NoStop}%
\bibitem [{\citenamefont {{Chabanier}}\ \emph {et~al.}(2024)\citenamefont
  {{Chabanier}}, \citenamefont {{Ravoux}}, \citenamefont {{Latrille}},
  \citenamefont {{Sexton}}, \citenamefont {{Armengaud}}, \citenamefont
  {{Bautista}}, \citenamefont {{Dumerchat}},\ and\ \citenamefont
  {{Luki{\'c}}}}]{chabanierAccel2Simulations2024}%
  \BibitemOpen
  \bibfield  {author} {\bibinfo {author} {\bibfnamefont {S.}~\bibnamefont
  {{Chabanier}}}, \bibinfo {author} {\bibfnamefont {C.}~\bibnamefont
  {{Ravoux}}}, \bibinfo {author} {\bibfnamefont {L.}~\bibnamefont
  {{Latrille}}}, \bibinfo {author} {\bibfnamefont {J.}~\bibnamefont
  {{Sexton}}}, \bibinfo {author} {\bibfnamefont {{\'E}.}~\bibnamefont
  {{Armengaud}}}, \bibinfo {author} {\bibfnamefont {J.}~\bibnamefont
  {{Bautista}}}, \bibinfo {author} {\bibfnamefont {T.}~\bibnamefont
  {{Dumerchat}}},\ and\ \bibinfo {author} {\bibfnamefont {Z.}~\bibnamefont
  {{Luki{\'c}}}},\ }\bibfield  {title} {\bibinfo {title} {{The ACCEL$^{2}$
  project: simulating Lyman-{\ensuremath{\alpha}} forest in large-volume
  hydrodynamical simulations}},\ }\href
  {https://doi.org/10.1093/mnras/stae2255} {\bibfield  {journal} {\bibinfo
  {journal} {\mnras}\ }\textbf {\bibinfo {volume} {534}},\ \bibinfo {pages}
  {2674} (\bibinfo {year} {2024})},\ \Eprint {https://arxiv.org/abs/2407.04473}
  {arXiv:2407.04473 [astro-ph.CO]} \BibitemShut {NoStop}%
\bibitem [{\citenamefont {Kirkby}\ \emph {et~al.}(2013)\citenamefont {Kirkby},
  \citenamefont {Margala}, \citenamefont {Slosar}, \citenamefont {Bailey},
  \citenamefont {Busca}, \citenamefont {Delubac}, \citenamefont {{James Rich}},
  \citenamefont {Bautista}, \citenamefont {Blomqvist}, \citenamefont
  {Brownstein}, \citenamefont {Carithers}, \citenamefont {Croft}, \citenamefont
  {Dawson}, \citenamefont {{Font-Ribera}}, \citenamefont
  {{Miralda-Escud{\'e}}}, \citenamefont {Myers}, \citenamefont {Nichol},
  \citenamefont {{Nathalie Palanque-Delabrouille}}, \citenamefont {P{\^a}ris},
  \citenamefont {Petitjean}, \citenamefont {Rossi}, \citenamefont {Schlegel},
  \citenamefont {Schneider}, \citenamefont {Viel}, \citenamefont {Weinberg},\
  and\ \citenamefont {Y{\`e}che}}]{kirkbyFittingMethodsBaryon2013}%
  \BibitemOpen
  \bibfield  {author} {\bibinfo {author} {\bibfnamefont {D.}~\bibnamefont
  {Kirkby}}, \bibinfo {author} {\bibfnamefont {D.}~\bibnamefont {Margala}},
  \bibinfo {author} {\bibfnamefont {A.}~\bibnamefont {Slosar}}, \bibinfo
  {author} {\bibfnamefont {S.}~\bibnamefont {Bailey}}, \bibinfo {author}
  {\bibfnamefont {N.~G.}\ \bibnamefont {Busca}}, \bibinfo {author}
  {\bibfnamefont {T.}~\bibnamefont {Delubac}}, \bibinfo {author} {\bibnamefont
  {{James Rich}}}, \bibinfo {author} {\bibfnamefont {J.~E.}\ \bibnamefont
  {Bautista}}, \bibinfo {author} {\bibfnamefont {M.}~\bibnamefont {Blomqvist}},
  \bibinfo {author} {\bibfnamefont {J.~R.}\ \bibnamefont {Brownstein}},
  \bibinfo {author} {\bibfnamefont {B.}~\bibnamefont {Carithers}}, \bibinfo
  {author} {\bibfnamefont {R.~A.~C.}\ \bibnamefont {Croft}}, \bibinfo {author}
  {\bibfnamefont {K.~S.}\ \bibnamefont {Dawson}}, \bibinfo {author}
  {\bibfnamefont {A.}~\bibnamefont {{Font-Ribera}}}, \bibinfo {author}
  {\bibfnamefont {J.}~\bibnamefont {{Miralda-Escud{\'e}}}}, \bibinfo {author}
  {\bibfnamefont {A.~D.}\ \bibnamefont {Myers}}, \bibinfo {author}
  {\bibfnamefont {R.~C.}\ \bibnamefont {Nichol}}, \bibinfo {author}
  {\bibnamefont {{Nathalie Palanque-Delabrouille}}}, \bibinfo {author}
  {\bibfnamefont {I.}~\bibnamefont {P{\^a}ris}}, \bibinfo {author}
  {\bibfnamefont {P.}~\bibnamefont {Petitjean}}, \bibinfo {author}
  {\bibfnamefont {G.}~\bibnamefont {Rossi}}, \bibinfo {author} {\bibfnamefont
  {D.~J.}\ \bibnamefont {Schlegel}}, \bibinfo {author} {\bibfnamefont {D.~P.}\
  \bibnamefont {Schneider}}, \bibinfo {author} {\bibfnamefont {M.}~\bibnamefont
  {Viel}}, \bibinfo {author} {\bibfnamefont {D.~H.}\ \bibnamefont {Weinberg}},\
  and\ \bibinfo {author} {\bibfnamefont {C.}~\bibnamefont {Y{\`e}che}},\
  }\bibfield  {title} {\bibinfo {title} {Fitting methods for baryon acoustic
  oscillations in the {{Lyman-$\alpha$}} forest fluctuations in {{BOSS}} data
  release 9},\ }\href {https://doi.org/10.1088/1475-7516/2013/03/024}
  {\bibfield  {journal} {\bibinfo  {journal} {\jcap}\ }\textbf {\bibinfo
  {volume} {2013}},\ \bibinfo {pages} {024} (\bibinfo {year}
  {2013})}\BibitemShut {NoStop}%
\bibitem [{\citenamefont {Talman}(1978)}]{talmanNumericalFourierBessel1978}%
  \BibitemOpen
  \bibfield  {author} {\bibinfo {author} {\bibfnamefont {J.~D.}\ \bibnamefont
  {Talman}},\ }\bibfield  {title} {\bibinfo {title} {Numerical {{Fourier}} and
  {{Bessel Transforms}} in {{Logarithmic Variables}}},\ }\href
  {https://doi.org/10.1016/0021-9991(78)90107-9} {\bibfield  {journal}
  {\bibinfo  {journal} {Journal of Computational Physics}\ }\textbf {\bibinfo
  {volume} {29}},\ \bibinfo {pages} {35} (\bibinfo {year} {1978})}\BibitemShut
  {NoStop}%
\bibitem [{\citenamefont
  {Hamilton}(2000)}]{hamiltonUncorrelatedModesNonlinear2000}%
  \BibitemOpen
  \bibfield  {author} {\bibinfo {author} {\bibfnamefont {A.~J.~S.}\
  \bibnamefont {Hamilton}},\ }\bibfield  {title} {\bibinfo {title}
  {Uncorrelated modes of the non-linear power spectrum},\ }\href
  {https://doi.org/10.1046/j.1365-8711.2000.03071.x} {\bibfield  {journal}
  {\bibinfo  {journal} {\mnras}\ }\textbf {\bibinfo {volume} {312}},\ \bibinfo
  {pages} {257} (\bibinfo {year} {2000})}\BibitemShut {NoStop}%
\bibitem [{\citenamefont {{Kara{\c{c}}ayl{\i}}}\ \emph
  {et~al.}(2020)\citenamefont {{Kara{\c{c}}ayl{\i}}}, \citenamefont
  {{Font-Ribera}},\ and\ \citenamefont
  {{Padmanabhan}}}]{karacayliOptimal1DLy2020}%
  \BibitemOpen
  \bibfield  {author} {\bibinfo {author} {\bibfnamefont {N.~G.}\ \bibnamefont
  {{Kara{\c{c}}ayl{\i}}}}, \bibinfo {author} {\bibfnamefont {A.}~\bibnamefont
  {{Font-Ribera}}},\ and\ \bibinfo {author} {\bibfnamefont {N.}~\bibnamefont
  {{Padmanabhan}}},\ }\bibfield  {title} {\bibinfo {title} {{Optimal 1D Ly
  {\ensuremath{\alpha}} forest power spectrum estimation - I. DESI-lite
  spectra}},\ }\href {https://doi.org/10.1093/mnras/staa2331} {\bibfield
  {journal} {\bibinfo  {journal} {\mnras}\ }\textbf {\bibinfo {volume} {497}},\
  \bibinfo {pages} {4742} (\bibinfo {year} {2020})},\ \Eprint
  {https://arxiv.org/abs/2008.06421} {arXiv:2008.06421 [astro-ph.CO]}
  \BibitemShut {NoStop}%
\bibitem [{\citenamefont {{Turner}}\ \emph {et~al.}(2024)\citenamefont
  {{Turner}}, \citenamefont {{Martini}}, \citenamefont {{Kara{\c{c}}ayl{\i}}},
  \citenamefont {{Aguilar}}, \citenamefont {{Ahlen}}, \citenamefont {{Brooks}},
  \citenamefont {{Claybaugh}}, \citenamefont {{de la Macorra}}, \citenamefont
  {{Dey}}, \citenamefont {{Doel}}, \citenamefont {{Fanning}}, \citenamefont
  {{Forero-Romero}}, \citenamefont {{Gontcho}}, \citenamefont
  {{Gonzalez-Morales}}, \citenamefont {{Gutierrez}}, \citenamefont {{Guy}},
  \citenamefont {{Herrera-Alcantar}}, \citenamefont {{Honscheid}},
  \citenamefont {{Juneau}}, \citenamefont {{Kisner}}, \citenamefont {{Kremin}},
  \citenamefont {{Lambert}}, \citenamefont {{Landriau}}, \citenamefont {{Le
  Guillou}}, \citenamefont {{Meisner}}, \citenamefont {{Miquel}}, \citenamefont
  {{Moustakas}}, \citenamefont {{Mueller}}, \citenamefont
  {{Mu{\~n}oz-Guti{\'e}rrez}}, \citenamefont {{Myers}}, \citenamefont {{Nie}},
  \citenamefont {{Niz}}, \citenamefont {{Poppett}}, \citenamefont {{Prada}},
  \citenamefont {{Rezaie}}, \citenamefont {{Rossi}}, \citenamefont {{Sanchez}},
  \citenamefont {{Schlafly}}, \citenamefont {{Schlegel}}, \citenamefont
  {{Schubnell}}, \citenamefont {{Seo}}, \citenamefont {{Sprayberry}},
  \citenamefont {{Tarl{\'e}}}, \citenamefont {{Weaver}},\ and\ \citenamefont
  {{Zou}}}]{turnerLyaForestMeanFluxFromDesiY12024}%
  \BibitemOpen
  \bibfield  {author} {\bibinfo {author} {\bibfnamefont {W.}~\bibnamefont
  {{Turner}}}, \bibinfo {author} {\bibfnamefont {P.}~\bibnamefont {{Martini}}},
  \bibinfo {author} {\bibfnamefont {N.~G.}\ \bibnamefont
  {{Kara{\c{c}}ayl{\i}}}}, \bibinfo {author} {\bibfnamefont {J.}~\bibnamefont
  {{Aguilar}}}, \bibinfo {author} {\bibfnamefont {S.}~\bibnamefont {{Ahlen}}},
  \bibinfo {author} {\bibfnamefont {D.}~\bibnamefont {{Brooks}}}, \bibinfo
  {author} {\bibfnamefont {T.}~\bibnamefont {{Claybaugh}}}, \bibinfo {author}
  {\bibfnamefont {A.}~\bibnamefont {{de la Macorra}}}, \bibinfo {author}
  {\bibfnamefont {A.}~\bibnamefont {{Dey}}}, \bibinfo {author} {\bibfnamefont
  {P.}~\bibnamefont {{Doel}}}, \bibinfo {author} {\bibfnamefont
  {K.}~\bibnamefont {{Fanning}}}, \bibinfo {author} {\bibfnamefont {J.~E.}\
  \bibnamefont {{Forero-Romero}}}, \bibinfo {author} {\bibfnamefont {S.~G.~A.}\
  \bibnamefont {{Gontcho}}}, \bibinfo {author} {\bibfnamefont {A.~X.}\
  \bibnamefont {{Gonzalez-Morales}}}, \bibinfo {author} {\bibfnamefont
  {G.}~\bibnamefont {{Gutierrez}}}, \bibinfo {author} {\bibfnamefont
  {J.}~\bibnamefont {{Guy}}}, \bibinfo {author} {\bibfnamefont {H.~K.}\
  \bibnamefont {{Herrera-Alcantar}}}, \bibinfo {author} {\bibfnamefont
  {K.}~\bibnamefont {{Honscheid}}}, \bibinfo {author} {\bibfnamefont
  {S.}~\bibnamefont {{Juneau}}}, \bibinfo {author} {\bibfnamefont
  {T.}~\bibnamefont {{Kisner}}}, \bibinfo {author} {\bibfnamefont
  {A.}~\bibnamefont {{Kremin}}}, \bibinfo {author} {\bibfnamefont
  {A.}~\bibnamefont {{Lambert}}}, \bibinfo {author} {\bibfnamefont
  {M.}~\bibnamefont {{Landriau}}}, \bibinfo {author} {\bibfnamefont
  {L.}~\bibnamefont {{Le Guillou}}}, \bibinfo {author} {\bibfnamefont
  {A.}~\bibnamefont {{Meisner}}}, \bibinfo {author} {\bibfnamefont
  {R.}~\bibnamefont {{Miquel}}}, \bibinfo {author} {\bibfnamefont
  {J.}~\bibnamefont {{Moustakas}}}, \bibinfo {author} {\bibfnamefont
  {E.}~\bibnamefont {{Mueller}}}, \bibinfo {author} {\bibfnamefont
  {A.}~\bibnamefont {{Mu{\~n}oz-Guti{\'e}rrez}}}, \bibinfo {author}
  {\bibfnamefont {A.~D.}\ \bibnamefont {{Myers}}}, \bibinfo {author}
  {\bibfnamefont {J.}~\bibnamefont {{Nie}}}, \bibinfo {author} {\bibfnamefont
  {G.}~\bibnamefont {{Niz}}}, \bibinfo {author} {\bibfnamefont
  {C.}~\bibnamefont {{Poppett}}}, \bibinfo {author} {\bibfnamefont
  {F.}~\bibnamefont {{Prada}}}, \bibinfo {author} {\bibfnamefont
  {M.}~\bibnamefont {{Rezaie}}}, \bibinfo {author} {\bibfnamefont
  {G.}~\bibnamefont {{Rossi}}}, \bibinfo {author} {\bibfnamefont
  {E.}~\bibnamefont {{Sanchez}}}, \bibinfo {author} {\bibfnamefont {E.~F.}\
  \bibnamefont {{Schlafly}}}, \bibinfo {author} {\bibfnamefont
  {D.}~\bibnamefont {{Schlegel}}}, \bibinfo {author} {\bibfnamefont
  {M.}~\bibnamefont {{Schubnell}}}, \bibinfo {author} {\bibfnamefont
  {H.}~\bibnamefont {{Seo}}}, \bibinfo {author} {\bibfnamefont
  {D.}~\bibnamefont {{Sprayberry}}}, \bibinfo {author} {\bibfnamefont
  {G.}~\bibnamefont {{Tarl{\'e}}}}, \bibinfo {author} {\bibfnamefont {B.~A.}\
  \bibnamefont {{Weaver}}},\ and\ \bibinfo {author} {\bibfnamefont
  {H.}~\bibnamefont {{Zou}}},\ }\bibfield  {title} {\bibinfo {title} {{New
  Measurements of the Ly{\ensuremath{\alpha}} Forest Continuum and Effective
  Optical Depth with LyCAN and DESI Y1 Data}},\ }\href
  {https://doi.org/10.3847/1538-4357/ad8239} {\bibfield  {journal} {\bibinfo
  {journal} {\apj}\ }\textbf {\bibinfo {volume} {976}},\ \bibinfo {eid} {143}
  (\bibinfo {year} {2024})},\ \Eprint {https://arxiv.org/abs/2405.06743}
  {arXiv:2405.06743 [astro-ph.CO]} \BibitemShut {NoStop}%
\bibitem [{\citenamefont {Hoffman}\ and\ \citenamefont
  {Gelman}(2014)}]{hoffman_no-u-turn_2014}%
  \BibitemOpen
  \bibfield  {author} {\bibinfo {author} {\bibfnamefont {M.~D.}\ \bibnamefont
  {Hoffman}}\ and\ \bibinfo {author} {\bibfnamefont {A.}~\bibnamefont
  {Gelman}},\ }\bibfield  {title} {\bibinfo {title} {The {No}-{U}-{Turn}
  {Sampler}: {Adaptively} {Setting} {Path} {Lengths} in {Hamiltonian} {Monte}
  {Carlo}},\ }\href {http://jmlr.org/papers/v15/hoffman14a.html} {\bibfield
  {journal} {\bibinfo  {journal} {Journal of Machine Learning Research}\
  }\textbf {\bibinfo {volume} {15}},\ \bibinfo {pages} {1593} (\bibinfo {year}
  {2014})}\BibitemShut {NoStop}%
\bibitem [{\citenamefont {Bingham}\ \emph {et~al.}(2019)\citenamefont
  {Bingham}, \citenamefont {Chen}, \citenamefont {Jankowiak}, \citenamefont
  {Obermeyer}, \citenamefont {Pradhan}, \citenamefont {Karaletsos},
  \citenamefont {Singh}, \citenamefont {Szerlip}, \citenamefont {Horsfall},\
  and\ \citenamefont {Goodman}}]{bingham_pyro_2019}%
  \BibitemOpen
  \bibfield  {author} {\bibinfo {author} {\bibfnamefont {E.}~\bibnamefont
  {Bingham}}, \bibinfo {author} {\bibfnamefont {J.~P.}\ \bibnamefont {Chen}},
  \bibinfo {author} {\bibfnamefont {M.}~\bibnamefont {Jankowiak}}, \bibinfo
  {author} {\bibfnamefont {F.}~\bibnamefont {Obermeyer}}, \bibinfo {author}
  {\bibfnamefont {N.}~\bibnamefont {Pradhan}}, \bibinfo {author} {\bibfnamefont
  {T.}~\bibnamefont {Karaletsos}}, \bibinfo {author} {\bibfnamefont
  {R.}~\bibnamefont {Singh}}, \bibinfo {author} {\bibfnamefont
  {P.}~\bibnamefont {Szerlip}}, \bibinfo {author} {\bibfnamefont
  {P.}~\bibnamefont {Horsfall}},\ and\ \bibinfo {author} {\bibfnamefont
  {N.~D.}\ \bibnamefont {Goodman}},\ }\bibfield  {title} {\bibinfo {title}
  {Pyro: {Deep} {Universal} {Probabilistic} {Programming}},\ }\href
  {http://jmlr.org/papers/v20/18-403.html} {\bibfield  {journal} {\bibinfo
  {journal} {Journal of Machine Learning Research}\ }\textbf {\bibinfo {volume}
  {20}},\ \bibinfo {pages} {1} (\bibinfo {year} {2019})}\BibitemShut {NoStop}%
\bibitem [{\citenamefont {Phan}\ \emph {et~al.}(2019)\citenamefont {Phan},
  \citenamefont {Pradhan},\ and\ \citenamefont
  {Jankowiak}}]{phan_numpyro_2019}%
  \BibitemOpen
  \bibfield  {author} {\bibinfo {author} {\bibfnamefont {D.}~\bibnamefont
  {Phan}}, \bibinfo {author} {\bibfnamefont {N.}~\bibnamefont {Pradhan}},\ and\
  \bibinfo {author} {\bibfnamefont {M.}~\bibnamefont {Jankowiak}},\ }\href
  {https://doi.org/10.48550/arXiv.1912.11554} {\bibinfo {title} {Composable
  {Effects} for {Flexible} and {Accelerated} {Probabilistic} {Programming} in
  {NumPyro}}} (\bibinfo {year} {2019})\BibitemShut {NoStop}%
\bibitem [{\citenamefont {Bradbury}\ \emph {et~al.}(2018)\citenamefont
  {Bradbury}, \citenamefont {Frostig}, \citenamefont {Hawkins}, \citenamefont
  {Johnson}, \citenamefont {Leary}, \citenamefont {Maclaurin}, \citenamefont
  {Necula}, \citenamefont {Paszke}, \citenamefont {Vander{P}las}, \citenamefont
  {Wanderman-{M}ilne},\ and\ \citenamefont {Zhang}}]{jax2018github}%
  \BibitemOpen
  \bibfield  {author} {\bibinfo {author} {\bibfnamefont {J.}~\bibnamefont
  {Bradbury}}, \bibinfo {author} {\bibfnamefont {R.}~\bibnamefont {Frostig}},
  \bibinfo {author} {\bibfnamefont {P.}~\bibnamefont {Hawkins}}, \bibinfo
  {author} {\bibfnamefont {M.~J.}\ \bibnamefont {Johnson}}, \bibinfo {author}
  {\bibfnamefont {C.}~\bibnamefont {Leary}}, \bibinfo {author} {\bibfnamefont
  {D.}~\bibnamefont {Maclaurin}}, \bibinfo {author} {\bibfnamefont
  {G.}~\bibnamefont {Necula}}, \bibinfo {author} {\bibfnamefont
  {A.}~\bibnamefont {Paszke}}, \bibinfo {author} {\bibfnamefont
  {J.}~\bibnamefont {Vander{P}las}}, \bibinfo {author} {\bibfnamefont
  {S.}~\bibnamefont {Wanderman-{M}ilne}},\ and\ \bibinfo {author}
  {\bibfnamefont {Q.}~\bibnamefont {Zhang}},\ }\href
  {http://github.com/jax-ml/jax} {\bibinfo {title} {{JAX}: composable
  transformations of {P}ython+{N}um{P}y programs}} (\bibinfo {year}
  {2018})\BibitemShut {NoStop}%
\bibitem [{\citenamefont {{Neal}}(2011)}]{nealMcmcHamiltonianMonteCarlo2011}%
  \BibitemOpen
  \bibfield  {author} {\bibinfo {author} {\bibfnamefont {R.}~\bibnamefont
  {{Neal}}},\ }\bibfield  {title} {\bibinfo {title} {{MCMC Using Hamiltonian
  Dynamics}},\ }in\ \href {https://doi.org/10.1201/b10905} {\emph {\bibinfo
  {booktitle} {Handbook of Markov Chain Monte Carlo}}},\ \bibinfo {editor}
  {edited by\ \bibinfo {editor} {\bibfnamefont {S.}~\bibnamefont {Brooks}},
  \bibinfo {editor} {\bibfnamefont {A.}~\bibnamefont {Gelman}}, \bibinfo
  {editor} {\bibfnamefont {G.}~\bibnamefont {Jones}},\ and\ \bibinfo {editor}
  {\bibfnamefont {X.-L.}\ \bibnamefont {Meng}}}\ (\bibinfo  {publisher}
  {Chapman and Hall/CRC},\ \bibinfo {year} {2011})\ pp.\ \bibinfo {pages}
  {113--162}\BibitemShut {NoStop}%
\bibitem [{\citenamefont {Harris}\ \emph {et~al.}(2020)\citenamefont {Harris},
  \citenamefont {Millman}, \citenamefont {van~der Walt}, \citenamefont
  {Gommers}, \citenamefont {Virtanen}, \citenamefont {Cournapeau},
  \citenamefont {Wieser}, \citenamefont {Taylor}, \citenamefont {Berg},
  \citenamefont {Smith}, \citenamefont {Kern}, \citenamefont {Picus},
  \citenamefont {Hoyer}, \citenamefont {van Kerkwijk}, \citenamefont {Brett},
  \citenamefont {Haldane}, \citenamefont {del R{\'{i}}o}, \citenamefont
  {Wiebe}, \citenamefont {Peterson}, \citenamefont {G{\'{e}}rard-Marchant},
  \citenamefont {Sheppard}, \citenamefont {Reddy}, \citenamefont {Weckesser},
  \citenamefont {Abbasi}, \citenamefont {Gohlke},\ and\ \citenamefont
  {Oliphant}}]{numpy}%
  \BibitemOpen
  \bibfield  {author} {\bibinfo {author} {\bibfnamefont {C.~R.}\ \bibnamefont
  {Harris}}, \bibinfo {author} {\bibfnamefont {K.~J.}\ \bibnamefont {Millman}},
  \bibinfo {author} {\bibfnamefont {S.~J.}\ \bibnamefont {van~der Walt}},
  \bibinfo {author} {\bibfnamefont {R.}~\bibnamefont {Gommers}}, \bibinfo
  {author} {\bibfnamefont {P.}~\bibnamefont {Virtanen}}, \bibinfo {author}
  {\bibfnamefont {D.}~\bibnamefont {Cournapeau}}, \bibinfo {author}
  {\bibfnamefont {E.}~\bibnamefont {Wieser}}, \bibinfo {author} {\bibfnamefont
  {J.}~\bibnamefont {Taylor}}, \bibinfo {author} {\bibfnamefont
  {S.}~\bibnamefont {Berg}}, \bibinfo {author} {\bibfnamefont {N.~J.}\
  \bibnamefont {Smith}}, \bibinfo {author} {\bibfnamefont {R.}~\bibnamefont
  {Kern}}, \bibinfo {author} {\bibfnamefont {M.}~\bibnamefont {Picus}},
  \bibinfo {author} {\bibfnamefont {S.}~\bibnamefont {Hoyer}}, \bibinfo
  {author} {\bibfnamefont {M.~H.}\ \bibnamefont {van Kerkwijk}}, \bibinfo
  {author} {\bibfnamefont {M.}~\bibnamefont {Brett}}, \bibinfo {author}
  {\bibfnamefont {A.}~\bibnamefont {Haldane}}, \bibinfo {author} {\bibfnamefont
  {J.~F.}\ \bibnamefont {del R{\'{i}}o}}, \bibinfo {author} {\bibfnamefont
  {M.}~\bibnamefont {Wiebe}}, \bibinfo {author} {\bibfnamefont
  {P.}~\bibnamefont {Peterson}}, \bibinfo {author} {\bibfnamefont
  {P.}~\bibnamefont {G{\'{e}}rard-Marchant}}, \bibinfo {author} {\bibfnamefont
  {K.}~\bibnamefont {Sheppard}}, \bibinfo {author} {\bibfnamefont
  {T.}~\bibnamefont {Reddy}}, \bibinfo {author} {\bibfnamefont
  {W.}~\bibnamefont {Weckesser}}, \bibinfo {author} {\bibfnamefont
  {H.}~\bibnamefont {Abbasi}}, \bibinfo {author} {\bibfnamefont
  {C.}~\bibnamefont {Gohlke}},\ and\ \bibinfo {author} {\bibfnamefont {T.~E.}\
  \bibnamefont {Oliphant}},\ }\bibfield  {title} {\bibinfo {title} {Array
  programming with {NumPy}},\ }\href
  {https://doi.org/10.1038/s41586-020-2649-2} {\bibfield  {journal} {\bibinfo
  {journal} {Nature}\ }\textbf {\bibinfo {volume} {585}},\ \bibinfo {pages}
  {357} (\bibinfo {year} {2020})}\BibitemShut {NoStop}%
\bibitem [{\citenamefont {Hunter}(2007)}]{matplotlib}%
  \BibitemOpen
  \bibfield  {author} {\bibinfo {author} {\bibfnamefont {J.~D.}\ \bibnamefont
  {Hunter}},\ }\bibfield  {title} {\bibinfo {title} {Matplotlib: A 2d graphics
  environment},\ }\href {https://doi.org/10.1109/MCSE.2007.55} {\bibfield
  {journal} {\bibinfo  {journal} {Computing in Science \& Engineering}\
  }\textbf {\bibinfo {volume} {9}},\ \bibinfo {pages} {90} (\bibinfo {year}
  {2007})}\BibitemShut {NoStop}%
\bibitem [{\citenamefont {{McDonald}}\ \emph {et~al.}(2005)\citenamefont
  {{McDonald}}, \citenamefont {{Seljak}}, \citenamefont {{Cen}}, \citenamefont
  {{Shih}}, \citenamefont {{Weinberg}}, \citenamefont {{Burles}}, \citenamefont
  {{Schneider}}, \citenamefont {{Schlegel}}, \citenamefont {{Bahcall}},
  \citenamefont {{Briggs}}, \citenamefont {{Brinkmann}}, \citenamefont
  {{Fukugita}}, \citenamefont {{Ivezi{\'c}}}, \citenamefont {{Kent}},\ and\
  \citenamefont {{Vanden Berk}}}]{mcdonaldLinearTheoryLyaSdss2005}%
  \BibitemOpen
  \bibfield  {author} {\bibinfo {author} {\bibfnamefont {P.}~\bibnamefont
  {{McDonald}}}, \bibinfo {author} {\bibfnamefont {U.}~\bibnamefont
  {{Seljak}}}, \bibinfo {author} {\bibfnamefont {R.}~\bibnamefont {{Cen}}},
  \bibinfo {author} {\bibfnamefont {D.}~\bibnamefont {{Shih}}}, \bibinfo
  {author} {\bibfnamefont {D.~H.}\ \bibnamefont {{Weinberg}}}, \bibinfo
  {author} {\bibfnamefont {S.}~\bibnamefont {{Burles}}}, \bibinfo {author}
  {\bibfnamefont {D.~P.}\ \bibnamefont {{Schneider}}}, \bibinfo {author}
  {\bibfnamefont {D.~J.}\ \bibnamefont {{Schlegel}}}, \bibinfo {author}
  {\bibfnamefont {N.~A.}\ \bibnamefont {{Bahcall}}}, \bibinfo {author}
  {\bibfnamefont {J.~W.}\ \bibnamefont {{Briggs}}}, \bibinfo {author}
  {\bibfnamefont {J.}~\bibnamefont {{Brinkmann}}}, \bibinfo {author}
  {\bibfnamefont {M.}~\bibnamefont {{Fukugita}}}, \bibinfo {author}
  {\bibfnamefont {{\v{Z}}.}~\bibnamefont {{Ivezi{\'c}}}}, \bibinfo {author}
  {\bibfnamefont {S.}~\bibnamefont {{Kent}}},\ and\ \bibinfo {author}
  {\bibfnamefont {D.~E.}\ \bibnamefont {{Vanden Berk}}},\ }\bibfield  {title}
  {\bibinfo {title} {{The Linear Theory Power Spectrum from the
  Ly{\ensuremath{\alpha}} Forest in the Sloan Digital Sky Survey}},\ }\href
  {https://doi.org/10.1086/497563} {\bibfield  {journal} {\bibinfo  {journal}
  {\apj}\ }\textbf {\bibinfo {volume} {635}},\ \bibinfo {pages} {761} (\bibinfo
  {year} {2005})},\ \Eprint {https://arxiv.org/abs/astro-ph/0407377}
  {arXiv:astro-ph/0407377 [astro-ph]} \BibitemShut {NoStop}%
\bibitem [{\citenamefont {{Pedersen}}\ \emph {et~al.}(2021)\citenamefont
  {{Pedersen}}, \citenamefont {{Font-Ribera}}, \citenamefont {{Rogers}},
  \citenamefont {{McDonald}}, \citenamefont {{Peiris}}, \citenamefont
  {{Pontzen}},\ and\ \citenamefont {{Slosar}}}]{pedersenEmulator2021}%
  \BibitemOpen
  \bibfield  {author} {\bibinfo {author} {\bibfnamefont {C.}~\bibnamefont
  {{Pedersen}}}, \bibinfo {author} {\bibfnamefont {A.}~\bibnamefont
  {{Font-Ribera}}}, \bibinfo {author} {\bibfnamefont {K.~K.}\ \bibnamefont
  {{Rogers}}}, \bibinfo {author} {\bibfnamefont {P.}~\bibnamefont
  {{McDonald}}}, \bibinfo {author} {\bibfnamefont {H.~V.}\ \bibnamefont
  {{Peiris}}}, \bibinfo {author} {\bibfnamefont {A.}~\bibnamefont
  {{Pontzen}}},\ and\ \bibinfo {author} {\bibfnamefont {A.}~\bibnamefont
  {{Slosar}}},\ }\bibfield  {title} {\bibinfo {title} {{An emulator for the
  Lyman-{\ensuremath{\alpha}} forest in beyond-{\ensuremath{\Lambda}}CDM
  cosmologies}},\ }\href {https://doi.org/10.1088/1475-7516/2021/05/033}
  {\bibfield  {journal} {\bibinfo  {journal} {\jcap}\ }\textbf {\bibinfo
  {volume} {2021}},\ \bibinfo {eid} {033} (\bibinfo {year} {2021})},\ \Eprint
  {https://arxiv.org/abs/2011.15127} {arXiv:2011.15127 [astro-ph.CO]}
  \BibitemShut {NoStop}%
\bibitem [{\citenamefont {{Bird}}\ \emph {et~al.}(2011)\citenamefont {{Bird}},
  \citenamefont {{Peiris}}, \citenamefont {{Viel}},\ and\ \citenamefont
  {{Verde}}}]{birdMinimallyParametricReconstructionLya2011}%
  \BibitemOpen
  \bibfield  {author} {\bibinfo {author} {\bibfnamefont {S.}~\bibnamefont
  {{Bird}}}, \bibinfo {author} {\bibfnamefont {H.~V.}\ \bibnamefont
  {{Peiris}}}, \bibinfo {author} {\bibfnamefont {M.}~\bibnamefont {{Viel}}},\
  and\ \bibinfo {author} {\bibfnamefont {L.}~\bibnamefont {{Verde}}},\
  }\bibfield  {title} {\bibinfo {title} {{Minimally parametric power spectrum
  reconstruction from the Lyman {\ensuremath{\alpha}} forest}},\ }\href
  {https://doi.org/10.1111/j.1365-2966.2011.18245.x} {\bibfield  {journal}
  {\bibinfo  {journal} {\mnras}\ }\textbf {\bibinfo {volume} {413}},\ \bibinfo
  {pages} {1717} (\bibinfo {year} {2011})},\ \Eprint
  {https://arxiv.org/abs/1010.1519} {arXiv:1010.1519 [astro-ph.CO]}
  \BibitemShut {NoStop}%
\bibitem [{\citenamefont {{Ivanov}}(2024)}]{ivanovEffectiveLya2024}%
  \BibitemOpen
  \bibfield  {author} {\bibinfo {author} {\bibfnamefont {M.~M.}\ \bibnamefont
  {{Ivanov}}},\ }\bibfield  {title} {\bibinfo {title} {{Lyman alpha forest
  power spectrum in effective field theory}},\ }\href
  {https://doi.org/10.1103/PhysRevD.109.023507} {\bibfield  {journal} {\bibinfo
   {journal} {\prd}\ }\textbf {\bibinfo {volume} {109}},\ \bibinfo {eid}
  {023507} (\bibinfo {year} {2024})},\ \Eprint
  {https://arxiv.org/abs/2309.10133} {arXiv:2309.10133 [astro-ph.CO]}
  \BibitemShut {NoStop}%
\end{thebibliography}%

\end{document}